\documentclass[11pt]{article}
\usepackage{geometry}                
\usepackage{pifont}
\usepackage{xcolor}           
\usepackage{tikz}
\usepackage{soul}
\usepackage{enumitem}
\usepackage{marvosym}
\usepackage{tikzsymbols}
\usepackage{stfloats}
\usepackage{verbatim}
\usepackage{booktabs}
\usepackage{multirow,multicol}
\usepackage{adjustbox}
\usepackage{array}
\usetikzlibrary{patterns}
\usetikzlibrary{matrix}
\usetikzlibrary{tikzmark}
\usetikzlibrary{decorations.markings} 
\usetikzlibrary{calc,arrows} 
\usepackage{comment}
\tikzstyle{arrowmid}[0.75]=[decoration= 
{markings, 
	mark=at position #1 with {\arrow{>}}}, 
postaction={decorate}] 

\usetikzlibrary{fit}
\usetikzlibrary{shadows.blur}
\usetikzlibrary{shapes.symbols}
\usepackage{menukeys}
\usetikzlibrary{shapes.geometric, arrows}
\usepackage{cite}
\usepackage{caption}
\usepackage{mdframed}
\usepackage{subcaption}
\usetikzlibrary{positioning}
\usepackage{cancel}
\usepackage{amssymb, amsmath, amsthm}
\usepackage{dsfont}
\usepackage{epstopdf}
\usepackage{algorithmic}
\usepackage{textcomp}
\usepackage{xcolor}
\usepackage{comment}
\def\BibTeX{{\rm B\kern-.05em{\sc i\kern-.025em b}\kern-.08em
		T\kern-.1667em\lower.7ex\hbox{E}\kern-.125emX}}
\tikzstyle{arrow} = [thick,->,>=stealth]
\newtheoremstyle{mytheoremstyle}{0pt}{0pt}{\itshape}{}{\bfseries}{.}{.5em}{} 
\newtheorem{definition}{Definition}
\newtheorem{theorem}{Theorem}
\newtheorem{corollary}{Corollary}

\newtheorem{lemma}{Lemma}

\newtheorem{example}{Example}

\def\df{{\mbox{\tiny \normalfont DF}}}
\newcommand{\E}{\mathrm{E}}

\newcommand{\RN}[1]{%
	\textup{\uppercase\expandafter{\romannumeral#1}}%
}

\newcommand{\cmark}{\ding{51}}
\newcommand{\xmark}{\ding{55}}

\begin{document}
	\title{Sum-GDoF of Symmetric Multi-hop Interference Channel under Finite Precision CSIT using Aligned-Images Sumset Inequalities}
	\author{Junge Wang and Syed A. Jafar\\[0.3cm]
	Center for Pervasive Communications and Computing (CPCC), UC Irvine\\
	Email:\{jungew, syed\}@uci.edu\\[0.3cm]}
	\date{}
	\maketitle
	\begin{abstract}
Aligned-Images Sumset Inequalities are used in this work to study the Generalized Degrees of Freedom (GDoF) of the symmetric layered multi-hop interference channel under the robust assumption that the channel state information at the transmitters (CSIT) is limited to finite precision. First, the sum-GDoF value is characterized for the $2\times 2\times 2$ setting that is comprised of $2$ sources, $2$ relays, and $2$ destinations. It is shown that the sum-GDoF do not improve even if perfect CSIT is allowed in the first hop, as long as the CSIT in the second hop is limited to finite precision. The sum GDoF characterization is then generalized to the $2\times 2\times \cdots \times 2$ setting that is comprised of $L$ hops. Remarkably, for large $L$, the GDoF value approaches that of the one hop broadcast channel that is obtained by full cooperation among the two transmitters of the last hop, with finite precision CSIT. Previous  studies of multi-hop interference networks  either identified sophisticated GDoF optimal schemes under perfect CSIT, such as aligned interference neutralization and network diagonalization, that are  powerful in theory but too fragile to be practical, or studied robust achievable schemes like classical amplify/decode/compress-and-forward without  claims of information-theoretic optimality. In contrast, under finite precision CSIT, we show that  the benefits of fragile schemes are lost, while a combination of classical random coding schemes that are simpler and much more robust, namely a rate-splitting between decode-and-forward and amplify-and-forward, is shown to be GDoF optimal. As such, this work represents another step towards bridging the gap between theory (optimality) and practice (robustness) with the aid of Aligned-Images Sumset Inequalities. 
\end{abstract}
	\newpage
	\section{Introduction}
	There is much interest in multihop interference networks due to their essential role in expanding  coverage and enabling high data rates over underutilized (e.g., mm-wave/THz) frequency bands that suffer from high path loss and blockages. However, an information theoretic understanding of the robust fundamental limits of such networks remains elusive, even in the approximate or asymptotic (high SNR) sense. Information theoretic studies of multihop interference networks, such as those in \cite{Cao_Chen,Simeone_mesh, Mohajer_Diggavi_Fragouli_Tse, Issa_Fong_Avestimehr_linear_twohop, Shomorony_Avestimehr,  Gou_Wang_Jafar_Jeon_Chung,  Wang_Gou_Jafar_MU, Gou_Wang_Jafar_NL, Shomorony_Avestimehr_K, MMK, Jafar_Shamai, Cadambe_Jafar_X, Cadambe_Jafar_XFB}, have focused primarily on the idealized setting where the channel state information at the transmitters (CSIT) is perfect. The search for optimal solutions under idealized assumptions leads to ideas like Interference Neutralization \cite{Cao_Chen,Simeone_mesh, Mohajer_Diggavi_Fragouli_Tse, Issa_Fong_Avestimehr_linear_twohop, Shomorony_Avestimehr}, Aligned Interference Neutralization \cite{Gou_Wang_Jafar_Jeon_Chung, Wang_Gou_Jafar_MU, Gou_Wang_Jafar_NL} and Network Diagonalization \cite{Shomorony_Avestimehr_K} that are  powerful in theory (e.g., everyone gets all the cake), but  too fragile to be relevant in practice, where CSIT is only available to finite precision. For example, Gou et al. introduced in \cite{Gou_Wang_Jafar_Jeon_Chung}  an aligned interference neutralization scheme for the layered $2\times 2\times 2$ interference channel which is comprised of two source nodes, two relay nodes and two destination nodes,  that achieves the sum Degrees of Freedom (DoF) value of $2$ under perfect CSIT. This is trivially optimal because even if all interference is eliminated, each user by itself cannot achieve more than $1$ DoF --- a straightforward consequence of the min-cut max-flow bound. The result is generalized to the $K\times K\times K$ setting in \cite{Shomorony_Avestimehr_K} where a network diagonalization scheme is shown to achieve $K$ DoF, also trivially optimal for the same reason. Such schemes,  that are based on precise alignment and/or neutralization of signals, are difficult to translate to practice because the residual interference due to imperfections in CSIT can be severely detrimental. Under perfect CSIT, even constrained alternatives like decode-and-forward, which can achieve $4/3$ DoF by treating each hop as a $2\times 2$ $X$ channel \cite{MMK, Jafar_Shamai, Cadambe_Jafar_X,  Cadambe_Jafar_XFB}, are too fragile as they rely strongly on infinitely precise CSIT to achieve perfect interference alignment.  Besides the assumption of perfect CSIT, another limitation of many of these works, e.g., \cite{Gou_Wang_Jafar_Jeon_Chung,  Wang_Gou_Jafar_MU, Gou_Wang_Jafar_NL, Shomorony_Avestimehr_K, MMK, Jafar_Shamai, Cadambe_Jafar_X, Cadambe_Jafar_XFB}, is that their focus is limited to the DoF metric which implicitly assumes that all non-zero channels are equally strong (every non-zero link can carry exactly $1$ DoF). To overcome this limitation, the Generalized Degrees of Freedom (GDoF) framework was introduced in \cite{Etkin_Tse_Wang}, which is  capable of representing weak and strong interference conditions and is the critical stepping stone to approximate capacity characterizations \cite{Etkin_Tse_Wang, Avestimehr_Diggavi_Tse, ADT_FnT, Geng_TIN, Avestimehr_Sezgin_Tse, Karmakar_Varanasi_gap, Niesen_Maddah_Ali_X}. Evidently, for a robust information-theoretic understanding of multihop interference networks it is important to study their GDoF under finite precision CSIT.

Despite the early recognition of their importance \cite{Lapidoth_Shamai_Wigger_BC}, network GDoF characterizations under finite precision CSIT have been generally intractable until recently, mainly due to the difficulty of obtaining tight information theoretic outer bounds under CSIT limitations. Indeed, the strongest achievable schemes under finite precision CSIT tend to be robust random coding schemes that are relatively well understood. Note that this is in sharp contrast to GDoF studies under \emph{perfect} CSIT, where the outer bounds tend to be relatively straightforward (e.g., min-cut max-flow bounds) and the main challenge is the construction of sophisticated achievable schemes based on alignment and neutralization of signals. Under finite precision CSIT, it is the outer bounds that tend to be challenging because they need to rule out the potential benefits of all forms of signal alignments  that are possible under perfect CSIT but  fail under limited CSIT. Since received signals in an interference network are sums (linear combinations) of transmitted signals up to noise distortion, bounding the potential benefits of signal alignments amounts to bounding the size (entropy) of sum-sets (received signals), an inherently combinatorial endeavor that marks a seemingly necessary departure from  the elegance of classical information theoretic arguments. This is indeed the approach taken by the so called Aligned Images (AI, in short) bounds that were introduced in \cite{Arash_Jafar} and recently expanded significantly  in scope to a broad class of sumset inequalities in \cite{Arash_Jafar_sumset}. AI Sumset Inequalities have been applied successfully  to find  GDoF characterizations under finite precision CSIT for a variety of single-hop interference and broadcast settings \cite{Arash_Jafar_IC, Arash_Jafar_MIMOICGDoF, Arash_Jafar_SLS, Arash_Bofeng_Jafar_BC, Arash_Jafar_MIMOBC, Arash_Jafar_cooperation, Yoga_Junge_Jafar,  Junge_Yuan_Huang_Jafar}. On the other hand, recent observations in \cite{Chan_Jafar_3to1}  indicate that further generalizations of the AI sumset inequalities may be needed beyond  \cite{Arash_Jafar_sumset}. Given this relatively new but limited set of tools that have yet to be applied to multihop settings, the extent of their utility for multihop interference networks in particular remains an interesting open question. It is this open question that motivates our work in this paper. An overview of our results is provided next.

To avoid the curse of dimensionality we begin our GDoF study with a symmetric, layered,  $2$-hop interference network, denoted as a $2\times 2\times 2$ setting, which is comprised of $2$ source nodes, $2$ relay nodes, and $2$ destination nodes. Each hop is a $2\times 2$ network, where the direct links are capable of carrying $1$ GDoF, and the cross-links are capable of carrying $\alpha$ GDoF. Since well-designed networks invariably operate in the weak-interference regime, our primary focus is on the weak interference regime ($\alpha<1$), although extensions to strong-interference are straightforward in this case.  As we  apply AI sumset inequalities to this setting, an immediate challenge manifests in the critical first step. All prior applications of AI bounds begin by transforming the channel to a deterministic model by a sequence of steps that include removing the Additive White Gaussian Noise (AWGN) and  quantizing the noise-less received signals. This transformation works well for one-hop settings because it can be shown that all the steps involved can collectively only contribute a bounded distortion that is inconsequential in the GDoF sense. However, the same deterministic transformation is difficult to justify in a multi-hop setting. This is because  the relays are free to choose \emph{arbitrary} mappings from their input signals to their output signals, but for arbitrary mappings, a bounded distortion of inputs does not necessarily correspond to a bounded distortion of their corresponding outputs. Fortunately, we are able to overcome this obstacle by realizing that a valid outer bound is obtained if we allow perfect CSIT for the first hop and finite precision CSIT for just the second hop. This requires the deterministic transformation only for the second hop, i.e., only the \emph{outputs} of the relays are distorted and not their inputs. Surprisingly, this outer bound is found to be achievable even with only finite precision CSIT for both hops. Specifically, using the compact GDoF expression available from the outer bound for insights, we are able to construct an achievable scheme that uses rate-splitting between amplify-and-forward and decode-and-forward strategies to match the outer bound. This settles the GDoF of the $2\times 2\times 2$ setting with finite precision CSIT in both hops, and also shows as a byproduct that the GDoF do not improve even if perfect CSIT is allowed in the first hop. While the proof is non-trivial, it is notable the AI Sumset Inequalities of \cite{Arash_Jafar_sumset} turn out to be sufficient for a tight GDoF characterization in this case. The optimal sum-GDoF value  appears in Theorem \ref{thm:2hop} in Section \ref{sec:2hopweak}. Also notable is that the results automatically translate to strong interference settings simply by switching the labels of the relays. This extension appears as Corollary \ref{cor:2hop} in Section \ref{sec:2hopstrong}. 

Next we generalize the setting to a symmetric  layered $L$-hop interference network, denoted as a $2\times 2\times \cdots\times 2$ setting. Here the idea of allowing perfect CSIT in all but the last hop does not work because the resulting bound would be loose for $L>2$. Instead, a recursive approach is taken, that bounds the maximum mutual information that can be delivered from the source nodes to the nodes in the $\ell^{th}$ hop, given the maximum mutual information that can be delivered to the nodes in the $(\ell-1)^{th}$ hop. As $\ell$ increases from $2$ to $L$, at each stage of this recursive expansion, the deterministic transformation is used only for the last ($\ell^{th}$) hop for that stage. This recursive approach, combined with the AI Sumset Inequalities and the insights from the $L=2$ setting, turns out to be sufficient to characterize the sum-GDoF value for the $L$-hop setting. As before, our focus is on the weak interference setting, for which the sum GDoF value is presented in Theorem \ref{thm:Lhop} in Section \ref{sec:Lhopweak}. The result can be immediately extended to strong interference by switching the labels of the relays in every other hop, provided that the number of hops, $L$ is even, thus giving us Corollary \ref{cor:Lhop} in Section \ref{sec:Lhopstrong}. Another remarkable aspect of this result is that as $L$ approaches infinity, the sum-GDoF value approaches the sum-GDoF of the corresponding one-hop broadcast channel where the $2$ sources are allowed to cooperate fully, under finite precision CSIT. From the achievability perspective, this happens because of a successive onion peeling approach that allows the relays in each successive stage to decode one more layer of interference, so that the common information accumulated asymptotically at the relays as $L$ approaches infinity, is enough to match the broadcast channel where the transmitters cooperate fully.

	The rest of this paper is organized as follows. The system model is presented in the next section. The main results appear in Section \ref{sec:res}. Proofs of converse (outer bounds) are provided in Section \ref{sec:upper},  and achievability results (inner bounds) are proved in Section \ref{sec:lower}. Section \ref{sec:con} concludes the paper.
	
	\textit{Notation:} The notation $(x)^+$ represents $\max(x,0)$. For integers $i,j$, the notation $[i:j]$ represents the set $\{i, i+1, \cdots, j\}$ if $i<j$ and the empty set otherwise. $X^{[N]}$ denotes the sequence $\{X(1),X(2),\cdots,X(N)\}$. $f(x)=o(g(x))$ denotes that $\limsup_{x\rightarrow\infty}\frac{|f(x)|}{|g(x)|}=0$. Define $\lfloor x \rfloor$ as the largest integer that is smaller than or equal to $x$ when $x$ is non-negative, and the smallest integer that is larger than or equal to $x$ when $x$ is negative.
	
	\section{System Model: Layered Symmetric $L$-hop Interference Channel}
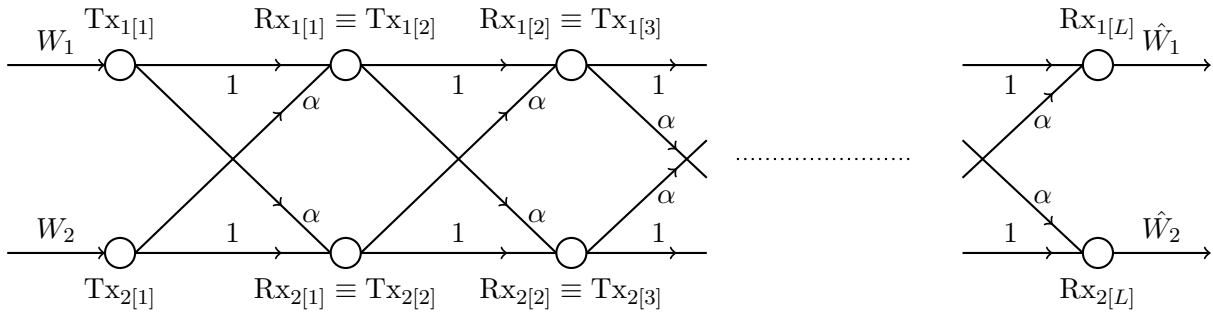
\begin{figure}[ht]
		\centering
		\begin{tikzpicture}[scale=1.0]
		\draw[thick] (2,1) circle(2mm)node[above=0.2]{$\text{Tx}_{1[1]}$};
		\draw[thick] (5,1) circle(2mm)node[above=0.2]{$\text{Rx}_{1[1]}\equiv \text{Tx}_{1[2]}$};
		\draw[thick] (2,-1.5) circle(2mm)node[below=0.2]{$\text{Tx}_{2[1]}$};
		\draw[thick] (5,-1.5) circle(2mm)node[below=0.2]{$\text{Rx}_{2[1]}\equiv \text{Tx}_{2[2]}$};

		\draw[thick, ->] (0.5,1)--(1.8,1)node[above,pos=0.5]{$W_1$};
		\draw[thick, ->] (0.5,-1.5)--(1.8,-1.5)node[above,pos=0.5]{$W_2$};
		
		\draw[ thick,arrowmid, ] (2.2,1)--(4.8,1)node[below,pos=0.5, black]{$1$};
		\draw[ thick,arrowmid, ] (2.2,-1.5)--(4.8,-1.5)node[above,pos=0.5, black]{$1$};
		\draw[ thick,arrowmid, ] (2.2,1)--(4.8,-1.5)node[right,pos=0.8, black]{$\alpha$};
		\draw[ thick,arrowmid, ] (2.2,-1.5)--(4.8,1)node[right,pos=0.8, black]{$\alpha$};
		
		\draw[thick] (8,1) circle(2mm)node[above=0.2]{$\text{Rx}_{1[2]}\equiv \text{Tx}_{1[3]}$};
		\draw[thick] (8,-1.5) circle(2mm)node[below=0.2]{$\text{Rx}_{2[2]}\equiv \text{Tx}_{2[3]}$};
		
		\draw[ thick,arrowmid, ] (5.2,1)--(7.8,1)node[below,pos=0.5, black]{$1$};
		\draw[thick,arrowmid] (5.2,-1.5)--(7.8,-1.5)node[above,pos=0.5, black]{$1$};
		\draw[thick,arrowmid] (5.2,1)--(7.8,-1.5)node[right,pos=0.8]{$\alpha$};
		\draw[ thick,arrowmid, ] (5.2,-1.5)--(7.8,1)node[right,pos=0.8, black]{$\alpha$};
		
		\draw[thick,arrowmid] (8.2,1)--(9.8,1)node[below,pos=0.6]{$1$};
		\draw[thick,arrowmid] (8.2,-1.5)--(9.8,-1.5)node[above,pos=0.6]{$1$};
		\draw[thick,arrowmid] (8.2,1)--(9.8,-0.5)node[right,pos=0.5]{$\alpha$};
		\draw[thick,arrowmid] (8.2,-1.5)--(9.8,0)node[right,pos=0.5]{$\alpha$};
		
		\draw[thick,dotted] (10.2,-0.25)--(12.5,-0.25); 
		
		\draw[thick,arrowmid] (13.2,1)--(14.8,1)node[below,pos=0.4]{$1$};
		\draw[thick,arrowmid] (13.2,-1.5)--(14.8,-1.5)node[above,pos=0.4]{$1$};
		\draw[thick,arrowmid] (13.2,0)--(14.8,-1.5)node[right,pos=0.5]{$\alpha$};
		\draw[thick,arrowmid] (13.2,-0.5)--(14.8,1)node[right,pos=0.5]{$\alpha$};
		\draw[thick] (15,1) circle(2mm)node[above=0.2]{$\text{Rx}_{1[L]}$};
		\draw[thick] (15,-1.5) circle(2mm)node[below=0.2]{$\text{Rx}_{2[L]}$};
		
		\draw[thick, ->] (15.2,1)--(16.5,1)node[above,pos=0.5]{$\hat{W_1}$};
		\draw[thick, ->] (15.2,-1.5)--(16.5,-1.5)node[above,pos=0.5]{$\hat{W_2}$};
		\end{tikzpicture}
		\caption{\it\small Layered Symmetric $L$-hop  Interference Channel model.}\label{fig:Lhop}
	\end{figure}

	Figure \ref{fig:Lhop} depicts the layered symmetric $L$-hop interference channel model. Each hop is a $2\times 2$ topology, comprised of $2$ transmitters and $2$ receivers. For the $\ell^{th}$ hop, $\ell\in[1:L]$, the two transmitters are labeled as $\text{Tx}_{1[\ell]}$, $\text{Tx}_{2[\ell]}$, and the corresponding receivers are labeled as $\text{Rx}_{1[\ell]}, \text{Rx}_{2[\ell]}$, respectively. The receivers for the $\ell^{th}$ hop are the same physical nodes  that act as the transmitters for the $(\ell+1)^{th}$ hop, i.e., $\text{Rx}_{i[\ell]}\equiv \text{Tx}_{i[\ell+1]}$, $\ell\in[1:L-1], i\in[1:2]$. The transmitters for the first hop, $\text{Tx}_{1[1]}, \text{Tx}_{2[1]}$ are  also referred to as \emph{sources}, the receivers for the last hop, $\text{Rx}_{1[L]}, \text{Rx}_{2[L]}$ are also referred to as \emph{destinations}, and the remaining nodes are also referred to as \emph{relays}.

	Suppose the communication takes place over $N$ channel uses. There are two independent messages, $W_1\in[1:\lceil 2^{NR_1}\rceil], W_2\in[1:\lceil 2^{NR_2}\rceil]$, such that for $i\in[1:2]$, message $W_i$ originates from  Source $\text{Tx}_{i[1]}$ and is intended for Destination $\text{Rx}_{i[L]}$ respectively. Following the GDoF formulation, under the $n^{th}$ channel use, $n\in[1:N]$, the inputs and outputs of the $\ell^{th}$ hop are related as follows,
	\begin{align}
	Y_{1[\ell]}(n)&=\sqrt{P}G_{11[\ell]}(n)X_{1[\ell]}(n)+\sqrt{P^{\alpha}}G_{12[\ell]}(n)X_{2[\ell]}(n)+Z_{1[\ell]}(n)\label{eq:inputoutput1}\\
	Y_{2[\ell]}(n)&=\sqrt{P^{\alpha}}G_{21[\ell]}(n)X_{1[\ell]}(n)+\sqrt{P}G_{22[\ell]}(n)X_{2[\ell]}(n)+Z_{2[\ell]}(n)\label{eq:inputoutput2}
	\end{align}
such that the signal sent from the transmitter $\text{Tx}_{i[\ell]}$ is denoted as $X_{i[\ell]}(n)$, the signal observed by the receiver $\text{Rx}_{i[\ell]}$ is denoted as $Y_{i[\ell]}(n)$, the channel coefficient between $\text{Tx}_{i[\ell]}$ and $\text{Rx}_{k[\ell]}$ is denoted as $G_{ki[\ell]}(n)$,  the additive noise observed by  the receiver $\text{Rx}_{i[\ell]}$ is denoted as $Z_{i[\ell]}(n)$, and $i,k\in[1:2], n\in[1:N]$. All symbols are complex, the noise terms represent i.i.d. zero mean unit variance circularly symmetric Additive White Gaussian Noise (AWGN), and the transmitted symbols $X_{1[\ell]}(n)$, $X_{2[\ell]}(n)$ are each subject to a unit transmit power constraint. 

We assume that the channel coefficients $G_{ki[\ell]}(n)$ follow the \emph{bounded density assumption} of \cite{Arash_Jafar}, i.e., all joint and conditional probability density functions exist and are bounded. To make this precise, let  $\mathcal{G}$ be a set of real-valued random variables, that satisfies the following two conditions:
	\begin{itemize}
		\item All random variables in $\mathcal{G}$ are bounded away from zero and infinity, i.e., $g\in\mathcal{G}\implies |g|\in[1/\Delta,\Delta]$ for some positive finite constant $\Delta$.
		\item There exists a finite positive constant $f_{max}$, such that for all finite cardinality disjoint subsets $\mathcal{G}_1,\mathcal{G}_2\subset\mathcal{G}$, the conditional probability density function $f_{\mathcal{G}_1|\mathcal{G}_2}$ exists and is bounded above by $f_{max}^{|\mathcal{G}_1|}$.
	\end{itemize}
Now, if we represent each channel coefficient  in terms of its real and imaginary components, $G_{ki[\ell]}(n)=G_{ki[\ell],R}(n)+jG_{ki[\ell],I}(n)$, then the bounded density assumption means that we require that  $G_{ki[\ell],R}(n),G_{ki[\ell],I}(n)$ are distinct elements of $\mathcal{G}$ for $k,i\in[1:2], \ell\in[1:L],n\in[1:N]$. 

Next, in order to specify the channel knowledge assumptions, let us define 
\begin{align}
\mathcal{G}_{[\ell]}&=\{G_{uv[\ell]}(n): u,v\in[1:2], n\in[1:N]\}
\end{align}
as the subset  of $\mathcal{G}$ comprised of only the channel coefficients associated with the $\ell^{th}$ hop. For simplicity,\footnote{This assumption is not strictly necessary for our results, but it will simplify the analysis.} and since this is a common assumption in practice, let us assume that the channels across different hops are independent. Also, channels are independent of messages and additive noise terms. 

Similarly, define $\mathcal{G}_{[\ell_1:\ell_2]}=\bigcup_{l=\ell_1}^{\ell_2}\mathcal{G}_{[l]}$ as the subset of $\mathcal{G}$ comprised of all channel coefficients across  hops $[\ell_1:\ell_2]$.  We assume that precise channel state information is available at the receivers (CSIR) for all channels in the \emph{preceding} hops. Specifically, the receivers in the $\ell^{th}$ hop, $\text{Rx}_{1[\ell]}, \text{Rx}_{2[\ell]}$, have precise knowledge of the realizations of all random variables in $\mathcal{G}_{[1:\ell]}$. Furthermore, since the receivers in the $\ell^{th}$ hop are the same as the transmitters in the $(\ell+1)^{th}$ hop, $\text{Rx}_{k[\ell]}\equiv\text{Tx}_{k[\ell+1]}$, we allow  that the same precise channel knowledge of $\mathcal{G}_{[1:\ell]}$ is available to $\text{Tx}_{1[\ell+1]}, \text{Tx}_{2[\ell+1]}$. The knowledge of all remaining channel coefficients is limited to their joint probability density functions. The assumption that these probability density functions satisfy the bounded density assumption is what limits the CSIT to finite precision. Note that the CSIT assumptions imply that \begin{align}
I(X_{1[\ell]}^{[N]},X_{2[\ell]}^{[N]},\mathcal{G}_{[1:\ell-1]};  \mathcal{G}_{[\ell:L]})&=0.\label{indep}
\end{align}
This is because the transmitters over the $\ell^{th}$ hop have no knowledge of  channel \emph{realizations} beyond what can be passed to them from preceding hops. 

{\it Remark:} The assumption that CSIT is available for preceding hops at each node  strengthens the GDoF converse  bounds, because additional channel knowledge cannot hurt, but it is noteworthy that the achievable schemes presented in this paper that meet those bounds do not make use of this  CSIT at any encoder. The receivers do utilize the corresponding CSIR of all preceding hops.  Similarly, let us note that while we allow a receiver to have perfect CSIR for the channels associated with the \emph{other} receiver in the same hop, e.g., $\text{Rx}_{1[\ell]}$ has perfect knowledge of $G_{22[\ell]}(n)$, such knowledge is not used by the achievable scheme either. As such this assumption also serves mainly to strengthen the converse, and our GDoF results  hold both with and without it.

While the bounded density assumption allows fairly general distributions for the channel coefficients, an interesting perspective of the channel coefficients is to view them as small \emph{perturbations}, say i.i.d. uniform in a small interval around $1$, such that the length of that interval corresponds to the finite precision constraint --- the shorter the length of the perturbation interval, the more precisely the channels are revealed by their statistics, and the larger the peak value of the probability density function. This also explains the need for density functions  to be bounded in order to limit CSIT to finite precision. The channel coefficients typically represent physical phenomena like channel fading, but viewed as perturbations they can also represent artifacts that are deliberately introduced  into the GDoF model in order to filter out or eliminate the possibility of fragile schemes emerging as optimal solutions, thus allowing us to explore information theoretic optimality of  random coding based solutions that are also practically appealing for their robustness. 

Recall that $P$ is a nominal parameter that approaches infinity to define the GDoF limit, and the exponents that appear with $P$ in \eqref{eq:inputoutput1},\eqref{eq:inputoutput2} represent coarse channel strength parameters that are assumed globally known (equivalently, channel strengths in the absence of perturbations). Specifically, for our symmetric model, the direct links (between $\text{Tx}_{i[\ell]}$ and $\text{Rx}_{i[\ell]}$) have channel strength  corresponding to the exponent $1$ and cross links have coarse channel strength corresponding to the exponent $\alpha\in\mathbb{R}^+$. Because well-designed networks tend to operate in the weak interference regime, our  focus is on the setting $\alpha<1$, although some of our results  generalize to strong interference settings in a straightforward manner. Intuitively, we may think of each of these channel strength parameters as the (approximate) capacity of the corresponding point to point link in its original finite SNR setting, and think of $\log(P)$ as a uniform scaling factor that is simultaneously applied to the capacities of all the links. Since each link capacity is logarithmic in $P$, linear scaling of capacity corresponds to exponential scaling of $P$, and the original channel capacities $\alpha$ appear as exponents. The fundamental intuition behind GDoF is that if the capacity of every link in a network is scaled by the same constant factor ($\log(P)$), then the network capacity should also scale (approximately) by the same factor. So normalizing the sum-capacity of the network by $\log(P)$ should produce an approximation to the capacity of the original network. This is indeed why we see normalizations by $\log(P)$ in the definition of GDoF, as presented next.

	The rate pair $(R_1,R_2)$ is said to be achievable if there exists a scheme, comprised of encoding functions at the sources, mappings from inputs to outputs at each of the relays, and decoding functions at the destinations, under which $\text{Rx}_{1[L]},\text{Rx}_{2[L]}$ can decode $W_1,W_2$ respectively with arbitrarily small error probability in the standard Shannon-theoretic sense\cite{NIT}. The closure of achievable rate tuples is the capacity region $\mathcal{C}(P)$. The GDoF region is defined as
	\begin{align}
		\mathcal{D}^{{\tiny f.p.}}=\left\{
		\begin{array}{ll}
		(d_1,d_2):&\exists(R_1(P),R_2(P))\in \mathcal{C}(P) \\ 
		&s.t.\quad d_1=\lim\limits_{P\rightarrow\infty}\frac{R_1(P)}{\log(P)}, ~~~d_2=\lim\limits_{P\rightarrow\infty}\frac{R_2(P)}{\log(P)}
		\end{array}
		\right\}\label{def:GDoF}
	\end{align}
	The superscript '\emph{f.p.}' highlights the finite precision CSIT constraint. 	Finally, the  sum-GDoF value is defined as  $\mathcal{D}_{\Sigma}^{\tiny f.p.}=\max\limits_{(d_1,d_2)\in\mathcal{D}^{\tiny f.p.}}(d_1+d_2)$.

    \section{Results}\label{sec:res}
    Following the information theoretic mindset of starting from the elemental scenarios, the simplest multihop setting, where $L=2$, i.e., the $2$-hop interference channel (especially in the weak interference regime, $\alpha\leq 1$) is our main focus in this paper. Our main result is the sum-GDoF characterization for this channel under finite precision CSIT, presented in Section \ref{sec:2hop}. Due to its relative simplicity the $2$-hop setting is also instructive  to introduce the main ideas in their simplest form, whose generalizations eventually allow us to find the sum-GDoF for arbitrary $L$, as presented in Section \ref{sec:Lhop}.
    
\subsection{Sum-GDoF of the $2$-hop Layered Symmetric Interference Channel under Finite Precision CSIT}\label{sec:2hop}
\subsubsection{Weak Interference Regime: $\alpha\leq 1$}\label{sec:2hopweak}
     \begin{theorem}\label{thm:2hop}
    	For the $2$-hop layered symmetric interference channel under finite precision CSIT, in the weak interference regime $\alpha\leq 1$, the sum-GDoF value is  given by,
    	\begin{align}
    	\mathcal{D}_{\Sigma}^{\tiny f.p.}=\left\{
    	\begin{array}{lcl}
    	2-4\alpha/3, &&0\leq \alpha\leq1/2,\\	
    	2/3+4\alpha/3,&&1/2\leq\alpha\leq 4/7,\\		
    	2-\alpha, &&4/7\leq\alpha\leq 1.\\
    	\end{array}
    	\right.
    	\end{align}
    \end{theorem}
    The converse proof for Theorem \ref{thm:2hop} is provided in Section \ref{sec:2hopcon}. The converse for the regime $4/7\leq \alpha\leq 1$ is already available because it corresponds to the GDoF value established in  \cite{Arash_Jafar} under finite precision CSIT for the MISO broadcast channel that is obtained by allowing full cooperation (which cannot hurt) among all nodes except  the two destination nodes. However, the converse for the remaining regime, $0\leq \alpha\leq\frac{4}{7}$, is non-trivial  and is obtained in this work based on the  sum-set inequalities of \cite{Arash_Jafar_sumset}. One of the challenging aspects of the converse is that the deterministic transformation that is the starting point of all prior applications of Aligned Images bounds \cite{Arash_Jafar_IC, Arash_Jafar_MIMOICGDoF, Arash_Jafar_SLS, Arash_Bofeng_Jafar_BC, Arash_Jafar_MIMOBC, Arash_Jafar_cooperation, Yoga_Junge_Jafar,  Junge_Yuan_Huang_Jafar}, is not directly applicable to the multi-hop setting as explained in the introduction. This challenge is overcome essentially by allowing perfect CSIT in the first hop (which cannot hurt) and only using the deterministic transformation for the second hop. Since this produces a tight converse bound that is achievable with finite precision CSIT in both hops, evidently the sum-GDoF value is the same (given by Theorem \ref{thm:2hop}) whether the CSIT in the first hop is perfect or restricted to finite precision, as long as the CSIT in the second hop is limited to finite precision.

The achievability for Theorem \ref{thm:2hop} is proved in Section \ref{sec:2hopach}. The achievable scheme is straightforward when $2/3\leq \alpha\leq 1$, because it corresponds to a concatenation of two interference channels \cite{Etkin_Tse_Wang, Arash_Jafar_cooperation} where the intermediate nodes (the relays) simply employ a decode-and-forward strategy. The achievable scheme is non-trivial for the remaining regime $0\leq \alpha\leq\frac{2}{3}$ and relies on a rate-splitting approach that is comprised of amplify-and-forward and decode-and-forward schemes. Specifically, the sources split their messages into sub-messages, the relays are able to decode-and-forward some of the sub-messages, while they amplify-and-forward the remaining superposition of codewords that they are not able to decode. The relays further split the  sub-messages that they are able to decode and then use a different superposition approach (assigning different powers) to transmit the decoded sub-messages. With proper choice of rate-splitting and superposition parameters, the destinations are able to decode their desired messages by a successive decoding approach. 
 
 \subsubsection{Extension to Strong Interference Regime: $\alpha \geq 1$}\label{sec:2hopstrong}
	As noted previously our main focus is on the weak interference regime, $\alpha\leq 1$. However, the extension of Theorem \ref{thm:2hop} to the strong interference regime, where $\alpha \geq 1$, turns out to be straightforward for the $2$-hop setting, as stated in the following  corollary.
	\begin{corollary}\label{cor:2hop}
		For the $2$-hop layered symmetric interference channel under finite precision CSIT, in the strong interference regime $\alpha\geq 1$, the sum-GDoF value is  given by,
		\begin{align}
			\mathcal{D}_{\Sigma}^{f.p.}=\left\{
			\begin{array}{lcl}
			2\alpha-1, && 1\leq\alpha\leq 7/4,\\
			2\alpha/3 + 4/3,&& 7/4\leq\alpha\leq 2,\\
			2\alpha -4/3, &&\alpha\geq 2.
			\end{array}
			\right.
		\end{align}
	\end{corollary}
The corollary follows from Theorem \ref{thm:2hop} directly, because switching the labels of the two relays immediately converts the weak interference setting into a strong interference setting. Specifically, switching the relays gives us a channel where the direct channels have strength $\alpha$ and the cross-channels have strength $1$. Now, let us scale all channel strength parameters by $1/\alpha$, so that we have direct channels with strength $\alpha\times1/\alpha =1$ and cross-channels with strength $\alpha' = 1\times 1/\alpha > 1$. It follows from the definition of GDoF that if all channel strength parameters are scaled by the same constant,\footnote{Essentially this corresponds to defining $P' = P^\alpha$ and substituting for $P$ with $P'$, so that $P\rightarrow {P'}^{1/\alpha}=P'^{\alpha'}$, and $P^\alpha \rightarrow P'$. The normalization factor in the definition of GDoF similarly maps as $\log(P)\rightarrow 1/\alpha \log(P')$.} then the GDoF value will be scaled by precisely the same constant as well. Thus, if we denote the sum-GDoF as a function of $\alpha$ as $\mathcal{D}_{\Sigma}^{f.p.}(\alpha)$, then we must have $\mathcal{D}_{\Sigma}^{f.p.}(1/\alpha)=1/\alpha\mathcal{D}_{\Sigma}^{f.p.}(\alpha)$ for $\alpha\leq 1$, or equivalently, $\mathcal{D}_{\Sigma}^{f.p.}(\alpha')=\alpha'\mathcal{D}_{\Sigma}^{f.p.}(1/\alpha')$ for $\alpha'>1$, which gives us Corollary \ref{cor:2hop}. Thus, Theorem \ref{thm:2hop} and Corollary \ref{cor:2hop} together fully characterize the sum-GDoF value of the $2$-hop layered symmetric interference channel under finite precision CSIT, for all $\alpha$.
	
	     \subsubsection{Comparisons}
	       	\begin{figure}[t]
    		\centering
    		\includegraphics[width=15cm]{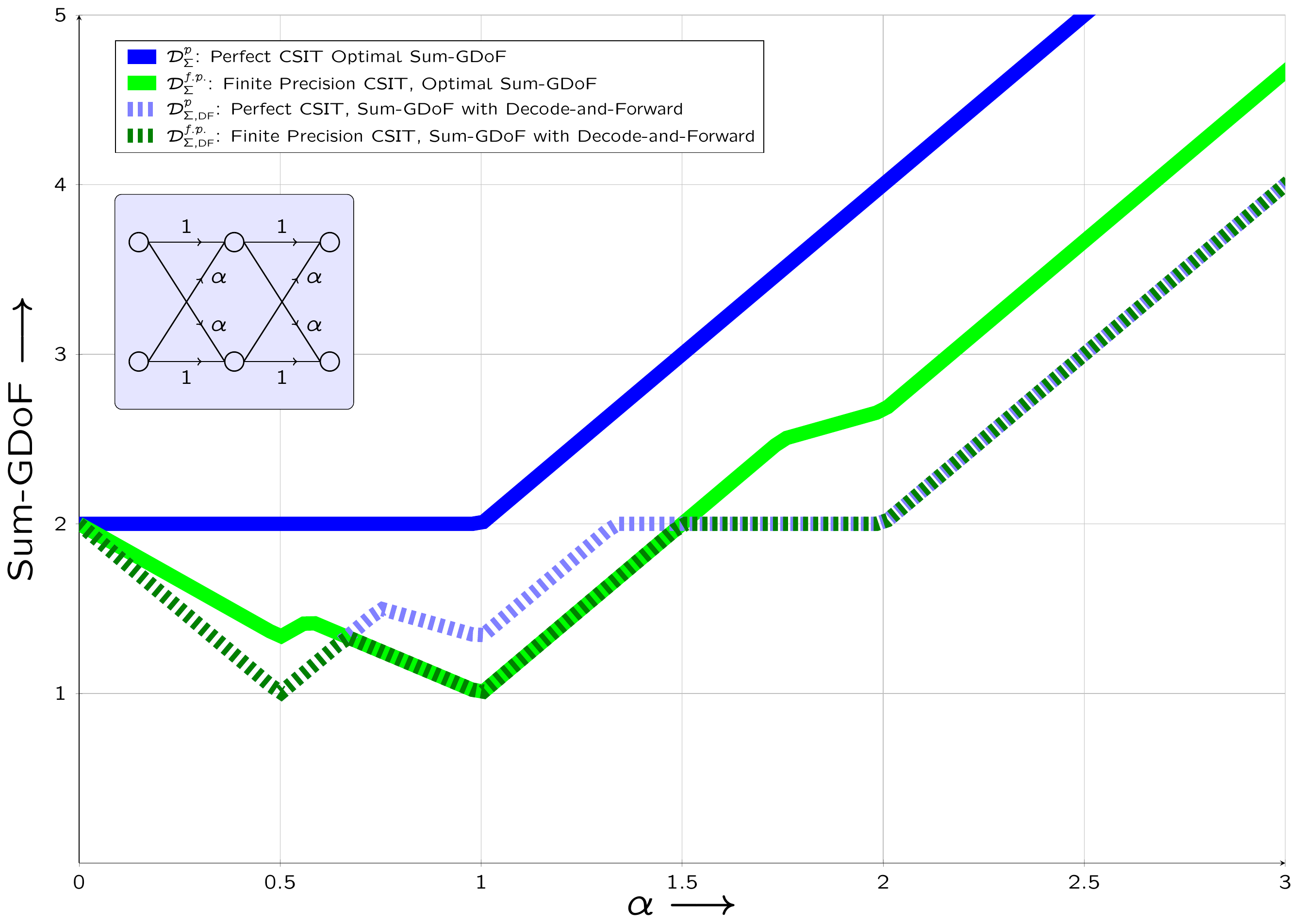}
    		\caption{\it\small Sum-GDoF comparisons for the layered symmetric $2$-hop interference channel.}
    		\label{fig:sgdofplot}
    	\end{figure}
To place the sum-GDoF result in perspective, let us compare it against a few benchmarks, as illustrated in Figure \ref{fig:sgdofplot}. These benchmarks are explained below.  
    \begin{itemize}
        	\item \textbf{Optimal Sum-GDoF under Perfect CSIT}: Recall the aligned interference neuralization scheme  introduced by Gou et al. in  \cite{Gou_Wang_Jafar_Jeon_Chung}, which was originally used to show that a sum-DoF value of $2$ is achievable for the $2$ hop interference channel under perfect CSIT. It is not difficult to apply the same scheme to find the sum-\emph{GDoF} value, $\mathcal{D}_{\Sigma}^{\tiny p}$ under perfect CSIT (the  `\emph{p}' in the superscript stands for `\emph{perfect}' CSIT), which turns out to be equal to the sum-GDoF value of the one hop MISO broadcast channel with perfect CSIT.
    	\begin{align}
    	\mathcal{D}_{\Sigma}^{\tiny p}=\left\{
    	\begin{array}{lcl}
    	2,&&\alpha\leq 1,\\	
    	2\alpha,&& \alpha\geq1.
    	\end{array}
    	\right.
    	\end{align}

    	\item \textbf{Sum-GDoF with Decode-and-Forward under Finite Precision CSIT}: A decode-and-forward solution for the $2$-hop interference channel  corresponds to treating each hop as an $X$ channel. Using the sum-GDoF of the $X$ channel under finite precision CSIT as characterized in \cite{Arash_Jafar_cooperation} we obtain the sum-GDoF value of the $2$ hop layered symmetric interference channel with decode-and-forward under  finite precision CSIT as follows.
    	\begin{align}
    	\mathcal{D}_{\Sigma, \df}^{\tiny f.p.}=\left\{
    	\begin{array}{lcl}
    	2-2\alpha &&\alpha\leq\frac{1}{2}\\	
    	2\alpha &&\frac{1}{2}\leq\alpha\leq \frac{2}{3}\\		
    	2-\alpha &&\frac{2}{3}\leq\alpha\leq 1\\
    	2\alpha-1 && 1\leq\alpha\leq \frac{3}{2}\\
    	2 && \frac{3}{2}\leq\alpha\leq 2\\
    	2\alpha-2 &&\alpha\geq 2\\
    	\end{array}
    	\right.
    	\end{align}
    	\item \textbf{Sum-GDoF with Decode-and-Forward under Perfect CSIT}: Using  the  sum-GDoF value of the $X$ channel under perfect CSIT as characterized in \cite{Huang_Cadambe_Jafar, Cadambe_Jafar_X, Jafar_Shamai, MMK}, we obtain the sum-GDoF value of the $2$ hop layered symmetric interference channel with decode-and-forward under perfect CSIT as follows.
    	\begin{align}
    	\mathcal{D}_{\Sigma,\df}^{\tiny p}=\left\{
    	\begin{array}{lcl}
    	2-2\alpha &&\alpha\leq\frac{1}{2}\\	
    	2\alpha &&\frac{1}{2}\leq\alpha\leq \frac{3}{4}\\		
    	\frac{6-2\alpha}{3} &&\frac{3}{4}\leq\alpha\leq 1\\
    	\frac{6\alpha-2}{3} && 1\leq\alpha\leq \frac{4}{3}\\
    	2&& \frac{4}{3}\leq\alpha\leq 2\\
    	2\alpha-2&&\alpha\geq 2\\
    	\end{array}
    	\right.
    	\end{align}		

    \end{itemize}    	

From Figure \ref{fig:sgdofplot} we note that except for the degenerate case of $\alpha=0$, there is always a significant loss of sum-GDoF relative to its optimal value under perfect CSIT, i.e., the GDoF benefits of aligned interference neutralization \cite{Gou_Wang_Jafar_Jeon_Chung} are pervasive and powerful under perfect CSIT but too fragile to survive under finite precision CSIT. Remarkably, we note that $\mathcal{D}_{\Sigma, \df}^{\tiny f.p.} = \min\left(\mathcal{D}_{\Sigma,\df}^{\tiny p}, \mathcal{D}_{\Sigma}^{\tiny f.p.}\right)$ even though $\mathcal{D}_{\Sigma,\df}^{\tiny p}$ and  $\mathcal{D}_{\Sigma}^{\tiny f.p.}$ are almost always different values (with the exception of cross-overs that occur at $\alpha=2/3, 3/2$). Thus, relative to the baseline of robust (finite precision) decode-and-forward, the robust (finite precision) gains of multi-hopping appear in the regimes where cross-channels are significantly weaker or stronger, i.e., $\alpha\leq 2/3, \alpha\geq 3/2$,  whereas the fragile gains of interference alignment  under perfect CSIT appear precisely in the complementary regime $2/3\leq \alpha\leq 3/2$ where the cross-channels are relatively of similar strength as direct channels.

\subsection{Sum-GDoF of the $L$-hop Layered Symmetric Interference Channel under Finite Precision CSIT}\label{sec:Lhop}
	Building on the insights from the $2$-hop setting, in this section we generalize the sum-GDoF results to the $L$ hop case under finite precision CSIT. As before we start with the weak interference regime.
	
	\subsubsection{Weak Interference Regime: $\alpha\leq 1$}\label{sec:Lhopweak}
     \begin{theorem}\label{thm:Lhop}
    	For the $L$-hop layered symmetric interference channel under finite precision CSIT, in the weak interference regime $\alpha\leq 1$, the sum-GDoF value is  given by,
		\begin{align}
		\mathcal{D}_{\Sigma}^{\tiny f.p.}=\left\{
			\begin{array}{lcl}
			2-\alpha-\alpha/(2^L-1), &&0\leq \alpha\leq1/2,\\	
			1+\alpha-(1-\alpha)/(2^L-1), &&1/2\leq\alpha\leq 2^L/(2^{L+1}-1),\\				2-\alpha, &&2^L/(2^{L+1}-1) \leq\alpha\leq 1\\
			\end{array}
			\right.\label{eq:Lhopweak}
		\end{align}
		\end{theorem}
		
						       	\begin{figure}[h]
    		\centering
    		\includegraphics[width=10cm]{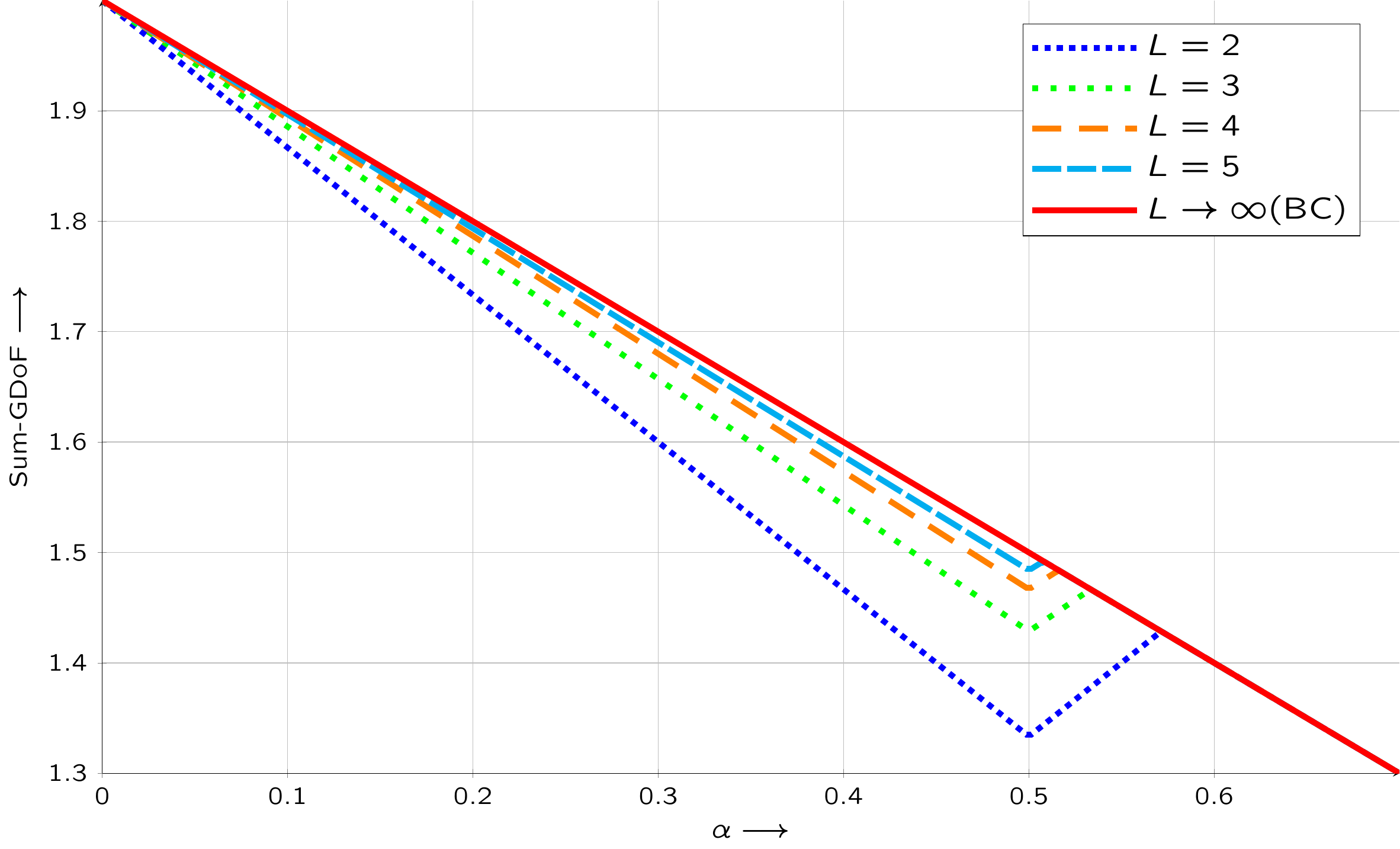}
    		\caption{\it\small Sum-GDoF of the layered symmetric $L$-hop interference channel in a subinterval of the weak-interference regime.}
    		\label{fig:Lhop_plot}
    	\end{figure}

		The sum-GDoF value specified in Theorem \ref{thm:Lhop} is illustrated in Figure \ref{fig:Lhop_plot} over a subinterval of the weak interference regime (since the plots are close together, the figure is zoomed in for clarity) for various $L$. 
		
		The  converse for Theorem \ref{thm:Lhop} is proved in Section \ref{sec:Lhopcon}. For the converse proof, the regime $2^L/(2^{L+1}-1) \leq\alpha\leq 1$ is straightforward because this is simply the GDoF value of the broadcast channel \cite{Arash_Jafar} that is obtained by allowing full cooperation among all nodes except the destination nodes. For the remaining regimes, as with the $2$-hop case, a challenging aspect is the deterministic transformation. Whereas in the $2$-hop case it was sufficient to enforce finite precision CSIT only in the last hop, the same idea does not work directly in the $L$-hop setting. Instead the problem is circumvented by first considering only $\ell$ hops at a time, as in Lemma \ref{lemma:deter} that appears in Section \ref{sec:lemmas}, and enforcing finite precision CSIT in the `last' (i.e., the $\ell^{th}$ hop) to bound the mutual information that can be transferred from the sources to the receivers in the $\ell^{th}$ hop.  Then a recursive argument is developed in Lemma \ref{lemma:recur} in  Section \ref{sec:lemmas} to obtain a bound for $\ell$ hops based on the bound for $\ell-1$ hops. 
		
		The proof of achievability for Theorem \ref{thm:Lhop} appears in Section \ref{sec:Lhopach}. The regime $2/3\leq \alpha\leq 1$ is straightforward as in the $2$-hop case, because it corresponds to a concatenation of $L$ interference channels \cite{Etkin_Tse_Wang, Arash_Jafar_cooperation} and a simple decode-and-forward strategy suffices. In other regimes however, the achievable scheme for the $L$-hop setting is a non-trivial extension of the $2$-hop case. While in principle the construction  is still based on rate-splitting between amplify-and-forward and decode-and-forward schemes, there is an important element of onion-peeling which allows each successive relay stage to decode one more layer of interference, so that with each hop the nodes acquire more common information and are closer to acting as a broadcast channel. Indeed, as the number of hops $L\rightarrow\infty$, the sum-GDoF value does approach that of a broadcast channel where all information is shared between the two transmitters of the last hop.

\subsubsection{Extension to Strong Interference Regime: $\alpha\geq 1$}\label{sec:Lhopstrong}
As in the $2$-hop case, the sum-GDoF result in Theorem \ref{thm:Lhop}  for the weak interference regime immediately implies an extension to the strong interference regime by the same argument of switching relay positions, and is presented in the following corollary.

	 	\begin{corollary}\label{cor:Lhop}
		If $L$ is even, then for the $L$-hop layered symmetric interference channel under finite precision CSIT, in the strong interference regime $\alpha\geq 1$, the sum-GDoF value is  given by,
		\begin{align}
		\mathcal{D}_{\Sigma}^{\tiny f.p.}=\left\{
		\begin{array}{lcl}
		2\alpha-1, && 1\leq\alpha\leq 2-2^{-L},\\
		\alpha+1-(\alpha-1)/(2^L-1), && 2-2^{-L} \leq\alpha\leq 2,\\
		2\alpha-1-1/(2^L-1), &&\alpha\geq 2.
		\end{array}
		\right.\label{eq:Lhopstrong}
		\end{align}
	\end{corollary}

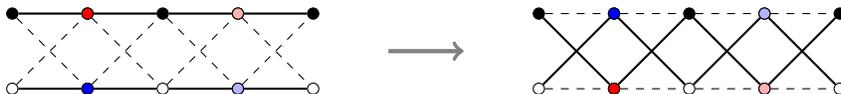
\begin{figure}[b]	
\begin{center}
\begin{tikzpicture}
\begin{scope}[shift={(0,0)}]

	\foreach \n in {1,2,3,4,5}
		{
		\node[circle, draw=black, fill=white, inner sep = 1.5] (N1\n) at (\n,1){};
		\node[circle, draw=black, fill=black, inner sep = 1.5] (N2\n) at (\n,2){};
		};
\foreach \n in {1,2,3,4}
{
	\pgfmathtruncatemacro{\nplusone}{\n + 1}
	\draw[thick, black](N1\n)--(N1\nplusone);
	\draw[thick, black](N2\n)--(N2\nplusone);
	\draw[dashed, black](N1\n)--(N2\nplusone);
	\draw[dashed, black](N2\n)--(N1\nplusone);
};

\node[circle, draw=black, fill=blue, inner sep = 1.5] at (2,1){};
\node[circle, draw=black, fill=red, inner sep = 1.5] at (2,2){};

\node[circle, draw=black, fill=blue!30, inner sep = 1.5] at (4,1){};
\node[circle, draw=black, fill=red!30, inner sep = 1.5] at (4,2){};
\end{scope}

\draw [->, ultra thick, gray] (6,1.5)--(7,1.5);

\begin{scope}[shift={(7,0)}]

	\foreach \n in {1,2,3,4,5}
		{
		\node[circle, draw=black, fill=white, inner sep = 1.5] (N1\n) at (\n,1){};
		\node[circle, draw=black, fill=black, inner sep = 1.5] (N2\n) at (\n,2){};
		};
\foreach \n in {1,2,3,4}
{
	\pgfmathtruncatemacro{\nplusone}{\n + 1}
	\draw[dashed, black](N1\n)--(N1\nplusone);
	\draw[dashed, black](N2\n)--(N2\nplusone);
	\draw[thick, black](N1\n)--(N2\nplusone);
	\draw[thick, black](N2\n)--(N1\nplusone);
};
\node[circle, draw=black, fill=red, inner sep = 1.5] at (2,1){};
\node[circle, draw=black, fill=blue, inner sep = 1.5] at (2,2){};

\node[circle, draw=black, fill=red!30, inner sep = 1.5] at (4,1){};
\node[circle, draw=black, fill=blue!30, inner sep = 1.5] at (4,2){};
\end{scope}

\end{tikzpicture}
\end{center}	
\caption{\it\small Two representations of the same network. Interchanging the positions of relay nodes in every \emph{other} hop changes the representation of the network from a weak interference setting to a strong interference setting. This works only when the number of hops, $L$, is even.}\label{fig:switch}
\end{figure}

A new constraint appears in Corollary \ref{cor:Lhop}, that $L$ must be even. This is because the idea of interchanging the positions of the relay nodes to convert weak interference into strong interference only works when  we switch relays in every \emph{other} hop, which can only be done if the number of hops is even. To see this explicitly, consider  Figure \ref{fig:switch} which shows an $L=4$ hop setting. The bold edges represent strong channels while the dashed edges represent weak channels. The original network topology is shown on the left side of Figure \ref{fig:switch}, where the cross-channels are weak and the direct channels are strong. Now, if we re-draw the \emph{same} network but switch the positions of the dark red relay with the dark blue relay, and the light red relay with the light blue relay, then we obtain the representation shown on the right side of Figure \ref{fig:switch}, where the direct channels are weak and cross channels are strong. However, this idea of switching the positions of relays in every alternate hop only works when $L$ is even.
	
\subsubsection{Sum-GDoF vs $L$: Non-Monotonicity}
 Corollary \ref{cor:Lhop} establishes the sum-GDoF in the strong-interference regime when $L$ is even, but leaves the sum-GDoF open for odd $L$ in the same regime. One might expect that the GDoF values for odd $L$ may be sandwiched between their even neighbors. The expectation is supported by the observation that the expressions in \eqref{eq:Lhopweak} and \eqref{eq:Lhopstrong} as well as the illustration in Figure \ref{fig:Lhop_plot} all seem to show that the Sum-GDoF  value monotonically increases with the number of hops, $L$. In fact, the gap between plots is rather small in Figure \ref{fig:Lhop_plot}, which suggests that the sum-GDoF values for odd $L$ may be estimated quite accurately from the neighboring even $L$ values. Somewhat surprisingly, this is not the case, as we show in this section. To highlight the non-monotonic behavior of sum-GDoF vs $L$, we characterize the sum-GDoF for odd $L$ in the \emph{very} strong interference regime, in the following theorem.
    
	     \begin{theorem}\label{thm:Lhopvstrong}
    	If $L$ is odd, then for the $L$-hop layered symmetric interference channel under finite precision CSIT, in the very strong interference regime where $\alpha\geq L+1$, the sum-GDoF value is  given by,
		\begin{align}
		\mathcal{D}_{\Sigma}^{\tiny f.p.}=2L = \mathcal{D}_{\Sigma}^{\tiny p}.\label{eq:Lhopvstrong}
		\end{align}
		\end{theorem}
		The converse proof of Theorem \ref{thm:Lhopvstrong} is presented in Section \ref{sec:Lhopvstrongcon} and the achievability is proved in Section \ref{sec:Lhopvstrongach}. Both are relatively straightforward. The converse is simply the min-cut bound, and achievability is a rate-splitting partitioning of multiple decode-and-forward schemes that require some filtering and rearrangement  of the superposition order of  codewords as they pass through the relays. Since the min-cut bound applies equally under perfect CSIT, the result of Theorem \ref{thm:Lhopvstrong} also holds under perfect CSIT.
		
\begin{figure}[h]
    		\centering
    		\includegraphics[width=10cm]{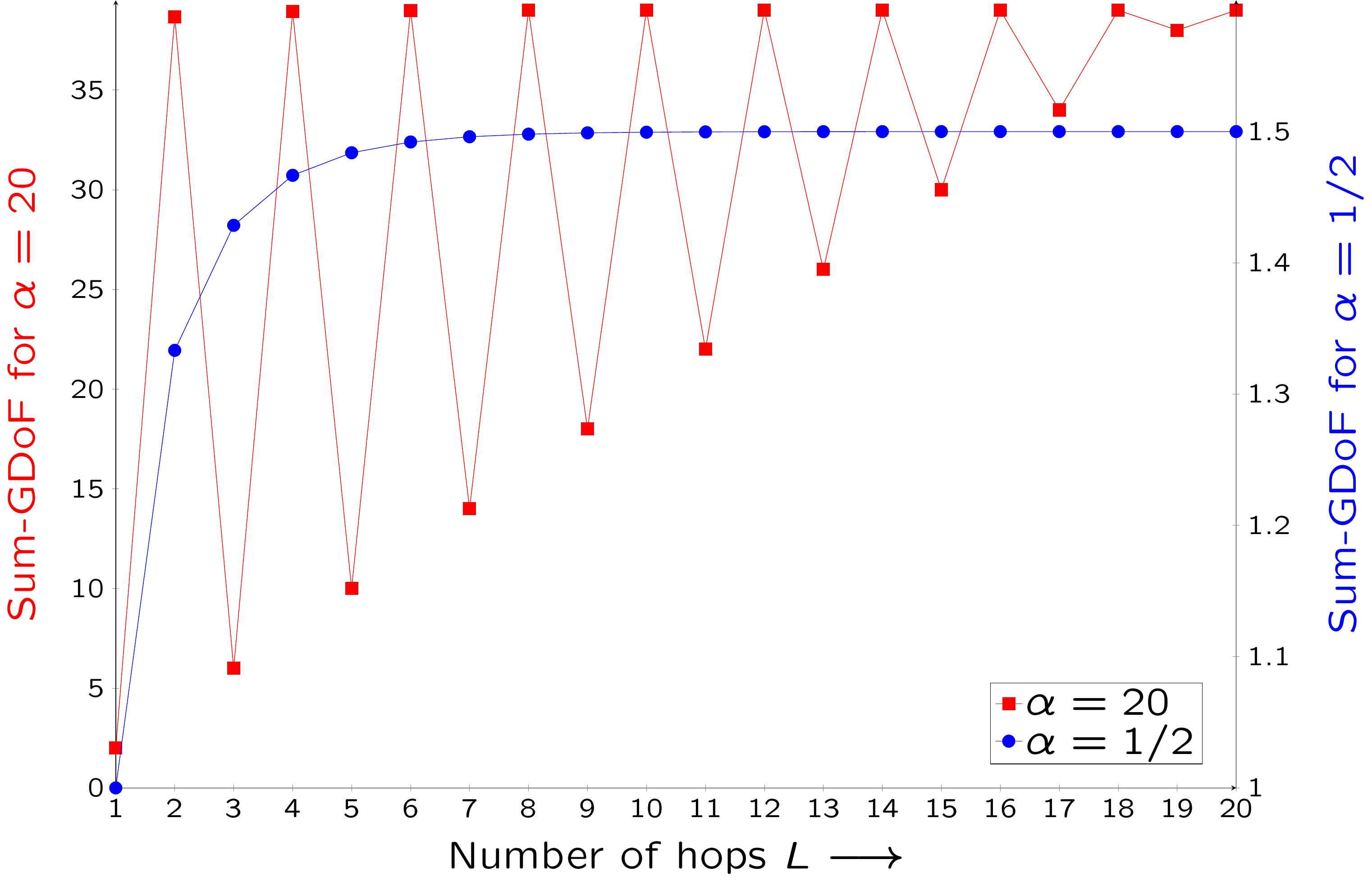}
    		\caption{\it\small Sum-GDoF of the layered symmetric $L$-hop interference channel vs the number of hops $L$ for $\alpha = 1/2$ (weak interference) shown in blue, and $\alpha=20$ (very strong interference) shown in red.}
    		\label{fig:mon}
    	\end{figure}

From Corollary \ref{cor:Lhop} and Theorem \ref{thm:Lhopvstrong} we note that as $\alpha\rightarrow\infty$ the sum-GDoF value of the $L$ hop layered symmetric interference channel approaches infinity if $L$ is even, but is only $2L$ if $L$ is odd, thus proving that the sum-GDoF value is not a monotonic function of $L$.  To see this intuitively, consider again the network shown on the right side of Figure \ref{fig:switch} and for this intuitive understanding assume that the dashed links are extremely weak (say, strength $0$) while the solid links are extremely strong (say, strength approaching infinity). In this $L=4$ hop network, consider the communication from Source $1$ to Destination $1$, for which there exists a very strong path, so the GDoF of this communication approach infinity. However, suppose the network had only $3$ hops, so Destination $1$ was the light blue node. Note that in this $L=3$ hop network there exists no path from Source $1$ to Destination $1$, i.e., the GDoF of this communication is $0$. This toy example intuitively shows  why we notice abrupt drops of sum-GDoF for odd $L$ in the very strong interference regime. Figure \ref{fig:mon} illustrates this fact as we note the different behaviors of sum-GDoF vs $L$ in the weak (monotonic) and very strong (non-monotonic) regimes.

To summarize our results for the $L$ hop layered symmetric interference channel, we have found the robust sum-GDoF for arbitrary number of hops $L$ if the network is in the weak interference regime $(\alpha\leq 1$) which is our main focus, or the very strong interference regime $(\alpha\geq L+1)$. For the remaining strong interference regime $(1\leq \alpha\leq L+1)$, we have found the sum-GDoF if $L$ is even, but the sum-GDoF for odd $L$ remain a non-trivial open problem in the sense that the answer may not even be approximated by sandwiching between adjacent even values of $L$. 
			
\section{Converse Proofs}\label{sec:upper}
In this section we provide the converse proofs for Theorem \ref{thm:2hop}, Theorem \ref{thm:Lhop} and Theorem \ref{thm:Lhopvstrong}. We start with the basic definitions, inherited from \cite{Arash_Jafar}\cite{Arash_Jafar_sumset}, that are essential for Aligned Images bounds.
\subsection{Definitions}
\begin{definition}[Power Levels]\label{def:plevel} An integer valued random variables $X_i$ with power level $\lambda_i$ takes values over the alphabet set $\mathcal{X}_{\lambda_i}$ defined as
	\begin{align}
	\mathcal{X}_{\lambda_i}\triangleq\{0,1,2,\cdots,\overline{P}^{\lambda_i}-1\}
	\end{align}
	where $\overline{P}^{\lambda_i}\triangleq\lfloor \sqrt{P^{\lambda_i}}\rfloor $. We are primarily interested in limits as $P\rightarrow\infty$, where $P\in \mathbb{R}_+$ is referred to as power. 
	\end{definition}

\begin{definition} For integer valued random variables  $X\in\mathcal{X}_{\lambda}$, and any non-negative real numbers $\lambda_1,\lambda_2$ such that $0\leq\lambda_1\leq\lambda_2\leq\lambda$, define
	\begin{align}
	(X)^{\lambda_2}&\triangleq \Bigl\lfloor\frac{X}{\overline{P}^{\lambda-\lambda_2}} \Bigr\rfloor,\label{def_top}\\
	(X)_{\lambda_1}&\triangleq X-\overline{P}^{\lambda_1}\Big\lfloor\frac{X}{\overline{P}^{\lambda_1}}\Big\rfloor,\label{def_bottom}\\
	(X)^{\lambda_2}_{\lambda_1}&\triangleq\Big\lfloor\frac{(X)_{\lambda_2}}{\overline{P}^{\lambda_1}}\Big\rfloor\label{def_medium}.
	\end{align}
	In other words, $(X)^{\lambda_2}$ retrieves the top $\lambda_2$ power levels of $X$, $(X)_{\lambda_1}$ retrieves the bottom $\lambda_1$ power levels of $X$ and $(X)^{\lambda_2}_{\lambda_1}$ retrieves the partition of $X$ between power levels $\lambda_1$ and $\lambda_2$. As a somewhat oversimplified interpretation for intuitive  purposes, $X$ can be thought as a non-negative integer value represented in $\sqrt{P}$-ary alphabet expansion, as $X=x_{\lambda}x_{\lambda-1}\cdots x_2x_1$, and $(X)^{\lambda_2}$ retrieves the most significant $\lambda_2$  symbols, i.e., $(X)^{\lambda_2}=x_{\lambda}\cdots x_{\lambda-\lambda_2+1}$. Similarly, $(X)^{\lambda_2}_{\lambda_1}$ is the sub-string $x_{\lambda_2}\cdots x_{\lambda_1}$, $(X)_{\lambda_1}$ is the sub-string $x_{\lambda_1-1}\cdots x_{1}$. This is oversimplified because $\lambda_1,\lambda_2,\lambda$ are not restricted to take only integer values.  This is a generalization to the ADT models \cite{Avestimehr_Diggavi_Tse}, where binary expansions are used to study GDoF under perfect CSIT. 
\end{definition}

\begin{definition}[Sub-section, Interval, Level, Size, Disjoint] For $X\in\mathcal{X}_\lambda$, we define $(X)_{\lambda_1}^{\lambda_2}$ as a `sub-section' of $X$ if $0\leq\lambda_1\leq\lambda_2\leq\lambda$, where  $(\lambda_1,\lambda_2)$ is the corresponding `interval'.  Furthermore, we define the lower end of the interval $(\lambda_1,\lambda_2)$ as the `level' of the partition, denoted as $\ell((X)_{\lambda_1}^{\lambda_2})=\lambda_1$. The length of the interval $(\lambda_1,\lambda_2)$, denoted as $\mathcal{T}((X)_{\lambda_1}^{\lambda_2})=\lambda_2-\lambda_1$ is called the `size' of the partition. Sub-sections $(X)_{\lambda_1}^{\lambda_2}$ and $(X)_{\nu_1}^{\nu_2}$ of the same $X\in\mathcal{X}_\lambda$ are `disjoint' if the two intervals $(\lambda_1,\lambda_2)$ and $(\nu_1,\nu_2)$ are disjoint.
\end{definition}

Next we recall the definition of the particular deterministic transformation \cite{Arash_Jafar} that is used   for Aligned Images bounds. The transformation has thus far been used only in single-hop settings, and as noted previously, extensions to multihop settings are not immediate. Fortunately, for our purpose and for all our arguments we only need to apply the deterministic transformation to one of the $L$ hops at any time, say the $\ell^{th}$ hop. This transformation for the $\ell^{th}$ hop is defined next.

\allowdisplaybreaks
\begin{definition} [Deterministic Transformation of the $\ell^{th}$ hop]\label{def:det} 
	In the $\ell^{th}$ hop, define the mapping from the original input $X_{i[\ell]}$ to the deterministic input $\overline{X}_{i[\ell]}$ as 
	\begin{align}
	\overline{X}_{i[\ell]}=\lfloor X_{i[\ell]}\rfloor \mod \lceil \sqrt{P^{\max(1,\alpha)}} \rceil \label{eq:X2Xbar}
	\end{align}
	 such that $\overline{X}_{i[\ell]}(n)=\overline{X}_{iR[\ell]}(n)+j\overline{X}_{iI[\ell]}(n)$, $i\in[1:2]$ and $\overline{X}_{iR[\ell]}(n),\overline{X}_{iI[\ell]}(n)\in\{0,1,2,\cdots,$ $\lceil\sqrt{P^{\max(1,\alpha)}}\rceil -1\}$ for all $n\in[1:N]$. Then the deterministic transformation for the $\ell^{th}$ hop is represented as follows:
	\begin{align}
	\overline{Y}_{1[\ell]}(n)&=\lfloor\sqrt{P^{1-\max(1,\alpha)}}G_{11[\ell]}(n)\overline{X}_{1[\ell]}(n)\rfloor+\lfloor\sqrt{P^{\alpha-\max(1,\alpha)}}G_{12[\ell]}(n)\overline{X}_{2[\ell]}(n)\rfloor\\
	\overline{Y}_{2[\ell]}(n)&=\lfloor\sqrt{P^{\alpha-\max(1,\alpha)}}G_{21[\ell]}(n)\overline{X}_{1[\ell]}(n)\rfloor+\lfloor\sqrt{P^{1-\max(1,\alpha)}}G_{22[\ell]}(n)\overline{X}_{2[\ell]}(n)\rfloor
	\end{align}
	Note that $|\overline{Y}_{i[\ell]}(n)|\leq 4\sqrt{P}\Delta$, and since the real and imaginary parts of $\overline{Y}_{i[\ell]}(n)$ are both integer valued, we must have $H(\overline{Y}_{i[\ell]}(n))\leq 2\log(8\sqrt{P}\Delta)=\log(P)+o(\log(P)) $. Similarly over $N$ channel uses, we must have
	\begin{align}
	H(\overline{Y}_{i[\ell]}^{[N]})\leq N\log(P)+o(\log(P)).\label{eq:maxHY}
	\end{align}
	As noted in \cite{Junge_Yuan_Huang_Jafar}, we can also represent this as:
	\begin{align}
	\overline{Y}_{1[\ell]}(n)&=\lfloor G_{11[\ell]}(n)\left({\overline X}_{1[\ell]}(n)\right)^{1}\rfloor+\lfloor G_{12[\ell]}(n)\left({\overline X}_{2[\ell]}(n)\right)^{\alpha}\rfloor+\zeta_{1[\ell]}(n)\label{deticmodel1}\\
	\overline{Y}_{2[\ell]}(n)&=\lfloor G_{21[\ell]}(n)\left({\overline X}_{1[\ell]}(n)\right)^{\alpha}\rfloor+\lfloor G_{22[\ell]}(n)\left({\overline X}_{2[\ell]}(n)\right)^{1}\rfloor+\zeta_{2[\ell]}(n)\label{deticmodel2}
	\end{align}
	where $\zeta_{1[\ell]}(n), \zeta_{2[\ell]}(n)$ are complex random variables whose real and imaginary parts are integer valued and whose magnitude is bounded, so it does not scale with $P$. Specifically, $\max(|\zeta_{1[\ell]}(n)|, |\zeta_{2[\ell]}(n)|)\leq 2(2+\Delta)=o(\log(P))$.
\end{definition}
\subsection{Lemmas}\label{sec:lemmas}
Our first three lemmas are  inherited from prior works \cite{Arash_Jafar, Arash_Jafar_cooperation, Arash_Jafar_sumset, Junge_Yuan_Huang_Jafar} on Aligned Images bounds in single-hop scenarios, and specialized to our multihop setting where the deterministic transformation has been applied only to the $\ell^{th}$ hop.

\begin{lemma}[Deterministic Bound \cite{Arash_Jafar}] \label{lemma:deter} With the deterministic transformation applied only to the $\ell^{th}$ hop,  we have the following bound for $i\in[1:2]$,
	\begin{align}
	I(W_i;Y_{i[\ell]}^{[N]}\mid\mathcal{G}_{[1:\ell]})\leq I(W_i; \overline{Y}_{i[\ell]}^{[N]}\mid\mathcal{G}_{[1:\ell]})+No(\log(P))\label{eq:deter}.
	\end{align}
\end{lemma}
Lemma \ref{lemma:deter} above is obtained from Lemma 1 of \cite{Arash_Jafar} as applied to our setting. Recall that in \cite{Arash_Jafar} the same deterministic transformation  that we apply to the $\ell^{th}$ hop, is applied to the one-hop MISO BC, and it is shown that this cannot reduce the mutual information in the GDoF sense between the messages and the corresponding deterministic outputs, conditioned on the one-hop channels for which only finite precision CSIT is available to the transmitters. There are two key distinctions in our  setting. First, unlike the MISO BC where the transmitters cooperate fully, because the transmitters in the $\ell^{th}$ hop do not directly have access to the messages, their coding functions are more restricted. Second, because transmitters in the $\ell^{th}$ hop have knowledge of channel realizations of preceding hops, the coding functions may utilize this knowledge, which means that the transmitted symbols need not be independent of the $\mathcal{G}_{[1:\ell-1]}$ terms that are included in the conditioning in \eqref{eq:deter}, unlike the one-hop MISO BC where the transmitted symbols are independent of the channels that appear in the conditioning. However, neither of these distinctions affects the validity of Lemma \ref{lemma:deter} because upon  inspection of the proof of Lemma 1 of \cite{Arash_Jafar} it becomes evident that the proof holds for \emph{all} feasible coding functions in the MISO BC, which includes the restricted class of coding functions available to the transmitters in the $\ell^{th}$ hop in the multihop setting. Furthermore, it turns out that the proof of Lemma 1 of \cite{Arash_Jafar} also holds under the additional conditioning on the channels $\mathcal{G}_{[1:\ell-1]}$ which are not necessarily independent of the transmitted symbols; what matters for the proof is that these additional conditioning terms are independent of the additive noise encountered by the receivers of the $\ell^{th}$ hop. Thus, the proof of Lemma 1 of \cite{Arash_Jafar} carries over  to Lemma \ref{lemma:deter} in this paper. For the sake of completeness, the proof is summarized in the Appendix section.

{\it Remark: Note that because $\mathcal{G}_{[\ell+1:L]}$ is independent of all terms that appear in \eqref{eq:deter}, the result of Lemma \ref{lemma:deter} can also be stated with  additional conditioning on $\mathcal{G}_{[\ell+1:L]}$ as:}
	\begin{align}
	I(W_i;Y_{i[\ell]}^{[N]}\mid\mathcal{G}_{[1:L]})\leq I(W_i; \overline{Y}_{i[\ell]}^{[N]}\mid\mathcal{G}_{[1:L]})+No(\log(P))\label{eq:deterfull}.
	\end{align}

Next let us recall two sum-set inequalities that will be critical to our converse proofs, as applied to our setting. The first sum-set inequality, namely Sum-set Inequality $1$, originally shown in \cite[Theorem 1]{Arash_Jafar_cooperation}, is used to bound the entropy difference of two received signals in the GDoF sense. Intuitively, this sum-set inequality says that in the GDoF sense the entropy difference is upper bounded by the maximum difference  of the corresponding  link strengths. The inequality applies to our setting because, as explained for Lemma \ref{lemma:deter}, the original version in \cite{Arash_Jafar_cooperation} is proved for the MISO broadcast channel which allows arbitrary coding functions, including the ones available to the transmitters in the $\ell^{th}$ hop.
\begin{lemma} \label{lemma:sumset1}(Sum-set Inequality $1$)
	Let $\overline{U}_{i[\ell]}^{[N]}=\lfloor G_{i1[\ell]}^{[N]}(\overline{X}_{1[\ell]}^{[N]})^{\mu_i}\rfloor+\lfloor G_{i2[\ell]}^{[N]}(\overline{X}_{2[\ell]}^{[N]})^{\nu_i}\rfloor$, then
	\begin{align}
	H(\overline{U}_{1[\ell]}^{[N]}\mid\mathcal{W}_S,\mathcal{G}_{[\ell]})-H(\overline{U}_{2[\ell]}^{[N]}\mid\mathcal{W}_S,\mathcal{G}_{[\ell]})\leq\max(\mu_1-\mu_2,\nu_1-\nu_2)^+N\log(P)+No(\log(P)),\label{eq:sumset1}
	\end{align}
	where $\mathcal{W}_S$ is a set of random variables satisfying \begin{align}
	I(\overline{X}_{1[\ell]}^{[N]},\overline{X}_{2[\ell]}^{[N]},\mathcal{W}_S; \mathcal{G}_{[\ell]})=0.\label{eq:Wcondition}
	\end{align}
\end{lemma}

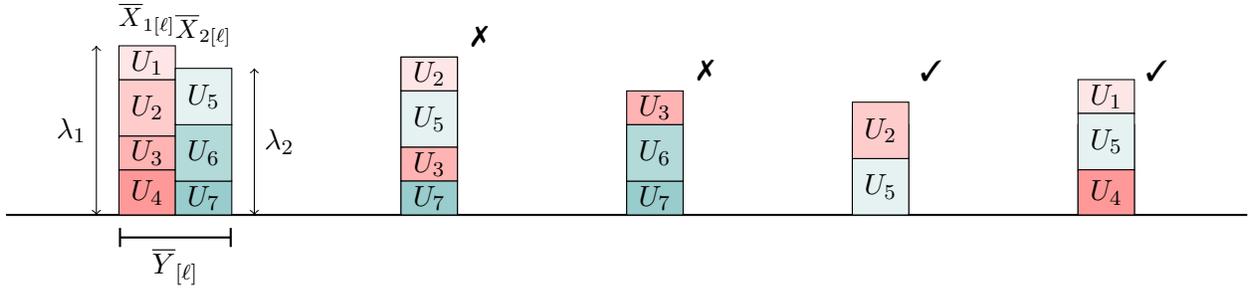
\begin{figure}[t]
	\centering
	\begin{tikzpicture}[scale = 1.5, baseline=(current bounding box.center)]
	\draw [thick] (1,-0.5)--(12,-0.5);
	\draw  [fill=red!40](2,-0.5) rectangle (2.5, -0.1) node[pos=.5] {$U_4$};
	\draw  [fill=red!30](2,-0.1) rectangle (2.5, 0.2) node[pos=.5] {$U_3$};
	\draw  [fill=red!20](2,0.2) rectangle (2.5, 0.7) node[pos=.5] {$U_2$};
	\draw  [fill=red!10](2,0.7) rectangle (2.5, 1) node[pos=.5] {$U_1$};
	\node at (2.25,1) [above]{\small $\overline{X}_{1[\ell]}$};
	\draw [<->, thin] (1.8, -0.5)--(1.8,1) node [midway, left] {$\lambda_1$};
	\node at (2.75,0.9) [above]{\small $\overline{X}_{2[\ell]}$};
	\draw  [fill=teal!40](2.5, -0.5) rectangle (3, -0.2) node[pos=.5] {$U_{7}$};
	\draw  [fill=teal!30](2.5, -0.2) rectangle (3, 0.3) node[pos=.5] {$U_{6}$};
	\draw [fill=teal!10] (2.5, 0.3)rectangle(3,0.8) node [pos=.5] {$U_{5}$};
	\draw [<->, thin] (3.2, -0.5)--(3.2,0.8) node [midway, right] {$\lambda_2$};
	\draw[thick, |-|] (2,-0.7)--(3,-0.7) node[midway, below]{$\overline{Y}_{[\ell]}$};
	
	\draw (4.5,-0.5) rectangle(5,0.9);
	\draw  [fill=red!10](4.5,0.6) rectangle (5, 0.9) node[pos=.5] {$U_2$};
	\draw [fill=teal!10] (4.5, 0.1)rectangle(5,0.6) node [pos=.5] {$U_{5}$};
	\draw  [fill=red!30](4.5,-0.2) rectangle (5, 0.1) node[pos=.5] {$U_3$};
	\draw  [fill=teal!40](4.5, -0.5) rectangle (5, -0.2) node[pos=.5] {$U_{7}$};
	\node at (5.2,0.9)[above]{\xmark};
	
	\draw (6.5,-0.5) rectangle(7,0.3);
	\draw  [fill=red!30](6.5,0.3) rectangle (7, 0.6) node[pos=.5] {$U_3$};
	\draw  [fill=teal!40](6.5, -0.5) rectangle (7, -0.2) node[pos=.5] {$U_{7}$};
	\draw  [fill=teal!30](6.5, -0.2) rectangle (7, 0.3) node[pos=.5] {$U_{6}$};
	\node at (7.2,0.6)[above]{\xmark};
	
	\draw (8.5,-0.5) rectangle(9,0.3);
	\draw [fill=teal!10] (8.5, 0)rectangle(9,-0.5) node [pos=.5] {$U_{5}$};
	\draw [fill=red!20] (8.5,0) rectangle (9, 0.5) node[pos=.5] {$U_2$};
	\node at (9.2,0.6)[above]{\cmark};
	
	\draw (10.5,-0.5) rectangle(11,0.3);
	\draw  [fill=red!10](10.5,0.7) rectangle (11, 0.4) node[pos=.5] {$U_1$};
	\draw [fill=teal!10] (10.5, 0.4)rectangle(11,-.1) node [pos=.5] {$U_5$};
	\draw  [fill=red!40](10.5, -0.1) rectangle (11, -0.5) node[pos=.5] {$U_4$};
	\node at (11.2,0.6)[above]{\cmark};
	\end{tikzpicture}
	\caption{\it\small An illustration  of Lemma \ref{lemma:sumset2}. Lemma \ref{lemma:sumset2} implies the sum-set inequalities $H(\overline{Y}_{[\ell]}^{[N]}\mid W_S,\mathcal{G}_{[\ell]})\geq H(U_2^{[N]},U_5^{[N]}\mid W_S,\mathcal{G}_{[\ell]})$ and $H(\overline{Y}^{[N]}\mid W_S,\mathcal{G}_{[\ell]})\geq H(U_1^{[N]},U_4^{[N]},U_5^{[N]}\mid W_S,\mathcal{G}_{[\ell]})$ in the GDoF sense because the boxes in these inequalities can be vertically stacked without elevating any sub-section of them above its original height in $\overline{Y}$. However, Lemma \ref{lemma:sumset2} implies neither $H(\overline{Y}^{[N]}\mid W_S,\mathcal{G}_{[\ell]})\geq H(U_2^{[N]},U_3^{[N]},U_5^{[N]},U_7^{[N]}\mid W_S,\mathcal{G}_{[\ell]})$ nor $H(\overline{Y}^{[N]}\mid W_S,\mathcal{G}_{[\ell]})\geq H(U_3^{[N]},U_6^{[N]},U_7^{[N]}\mid W_S,\mathcal{G}_{[\ell]})$, because it is impossible to vertically stack the corresponding boxes in any order without elevating at least one of them above its original position in $\overline{Y}$. }\label{fig:sumset2}
\end{figure}

The next sum-set inequality, Sum-set Inequality $2$, appeared originally  in a generalized form in \cite[Theorem 4]{Arash_Jafar_sumset}. The following simplified form, taken from \cite[Lemma 1]{Junge_Yuan_Huang_Jafar} and specialized to our setting, is sufficient for our purpose.
\begin{lemma}\label{lemma:sumset2} (Sum-set Inequality $2$)
	Let $\overline{Y}_{[\ell]}(n)=\sum_{k=1}^2 \lfloor G_{k[\ell]}(n)\overline{X}_{k[\ell]}(n)\rfloor$ for $\overline{X}_{k[\ell]}(n)\in\mathcal{X}_{\mu_{k[\ell]}}$, and let $G_{k[\ell]}(n)$ be distinct elements of $\mathcal{G}$ for all $k\in[2], n\in[1:N]$. For all $k\in[2]$, let $S_k$ be a set of finitely many disjoint sub-sections of $\overline{X}_{k[\ell]}$ (the same partitioning is applied to $\overline{X}_{k[\ell]}(n)$ for every $t$), and let $\{U_{1}, U_{2}, \cdots, U_{m}\}$ be a subset of $S_1 \cup S_2$. The  following sum-set inequality holds,
	\begin{align}
	H(&\overline{Y}_{[\ell]}^{[N]}\mid \mathcal{W}_S,\mathcal{G}_{[\ell]})
	\geq H(U_{1}^{[N]}, U_{2}^{[N]}, \cdots, U_{m}^{[N]}\mid \mathcal{W}_S,\mathcal{G}_{[\ell]})+N o(\log(P)),\label{eq:sumset2}
	\end{align}
	if both of the following conditions are satisfied.
	\begin{align}
	I\Big(\overline{X}_{1[\ell]}^{[N]},\overline{X}_{2[\ell]}^{[N]},\mathcal{W}_S~;~ \mathcal{G}_{[\ell]}\Big)&=0,\label{eq:indcond}\\
	 \sum_{j=1}^{i-1}\mathcal{T}(U_{j})&\leq \ell(U_{i}),&&\forall i\in[2:m].\label{eq:stack1}
	\end{align}
	
\end{lemma}
Condition \eqref{eq:stack1} can be visualized in terms of a vertical stacking of  $m$ boxes $U_{1}, \cdots, U_{m}$ in that order from bottom to top where the $j^{th}$ box has height $\mathcal{T}(U_j)$. Conditions \eqref{eq:stack1} simply means that the height at which the $i^{th}$ box appears in the vertical stacking (the LHS of \eqref{eq:stack1}) should not be higher than its original level in $\overline{Y}_{[\ell]}$, i.e., $\ell(U_i)$. In other words, if there exists \emph{any} ordering such that we can vertically stack all of the sub-sections 
without lifting up any one of them above its original height in $\overline{Y}_{[\ell]}$, then the sum-set inequality \eqref{eq:sumset2} holds. Figure \ref{fig:sumset2} presents a few examples that satisfy or violate Lemma \ref{lemma:sumset2}. 

Next we present our main lemma that is developed in this work specifically for the multihop setting, to capture a recursive bounding argument which will allow us to use the deterministic bounds for each hop,  one hop at a time.
	\begin{lemma}\label{lemma:recur}
	The following inequality holds for any $\ell\in[2:L]$.
	\begin{align}
	&NR_1+NR_2+2I(W_1;Y_{1[\ell]}^{[N]}\mid \mathcal{G}_{[1:L]})+2I(W_2;Y_{2[\ell]}^{[N]}\mid \mathcal{G}_{[1:L]}) \notag\\
	& \leq
	(2+2\max(1-\alpha,\alpha))N\log(P)+I(W_1;Y_{1[\ell-1]}^{[N]}\mid \mathcal{G}_{[1:L]})+I(W_2;Y_{2[\ell-1]}^{[N]}\mid \mathcal{G}_{[1:L]})\notag\\
	&\hspace{1cm} + No(\log(P))\label{inq:701}
	\end{align}
	\end{lemma}

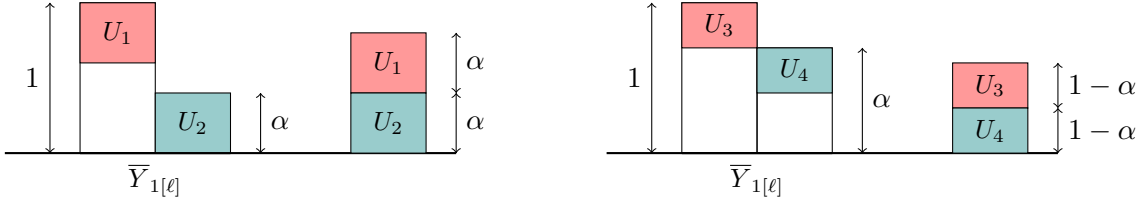
\begin{figure}
\centering
	\begin{tikzpicture}[scale=2]
	\draw [thick] (1.5,0)--(4.5,0);
	\draw  (2,0) rectangle (2.5, 1);
	\draw  [fill=red!40](2,0.6) rectangle (2.5, 1) node[pos=.5] {\small $U_1$};
	\draw [<->, thin] (1.8, 0)--(1.8,1) node [midway, left] {$1$};
	\draw  [fill=teal!40](2.5, 0) rectangle (3, 0.4) node[pos=.5] {\small$U_2$};
	\draw [<->, thin] (3.2, 0)--(3.2,0.4) node [midway, right] {$\alpha$};
	\draw[thick] (2,-0)--(3,-0) node[midway, below]{\small $\overline{Y}_{1[\ell]}$};
	
	\draw  [fill=red!40](3.8,0.4) rectangle (4.3, 0.8) node[pos=.5] {\small $U_1$};
	\draw [<->, thin] (4.5, 0.4)--(4.5,0.8) node [midway, right] {$\alpha$};
	\draw  [fill=teal!40](3.8, 0) rectangle (4.3, 0.4) node[pos=.5] {\small $U_2$};
	\draw [<->, thin] (4.5, 0)--(4.5,0.4) node [midway, right] {$\alpha$};
	
	\begin{scope}[shift={(4,0)}]
	\draw [thick] (1.5,0)--(4.5,0);
	\draw  (2,0) rectangle (2.5, 1);
	\draw  [fill=red!40](2,0.7) rectangle (2.5, 1) node[pos=.5] {\small $U_3$};
	\draw [<->, thin] (1.8, 0)--(1.8,1) node [midway, left] {$1$};
	\draw  (2.5,0) rectangle (3, 0.7);
	\draw  [fill=teal!40](2.5, 0.4) rectangle (3, 0.7) node[pos=.5] {\small $U_4$};
	\draw [<->, thin] (3.2, 0)--(3.2,0.7) node [midway, right] {$\alpha$};
	\draw[thick] (2,-0)--(3,-0) node[midway, below]{\small $\overline{Y}_{1[\ell]}$};
	
	\draw  [fill=red!40](3.8,0.3) rectangle (4.3, 0.6) node[pos=.5] {\small  $U_3$};
	\draw [<->, thin] (4.5, 0.6)--(4.5,0.3) node [midway, right] {$1-\alpha$};
	\draw  [fill=teal!40](3.8, 0) rectangle (4.3, 0.3) node[pos=.5] {\small $U_4$};
	\draw [<->, thin] (4.5, 0)--(4.5,0.3) node [midway, right] {$1-\alpha$};
	\end{scope}
	\end{tikzpicture}
	\caption{\it In the left figure, $\alpha\leq\frac{1}{2}, U_1=(\overline{X}_{1[\ell]})^\alpha,U_2=(\overline{X}_{2[\ell]})^\alpha$, evidently $U_1,U_2$ can be stacked without elevating either one of them above the level at which it appears in $\bar{Y}_{1[\ell]}$. On the right, $\alpha\geq\frac{1}{2}$, $ U_3=(\overline{X}_{1[\ell]})^{1-\alpha}$, $U_4=(\overline{X}_{2[\ell]})^{1-\alpha}$; $U_3,U_4$ can also be stacked without elevating either  of them.}\label{fig:stackproof}
	\end{figure}

	\proof  
	For compact notation, in this proof we will occasionally suppress $N o(\log(P))$ terms that are inconsequential for GDoF. Starting with \eqref{eq:deterfull}, we have 
	\begin{align}
		I(W_1;Y_{1[\ell]}^{[N]}\mid \mathcal{G}_{[1:L]})
		&\leq I(W_1; \overline{Y}_{1[\ell]}^{[N]}\mid \mathcal{G}_{[1:L]})\label{inq:707}\\
		I(W_2;Y_{2[\ell]}^{[N]}\mid \mathcal{G}_{[1:L]})&\leq I(W_2;\overline{Y}_{2[\ell]}^{[N]}\mid \mathcal{G}_{[1:L]})\label{inq:708}
	\end{align}
	Next, we have the Markov chain,
	\begin{align}
	(W_1,W_2)\leftrightarrow (Y_{1[\ell]}^{[N]},Y_{2[\ell]}^{[N]},  \mathcal{G}_{[1:\ell]})\leftrightarrow (Y_{1[L]}^{[N]},Y_{2[L]}^{[N]},  \mathcal{G}_{[1:L]})\label{eq:markov}
	\end{align}
	Using this Markov chain and the data-processing inequality, we proceed as follows.
	\begin{align}
		NR_1+NR_2&\leq I\left(W_1,W_2; Y_{1[L]}^{[N]},Y_{2[L]}^{[N]}, \mathcal{G}_{[1:L]}\right)\label{eq:explain0}\\
			&\leq I\left(W_1,W_2; Y_{1[\ell]}^{[N]},Y_{2[\ell]}^{[N]}, \mathcal{G}_{[1:\ell]}\right)\label{eq:explain1}\\
			&=I\left(W_1,W_2; Y_{1[\ell]}^{[N]},Y_{2[\ell]}^{[N]}\mid \mathcal{G}_{[1:\ell]}\right)\label{eq:explain2}\\
			&\leq I(W_1,W_2; \overline{Y}_{1[\ell]}^{[N]},\overline{Y}_{2[\ell]}^{[N]}\mid \mathcal{G}_{[1:\ell]})\label{ineq:vecdeter}\\
			&\leq I(W_1,W_2; \overline{X}_{1[\ell]}^{[N]},\overline{X}_{2[\ell]}^{[N]},\overline{Y}_{1[\ell]}^{[N]},\overline{Y}_{2[\ell]}^{[N]}\mid \mathcal{G}_{[1:\ell]})\label{eq:explain4}\\
			&= I(W_1,W_2; \overline{X}_{1[\ell]}^{[N]},\overline{X}_{2[\ell]}^{[N]}\mid \mathcal{G}_{[1:\ell]})\label{ineq:41}\\
			&\leq H(\overline{X}_{1[\ell]}^{[N]},\overline{X}_{2[\ell]}^{[N]}\mid \mathcal{G}_{[1:\ell]})\label{eq:explain6}\\
			&=H(\overline{X}_{1[\ell]}^{[N]},\overline{X}_{2[\ell]}^{[N]}\mid \mathcal{G}_{[1:L]})\label{inq:905}\\
			&=H(\overline{X}_{1[\ell]}^{[N]}\mid \mathcal{G}_{[1:L]})+H(\overline{X}_{2[\ell]}^{[N]}\mid \overline{X}_{1[\ell]}^{[N]},\mathcal{G}_{[1:L]})\label{ineq:1}
	\end{align}
	Step \eqref{eq:explain0} is obtained by Fano's inequality. Step \eqref{eq:explain1} follows from the Markov Chain in \eqref{eq:markov} and the data-processing inequality. Step \eqref{eq:explain2} uses the chain rule of mutual information and the fact that the messages are independent of the channels. Step \eqref{ineq:vecdeter} is proved in Appendix B. Step \eqref{eq:explain4} uses the property that $I(A;B\mid C)\leq I(A;B,D\mid C)$. Step \eqref{ineq:41} is because $(\overline{Y}_{1[\ell]}^{[N]},\overline{Y}_{2[\ell]}^{[N]})$ is determined by $(\overline{X}_{1[\ell]}^{[N]},\overline{X}_{2[\ell]}^{[N]},\mathcal{G}_{[\ell]})$. Step \eqref{eq:explain6} uses the definition of mutual information $I(A;B\mid C)=H(B\mid C)-H(B\mid A,C)$ and the non-negativity of entropy in dropping the negative term. Including $\mathcal{G}_{[\ell+1:L]}$ in the conditioning in \eqref{inq:905} is justified because these channels are independent of all the other terms that appear in the entropy expression. Step \eqref{ineq:1} is simply the chain rule of entropy.
	Next,
	\begin{align}
	I(W_1;\overline{Y}_{1[\ell]}^{[N]}\mid \mathcal{G}_{[1:L]})&=H(\overline{Y}_{1[\ell]}^{[N]}\mid \mathcal{G}_{[1:L]})-H(\overline{Y}_{1[\ell]}^{[N]}\mid W_1,\mathcal{G}_{[1:L]})\\
	&\leq H(\overline{Y}_{1[\ell]}^{[N]}\mid \mathcal{G}_{[1:L]})-H(\overline{X}_{1[\ell]}^{[N]}\mid W_1, \mathcal{G}_{[1:L]})\label{iq9}\\
	&= H(\overline{Y}_{1[\ell]}^{[N]}\mid \mathcal{G}_{[1:L]})+I(\overline{X}_{1[\ell]}^{[N]}; W_1\mid \mathcal{G}_{[1:L]})-H(\overline{X}_{1[\ell]}^{[N]}\mid \mathcal{G}_{[1:L]})\\
	&\leq H(\overline{Y}_{1[\ell]}^{[N]}\mid \mathcal{G}_{[1:L]})+I(Y_{1[\ell-1]}^{[N]}; W_1\mid \mathcal{G}_{[1:L]})-H(\overline{X}_{1[\ell]}^{[N]}\mid \mathcal{G}_{[1:L]})\label{ineq:2}\\
	&\leq N\log(P)+I(W_1; Y_{1[\ell-1]}^{[N]} \mid \mathcal{G}_{[1:L]})-H(\overline{X}_{1[\ell]}^{[N]}\mid \mathcal{G}_{[1:L]})\label{ineq:lm2_1}
	\end{align}
	Notably in \eqref{iq9}  we used Sum-set Inequality 1 from Lemma \ref{lemma:sumset1} as follows. Since  $H(\overline{Y}_{1[\ell]}^{[N]}\mid W_1, \mathcal{G}_{[1:L]})=H(\lfloor G_{11[\ell]}^{[N]}(\overline{X}_{1[\ell]}^{[N]})^1\rfloor+\lfloor G_{12[\ell]}^{[N]}(\overline{X}_{2[\ell]}^{[N]})^\alpha\rfloor\mid W_1, \mathcal{G}_{[1:L]})$ if we set $\mathcal{W}_S=(W_1, \mathcal{G}_{[1:L]\setminus[\ell]})$, then from Lemma \ref{lemma:sumset1} we obtain,
	\begin{align}
	No(\log(P))&=(1-1,0-\alpha)^++No(\log(P))\\
	&\geq H(\lfloor G_{11[\ell]}^{[N]}(\overline{X}_{1[\ell]}^{[N]})^1\rfloor+\lfloor G_{12[\ell]}^{[N]}(\overline{X}_{2[\ell]}^{[N]})^0\rfloor \mid W_1, \mathcal{G}_{[1:L]})\notag\\
	&\hspace{1cm}-H(\lfloor G_{11[\ell]}^{[N]}(\overline{X}_{1[\ell]}^{[N]})^1\rfloor+\lfloor G_{12[\ell]}^{[N]}(\overline{X}_{2[\ell]}^{[N]})^\alpha\rfloor \mid W_1,\mathcal{G}_{[1:L]})\\
	&\geq H(\overline{X}_{1[\ell]}^{[N]}\mid W_1, \mathcal{G}_{[1:L]}) - H(\overline{Y}_{1[\ell]}^{[N]}\mid W_1, \mathcal{G}_{[1:L]}) + No(\log(P))
	\end{align}
	Step \eqref{ineq:2} holds because the output of the relay node is a function of its input signal and the channels of the preceding hops. Specifically,  $\overline{X}_{1[\ell]}^{[N]}$ is a function of ${X}_{1[\ell]}^{[N]}$ according to \eqref{eq:X2Xbar}; and ${X}_{1[\ell]}^{[N]}$  is a function of $(Y_{1[\ell-1]}^{[N]},\mathcal{G}_{[1:\ell-1]})$ according to the relay mapping function, so it is also a function of $(Y_{1[\ell-1]}^{[N]},\mathcal{G}_{[1:L]})$. Step \eqref{ineq:lm2_1} follows from \eqref{eq:maxHY} which uses the fact that a uniform distribution maximizes entropy over a  discrete alphabet of bounded cardinality. By symmetry, it follows from \eqref{ineq:lm2_1}, that we must also have,
	\begin{align}
	I(W_2;\overline{Y}_{2[\ell]}^{[N]}\mid \mathcal{G}_{[1:L]})\leq N\log(P)+I(W_2; Y_{2[\ell-1]}^{[N]}\mid\mathcal{G}_{[1:L]})-H(\overline{X}_{2[\ell]}^{[N]}\mid \mathcal{G}_{[1:L]})\label{ineq:3}
	\end{align}
	Adding \eqref{ineq:1},\eqref{ineq:lm2_1},\eqref{ineq:3} together, we have 
	\begin{align}
	NR_1&+NR_2+I(W_1;\overline{Y}_{1[\ell]}^{[N]}\mid \mathcal{G}_{[1:L]})+I(W_2;\overline{Y}_{2[\ell]}^{[N]}\mid \mathcal{G}_{[1:L]})\notag\\
	&\leq 2N\log(P)+I(W_1;Y_{1[\ell-1]}^{[N]}\mid \mathcal{G}_{[1:L]})+I(W_2;Y_{2[\ell-1]}^{[N]}\mid \mathcal{G}_{[1:L]})\notag\\&\hspace{3cm}+H(\overline{X}_{2[\ell]}^{[N]}\mid \overline{X}_{1[\ell]}^{[N]},\mathcal{G}_{[1:L]})-H(\overline{X}_{2[\ell]}^{[N]}\mid \mathcal{G}_{[1:L]})\\
	&=2N\log(P)+I(W_1;Y_{1[\ell-1]}^{[N]}\mid \mathcal{G}_{[1:L]})+I(W_2;Y_{2[\ell-1]}^{[N]}\mid \mathcal{G}_{[1:L]})-I(\overline{X}_{1[\ell]}^{[N]};\overline{X}_{2[\ell]}^{[N]}\mid \mathcal{G}_{[1:L]})\label{ineq:4a}\\
	&\leq 2N\log(P)+I(W_1;Y_{1[\ell-1]}^{[N]}\mid \mathcal{G}_{[1:L]})+I(W_2;Y_{2[\ell-1]}^{[N]}\mid \mathcal{G}_{[1:L]})\notag\\&\hspace{3cm}-I((\overline{X}_{1[\ell]}^{[N]})^{\min(\alpha,1-\alpha)};(\overline{X}_{2[\ell]}^{[N]})^{\min(\alpha,1-\alpha)}\mid \mathcal{G}_{[1:L]})\label{ineq:4}
	\end{align}
The definition of mutual information was used to obtain \eqref{ineq:4a}, and in \eqref{ineq:4} we used the property that $I(A;B)\geq I(f(A);g(B))$ for any functions $f,g$. Next, we bound $I(W_1;\overline{Y}_{1[\ell]}^{[N]}\mid \mathcal{G}_{[1:L]})$ as follows.
	\begin{align}
	I(W_1;\overline{Y}_{1[\ell]}^{[N]}\mid \mathcal{G}_{[1:L]})&=H(\overline{Y}_{1[\ell]}^{[N]}\mid \mathcal{G}_{[1:L]})-H(\overline{Y}_{1[\ell]}^{[N]}\mid W_1, \mathcal{G}_{[1:L]})\\
	&\leq H(\overline{Y}_{1[\ell]}^{[N]}\mid \mathcal{G}_{[1:L]})-H((\overline{X}_{1[\ell]}^{[N]})^{\min(\alpha,1-\alpha)},(\overline{X}_{2[\ell]}^{[N]})^{\min(\alpha,1-\alpha)} \mid W_1,\mathcal{G}_{[1:L]})\label{ineq:5}
	\end{align}
	This step is significant, because it invokes Sum-set Inequality 2 from Lemma \ref{lemma:sumset2}, noting that $(\overline{X}_{1[\ell]}^{[N]})^{\min(1-\alpha,\alpha)},(\overline{X}_{2[\ell]}^{[N]})^{\min(1-\alpha,\alpha)}$ can be stacked vertically without elevating either of them above their original height in $\overline{Y}_{1[\ell]}^{[N]}$. See Figure \ref{fig:stackproof} for an illustration of the stacking.
	
	Similarly,
	\begin{align}
	I(W_2;\overline{Y}_{2[\ell]}^{[N]}\mid \mathcal{G}_{[1:L]})&\leq H(\overline{Y}_{2[\ell]}^{[N]}\mid \mathcal{G}_{[1:L]})-H((\overline{X}_{1[\ell]}^{[N]})^{\min(\alpha,1-\alpha)},(\overline{X}_{2[\ell]}^{[N]})^{\min(\alpha,1-\alpha)} \mid W_2, \mathcal{G}_{[1:L]})\label{ineq:6}
	\end{align}
	Adding \eqref{ineq:5} and \eqref{ineq:6}, we get
	\begin{align}
	&I(W_1;\overline{Y}_{1[\ell]}^{[N]}\mid \mathcal{G}_{[1:L]})+I(W_2;\overline{Y}_{2[\ell]}^{[N]}\mid \mathcal{G}_{[1:L]})\notag\\ &\leq H(\overline{Y}_{1[\ell]}^{[N]}\mid \mathcal{G}_{[1:L]})+H(\overline{Y}_{2[\ell]}^{[N]}\mid \mathcal{G}_{[1:L]})-H((\overline{X}_{1[\ell]}^{[N]})^{\min(\alpha,1-\alpha)},(\overline{X}_{2[\ell]}^{[N]})^{\min(\alpha,1-\alpha)} \mid W_1, \mathcal{G}_{[1:L]})\notag\\
	&\hspace{3cm}-H((\overline{X}_{1[\ell]}^{[N]})^{\min(\alpha,1-\alpha)},(\overline{X}_{2[\ell]}^{[N]})^{\min(\alpha,1-\alpha)} \mid W_2, \mathcal{G}_{[1:L]})\\
	&\leq H(\overline{Y}_{1[\ell]}^{[N]}\mid \mathcal{G}_{[1:L]})+H(\overline{Y}_{2[\ell]}^{[N]}\mid \mathcal{G}_{[1:L]})-H((\overline{X}_{1[\ell]}^{[N]})^{\min(\alpha,1-\alpha)},(\overline{X}_{2[\ell]}^{[N]})^{\min(\alpha,1-\alpha)} \mid \mathcal{G}_{[1:L]})\notag\\&\hspace{3cm}-H((\overline{X}_{1[\ell]}^{[N]})^{\min(\alpha,1-\alpha)},(\overline{X}_{2[\ell]}^{[N]})^{\min(\alpha,1-\alpha)} \mid W_1,W_2,\mathcal{G}_{[1:L]})\label{iq7}\\
	&\leq H(\overline{Y}_{1[\ell]}^{[N]}\mid \mathcal{G}_{[1:L]})+H(\overline{Y}_{2[\ell]}^{[N]}\mid \mathcal{G}_{[1:L]})-H((\overline{X}_{1[\ell]}^{[N]})^{\min(\alpha,1-\alpha)},(\overline{X}_{2[\ell]}^{[N]})^{\min(\alpha,1-\alpha)} \mid \mathcal{G}_{[1:L]})\label{ineq:7}
	\end{align}
	where \eqref{iq7} follows from the property that for any three random variables $A,B,C$, if $B,C$ are independent, then
	\begin{align}
		H(A\mid B)+H(A\mid C) \geq H(A) +H(A\mid B,C) 
	\end{align} 
	and \eqref{ineq:7} simply uses the fact that entropy is non-negative. 
	Adding $2\times\eqref{inq:707}+2\times\eqref{inq:708}+\eqref{ineq:4}+\eqref{ineq:7}$, we  obtain,
	\begin{align}
	NR_1&+NR_2+2I(W_1;Y_{1[\ell]}^{[N]}\mid \mathcal{G}_{[1:L]})+2I(W_2;Y_{2[\ell]}^{[N]}\mid \mathcal{G}_{[1:L]})\notag\\
	&\leq 2N\log(P)+I(W_1;Y_{1[\ell-1]}^{[N]}\mid \mathcal{G}_{[1:L]})+I(W_2;Y_{2[\ell-1]}^{[N]}\mid \mathcal{G}_{[1:L]})+H(\overline{Y}_{1[\ell]}^{[N]}\mid \mathcal{G}_{[1:L]})+H(\overline{Y}_{2[\ell]}^{[N]}\mid \mathcal{G}_{[1:L]})\notag\\&\hspace{1cm}-I((\overline{X}_{1[\ell]}^{[N]})^{\min(\alpha,1-\alpha)};(\overline{X}_{2[\ell]}^{[N]})^{\min(\alpha,1-\alpha)} \mid \mathcal{G}_{[1:L]})-H((\overline{X}_{1[\ell]}^{[N]})^{\min(\alpha,1-\alpha)},(\overline{X}_{2[\ell]}^{[N]})^{\min(\alpha,1-\alpha)} \mid \mathcal{G}_{[1:L]})\\
	&= 2N\log(P)+I(W_1;Y_{1[\ell-1]}^{[N]}\mid \mathcal{G}_{[1:L]})+I(W_2;Y_{2[\ell-1]}^{[N]}\mid \mathcal{G}_{[1:L]})+H(\overline{Y}_{1[\ell]}^{[N]}\mid \mathcal{G}_{[1:L]})+H(\overline{Y}_{2[\ell]}^{[N]}\mid \mathcal{G}_{[1:L]})\notag\\&\hspace{1cm}-H((\overline{X}_{1[\ell]}^{[N]})^{\min(\alpha,1-\alpha)}\mid \mathcal{G}_{[1:L]})-H((\overline{X}_{2[\ell]}^{[N]})^{\min(\alpha,1-\alpha)}\mid \mathcal{G}_{[1:L]})\label{eq:99b}\\
	&= 2N\log(P)+I(W_1;Y_{1[\ell-1]}^{[N]}\mid \mathcal{G}_{[1:L]})+I(W_2;Y_{2[\ell-1]}^{[N]}\mid \mathcal{G}_{[1:L]})\notag\\&\hspace{1cm}+\{H(\overline{Y}_{1[\ell]}^{[N]}\mid \mathcal{G}_{[1:L]})-H((\overline{X}_{1[\ell]}^{[N]})^{\min(\alpha,1-\alpha)}\mid \mathcal{G}_{[1:L]})\}\notag\\&\hspace{1cm}+\{H(\overline{Y}_{2[\ell]}^{[N]}\mid \mathcal{G}_{[1:L]})-H((\overline{X}_{2[\ell]}^{[N]})^{\min(\alpha,1-\alpha)}\mid \mathcal{G}_{[1:L})\}\label{eq:99a}\\
	&\leq 2N\log(P)+I(W_1;Y_{1[\ell-1]}^{[N]}\mid \mathcal{G}_{[1:L]})+I(W_2;Y_{2[\ell-1]}^{[N]}\mid \mathcal{G}_{[1:L]})+2\max(\alpha,1-\alpha)N\log(P)\label{inq100}\\
	&=(2+2\max(\alpha, 1-\alpha))N\log(P)+I(W_1;Y_{1[\ell-1]}^{[N]}\mid \mathcal{G}_{[1:L]})+I(W_2;Y_{2[\ell-1]}^{[N]}\mid \mathcal{G}_{[1:L]})\label{eq:matchlemma}
	\end{align} 
	where \eqref{eq:99b} follows from the definition of mutual information, and \eqref{eq:99a} is simply a rearrangement of terms. Step \eqref{inq100} is significant because it invokes Sumset Inequality 1 from Lemma \ref{lemma:sumset1} as follows. 
	\begin{align}
	&H(\overline{Y}_{1[\ell]}^{[N]}\mid \mathcal{G}_{[1:L]})-H((\overline{X}_{1[\ell]}^{[N]})^{\min(\alpha,1-\alpha)}\mid \mathcal{G}_{[1:L})\\
	&=H\left(\left.\left\lfloor G_{11[\ell]}(\overline{X}_{1[\ell]}^{[N]})^1\right\rfloor+\left\lfloor {G}_{12[\ell]}(\overline{X}_{2[\ell]}^{[N]})^\alpha\right\rfloor\right|\mathcal{G}_{[1:L]}\right)-H((\overline{X}_{1[\ell]}^{[N]})^{\min(\alpha,1-\alpha)}\mid \mathcal{G}_{[1:L]})\notag\\
	&\leq \max\left(1-\min(\alpha,1-\alpha), \alpha-0\right)^+N\log(P)\\
	&=\max\left(\alpha,1-\alpha\right)N\log(P) \\
	\intertext{and similarly,}
	&H(\overline{Y}_{2[\ell]}^{[N]}\mid \mathcal{G}_{[1:L]})-H((\overline{X}_{2[\ell]}^{[N]})^{\min(\alpha,1-\alpha)}\mid \mathcal{G}_{[1:L]})\leq\max(\alpha,1-\alpha)N\log(P) 
	\end{align}
Note that  \eqref{eq:matchlemma} matches the RHS of \eqref{inq:701}, so that the proof of Lemma \ref{lemma:recur} is complete. With the help of these lemmas, we are now ready to present the converse proof of Theorem \ref{thm:2hop}.

\subsection{Converse Proof for Theorem \ref{thm:2hop}}\label{sec:2hopcon}
	As noted previously,  the upper bound $\mathcal{D}_{\Sigma}\leq2-\alpha$ in the regime $\frac{4}{7}\leq\alpha\leq1$ is immediate, because it corresponds to the sum-GDoF of the MISO broadcast channel \cite{Arash_Jafar_cooperation} formed at the second hop  by allowing full cooperation among the relays, which cannot decrease the GDoF. Therefore, we will assume $\alpha\leq\frac{4}{7}$ in the following proof. As usual, we will sometimes suppress the $o(\log(P))$ terms  for simplicity as they are inconsequential for GDoF studies. 
	Starting with Fano's inequality, we have
	\begin{align}
		3NR_1+3NR_2&\leq2I(W_1;Y_{1[2]}^{[N]}\mid\mathcal{G}_{[1:2]})+2I(W_2;Y_{2[2]}^{[N]}\mid\mathcal{G}_{[1:2]})+NR_1+NR_2\\
		&\leq \big(2+2\max(1-\alpha,\alpha)\big)N\log(P)+I(W_1;Y_{1[1]}^{[N]}\mid\mathcal{G}_{[1:2]})+I(W_2;Y_{2[1]}^{[N]}\mid\mathcal{G}_{[1:2]})\label{iq:901}\\
		&\leq \big(2+2\max(1-\alpha,\alpha)+2\max(1-\alpha,\alpha)\big)N\log(P)\label{iq:902}\\
		&=\big(2+4\max(1-\alpha,\alpha)\big)N\log(P)
	\end{align}
	where \eqref{iq:901} is obtained from Lemma \ref{lemma:recur} by setting $\ell=2$, and \eqref{iq:902} is essentially the well-known sum-GDoF bound for the single-hop interference channel (corresponding to the first hop) that is obtained in \cite[Section \RN{3}. D]{Etkin_Tse_Wang} when $\alpha\leq\frac{2}{3}$ by using a genie-aided approach. Since $\frac{4}{7}\leq\frac{2}{3}$, the bound holds for $\alpha\leq\frac{4}{7}$. Normalizing both sides by $3N\log(P)$ and applying the GDoF limit ($P\rightarrow\infty$), we obtain the sum-GDoF bound $\mathcal{D}_{\Sigma}^{\tiny f.p.}\leq\frac{2+4\max(1-\alpha,\alpha)}{3}$. Therefore,  when $\alpha\leq\frac{1}{2}$, we get the bound $\mathcal{D}_{\Sigma}^{\tiny f.p.}\leq\frac{6-4\alpha}{3}$, and when  $\frac{1}{2}\leq\alpha\leq\frac{4}{7}$, we obtain the bound  $\mathcal{D}_{\Sigma}^{\tiny f.p.}\leq\frac{2+4\alpha}{3}$.
	
	\subsection{Converse Proof for Theorem \ref{thm:Lhop}}\label{sec:Lhopcon}
	In the symmetric multi-hop channel, the bound $2-\alpha$ for the regime $\frac{2^L}{2^{L+1}-1}\leq\alpha\leq1$ is also trivial, because it is the sum-GDoF of the broadcast channel  \cite{Arash_Jafar_cooperation} formed in the last hop under finite precision CSIT by allowing  full cooperation among the relays which cannot reduce the GDoF. Next, to derive the upper bounds for $\alpha\leq\frac{2^L}{2^{L+1}-1}$, we need to recursively apply Lemma \ref{lemma:recur}. Starting with Fano's inequality, we have
	\allowdisplaybreaks
	\begin{align}
	N(R_1+R_2)&\leq I(W_1;Y_{1[L]}^{[N]}\mid \mathcal{G}_{[1:L]})+I(W_2;Y_{2[L]}^{[N]}\mid \mathcal{G}_{[1:L]})\\
	&= \frac{2I(W_1;Y_{1[L]}^{[N]}\mid \mathcal{G}_{[1:L]})+2I(W_2;Y_{2[L]}^{[N]}\mid \mathcal{G}_{[1:L]})+NR_1+NR_2}{3}\\
	&\leq \frac{(2+2\max(1-\alpha,\alpha))N\log(P)}{3}+\frac{I(W_1;Y_{1[L-1]}^{[N]}\mid \mathcal{G}_{[1:L]})+I(W_2;Y_{2[L-1]}^{[N]}\mid \mathcal{G}_{[1:L]})}{3}\label{iq200}\\
	&= \frac{(2+2\max(1-\alpha,\alpha))N\log(P)}{3}+\frac{2I(W_1;Y_{1[L-1]}^{[N]}\mid \mathcal{G}_{[1:L]})+2I(W_2;Y_{2[L-1]}^{[N]}\mid \mathcal{G}_{[1:L]})}{6}\\
	&\leq \frac{(2+2\max(1-\alpha,\alpha))N\log(P)}{3}+\frac{(2+2\max(1-\alpha,\alpha))N\log(P)}{6}\notag\\&\hspace{3cm}+\frac{I(W_1;Y_{1[L-2]}^{[N]}\mid \mathcal{G}_{[1:L]})+I(W_2;Y_{2[L-2]}^{[N]}\mid \mathcal{G}_{[1:L]})}{6}-\frac{N(R_1+R_2)}{6}\label{iq202}\\
	&\leq\cdots\notag\\
	&\leq (2+2\max(1-\alpha,\alpha))\left(\sum_{m=0}^{M-1}\frac{1}{3\times 2^{m}}\right)N\log(P)\notag\\&+\frac{I(W_1;Y_{1[L-M]}^{[N]}\mid \mathcal{G}_{[1:L]})+I(W_2;Y_{2[L-M]}^{[N]}\mid \mathcal{G}_{[1:L]})}{3\times 2^{M-1}}-\left(\sum_{k=1}^{M-1} \frac{1}{3\times 2^k}\right)N(R_1+R_2)\label{iq203}\\
	&\leq\cdots\notag\\
	&\leq (2+2\max(1-\alpha,\alpha))\left(\sum_{m=0}^{L-3}\frac{1}{3\times 2^{m}}\right)N\log(P)+\frac{I(W_1;Y_{1[2]}^{[N]}\mid\mathcal{G}_{[1:L]})+I(W_2;Y_{2[2]}^{[N]}\mid\mathcal{G}_{[1:L]})}{3\times 2^{L-3}}\notag\\&\hspace{7cm}-\left(\sum_{k=1}^{L-3} \frac{1}{3\times 2^k}\right)N(R_1+R_2)\\
	&\leq (2+2\max(1-\alpha,\alpha))\left(\sum_{m=0}^{L-2}\frac{1}{3\times 2^{m}}\right)N\log(P)+\frac{I(W_1;Y_{1[1]}^{[N]}\mid\mathcal{G}_{[1:L]})+I(W_2;Y_{2[1]}^{[N]}\mid\mathcal{G}_{[1:L]})}{3\times 2^{L-2}}\notag\\&\hspace{7cm}-\left(\sum_{k=1}^{L-2} \frac{1}{3\times 2^k}\right)N(R_1+R_2)\label{iq204}\\
	&\leq (2+2\max(1-\alpha,\alpha))\left(\sum_{m=0}^{L-2}\frac{1}{3\times 2^{m}}\right)N\log(P)+\frac{I(X_{1[1]}^{[N]};Y_{1[1]}^{[N]}\mid\mathcal{G}_{[1:L]})+I(X_{2[1]}^{[N]};Y_{2[1]}^{[N]}\mid\mathcal{G}_{[1:L]})}{3\times 2^{L-2}}\notag\\&\hspace{7cm}-\left(\sum_{k=1}^{L-2} \frac{1}{3\times 2^k}\right)N(R_1+R_2) \label{iq205}
	\end{align}
	where \eqref{iq200} is obtained by setting $\ell=L$,  \eqref{iq202} is obtained by setting $\ell=L-1$, \eqref{iq203} is obtained by setting $\ell=3$ and \eqref{iq204} is obtained by setting $\ell=2$ at \eqref{inq:701}. Rearranging \eqref{iq205}, we  obtain
	\begin{align}
	\left(1+\frac{2^{L-2}-1}{3\times2^{L-2}}\right)N(R_1+R_2)&\leq (2+2\max(1-\alpha,\alpha))\left(\frac{2^{L-1}-1}{3\times2^{L-2}}\right)N\log(P)\notag\\&\hspace{1cm}+\frac{I(X_{1[1]}^{[N]};Y_{1[1]}^{[N]}\mid\mathcal{G}_{[1:L]})+I(X_{2[1]}^{[N]};Y_{2[1]}^{[N]}\mid \mathcal{G}_{[1:L]})}{3\times 2^{L-2}}
	\end{align}
	With the inequality $I(X_{1[1]}^{[N]};Y_{1[1]}^{[N]}\mid\mathcal{G}_{[1:L]})+I(X_{2[1]}^{[N]};Y_{2[1]}^{[N]}\mid \mathcal{G}_{[1:L]})\leq (2\max(1-\alpha,\alpha))N\log(P)$ obtained in \cite[Section \RN{3}. D]{Etkin_Tse_Wang}\footnote{The original proof assumes perfect CSIT, but since the availability of CSIT cannot hurt, such bounds also hold with finite precision CSIT.} when $\alpha\leq\frac{2}{3}$, we have,
	\begin{align}
	\left(1+\frac{2^{L-2}-1}{3\times2^{L-2}}\right)N(R_1+R_2)\leq (2+2\max(1-\alpha,\alpha))\left(\frac{2^{L-1}-1}{3\times2^{L-2}}\right)N\log(P)+\frac{2\max(1-\alpha,\alpha)}{3\times 2^{L-2}} N\log(P)
	\end{align}
	Dividing by $(1+\frac{2^{L-2}-1}{3\times2^{L-2}})$, we get
	\begin{align}
	N(R_1+R_2)&\leq \frac{(1+\max(\alpha,1-\alpha))\times(2^L-2)+2\max(\alpha,1-\alpha)}{2^L-1}N\log(P)\\
	&=\left(1+\max(\alpha,1-\alpha)-\frac{1-\max(\alpha,1-\alpha)}{2^L-1}\right)N\log(P)
	\end{align}
	Normalizing both sides by $N\log(P)$ and applying the GDoF limit, we obtain the bound,
	\begin{align}
	\mathcal{D}_{\Sigma}^{\tiny f.p.}\leq 1+\max(\alpha,1-\alpha)-\frac{1-\max(\alpha,1-\alpha)}{2^L-1}
	\end{align}
	Thus, when $\alpha\leq\frac{1}{2}$, we have $\mathcal{D}_{\Sigma}^{\tiny f.p.}\leq2-\alpha-\frac{\alpha}{2^L-1}$, when $\frac{1}{2}\leq\alpha\leq\frac{2^L}{2^{L+1}-1}$, we have $\mathcal{D}_{\Sigma}^{\tiny f.p.}\leq 1+\alpha-\frac{1-\alpha}{2^L-1}$.

\subsection{Converse Proof for Theorem \ref{thm:Lhopvstrong}}\label{sec:Lhopvstrongcon}
The upper bound is straightforward by using the cut-set bound and considering the multihop channel as two unicast channel.
\begin{align}
NR_1&\leq I(X_{1[1]}^{[N]};Y_{1[L]}^{[N]}\mid\mathcal{G}_{[1:L]})\\
&\leq I(X_{1[1]}^{[N]};Y_{1[1]}^{[N]},Y_{2[2]}^{[N]},Y_{1[3]}^{[N]},Y_{2[2]}^{[N]},\cdots,Y_{1[L]}^{[N]}\mid\mathcal{G}_{[1:L]})\\
&= I(X_{1[1]}^{[N]};Y_{1[1]}^{[N]}\mid\mathcal{G}_{[1:L]})+I(X_{1[1]}^{[N]};Y_{2[2]}^{[N]}\mid Y_{1[1]}^{[N]},\mathcal{G}_{[1:L]})+I(X_{1[1]}^{[N]};Y_{1[3]}^{[N]}\mid Y_{1[1]}^{[N]},Y_{2[2]}^{[N]},\mathcal{G}_{[1:L]})+\cdots \notag\\&\hspace{2cm}+I(X_{1[1]}^{[N]};Y_{1[L]}^{[N]}\mid Y_{1[1]}^{[N]},Y_{2[2]}^{[N]},\cdots,Y_{2[L-1]}^{[N]},\mathcal{G}_{[1:L]})\\
&\leq I(X_{1[1]}^{[N]};Y_{1[1]}^{[N]}\mid\mathcal{G}_{[1:L]})+h(Y_{2[2]}^{[N]}\mid Y_{1[1]}^{[N]},\mathcal{G}_{[1:L]})+h(Y_{1[3]}^{[N]}\mid Y_{1[1]}^{[N]},Y_{2[2]}^{[N]},\mathcal{G}_{[1:L]})+\cdots\notag\\ &\hspace{2cm}+h(Y_{1[L]}^{[N]}\mid Y_{1[1]}^{[N]},Y_{2[2]}^{[N]},\cdots,Y_{2[L-1]}^{[N]},\mathcal{G}_{[1:L]})\\
&\leq N\log(P)+h(Y_{2[2]}^{[N]},X_{1[2]}^{[N]}\mid Y_{1[1]}^{[N]},\mathcal{G}_{[1:L]})+h(Y_{1[3]}^{[N]},X_{2[3]}^{[N]}\mid Y_{1[1]}^{[N]},Y_{2[2]}^{[N]},\mathcal{G}_{[1:L]})+\cdots\notag\\ &\hspace{2cm}+h(Y_{1[L]}^{[N]},X_{2[L]}^{[N]}\mid Y_{1[1]}^{[N]},Y_{2[2]}^{[N]},\cdots,Y_{2[L-1]}^{[N]},\mathcal{G}_{[1:L]})\\
&=N\log(P)+h(Y_{2[2]}^{[N]}\mid X_{1[2]}^{[N]},Y_{1[1]}^{[N]},\mathcal{G}_{[1:L]})+h(Y_{1[3]}^{[N]}\mid X_{2[3]}^{[N]},Y_{1[1]}^{[N]},Y_{2[2]}^{[N]},\mathcal{G}_{[1:L]})+\cdots\notag\\ &\hspace{2cm}+h(Y_{1[L]}^{[N]}\mid X_{2[L]}^{[N]}, Y_{1[1]}^{[N]},Y_{2[2]}^{[N]},\cdots,Y_{2[L-1]}^{[N]},\mathcal{G}_{[1:L]})\label{inq:Lodd1}\\
&\leq L\times N\log(P)\label{inq:Lodd2}
\end{align}
where \eqref{inq:Lodd1} holds since $X_{i[\ell]}^{[N]}$ is completely determined by $(Y_{i[\ell-1]}^{[N]},\mathcal{G}_{[1:L]})$, \eqref{inq:Lodd2} holds because at $\ell^{th}$ hop, we have
\begin{align}
h(Y_{i[\ell]}^{[N]}\mid X_{\bar{i}[\ell]}^{[N]},& Y_{1[1]}^{[N]},Y_{2[2]}^{[N]},\cdots,Y_{\bar{i}[\ell-1]}^{[N]},\mathcal{G}_{[1:L]})\notag\\
&=h(\sqrt{P^1}G_{ii[\ell]}^{[N]}X_{i[\ell]}^{[N]}+\sqrt{P^{\alpha}}G_{i\bar{i}[\ell]}^{[N]}X_{\bar{i}[\ell]}^{[N]}\mid X_{\bar{i}[\ell]}^{[N]}, Y_{1[1]}^{[N]},Y_{2[2]}^{[N]},\cdots,Y_{\bar{i}[\ell-1]}^{[N]},\mathcal{G}_{[1:L]})\\
&=h(\sqrt{P^1}X_{i[\ell]}^{[N]}\mid X_{\bar{i}[\ell]}^{[N]}, Y_{1[1]}^{[N]},Y_{2[2]}^{[N]},\cdots,Y_{\bar{i}[\ell-1]}^{[N]},\mathcal{G}_{[1:L]})\\
&\leq h(\sqrt{P^1}X_{i[\ell]}^{[N]})\\
&\leq N\log(P)
\end{align}
Symmetrically, we get
\begin{align}
NR_2\leq L\times N\log(P)\label{inq:Lodd3}
\end{align}
Combining \eqref{inq:Lodd1}\eqref{inq:Lodd2} and dividing by $N\log(P)$ at both sides, we get $\mathcal{D}_{\Sigma}^{p}\leq 2L$. Since perfect CSIT cannot hurt the GDoF value, we also get $\mathcal{D}_{\Sigma}^{f.p.}\leq 2L$.

\section{Achievability}\label{sec:lower}
	\subsection{Proof of Achievability for Theorem \ref{thm:2hop}}\label{sec:2hopach}
	There are $4$ sub-cases for the achievable scheme.
	\begin{itemize}[leftmargin=*]
		\item $\alpha\leq\frac{1}{2}$\\
		The achievable scheme is as follows: message $W_i$ is split into $4$ sub-messages $W_i=(W_{i1},W_{i2},W_{i3},W_{i4})$. They carry $d_{i1}=\frac{2\alpha}{3},d_{i2}=\frac{\alpha}{3},d_{i3}=1-2\alpha,d_{i4}=\frac{\alpha}{3}$ GDoF respectively. They are encoded into independent Gaussian codebooks producing codewords $X_{i1},X_{i2},X_{i3},X_{i4}$ with powers $\E|X_{i1}|^2=1-P^{-d_{i1}},\E|X_{i2}|^2=P^{-d_{i1}}-P^{-d_{i1}-d_{i2}},\E|X_{i3}|^2=P^{-d_{i1}-d_{i2}}-P^{-1+d_{i4}},\E|X_{i4}|^2=P^{-1+d_{i4}}$. The transmitted signals at the sources are $X_{1[1]}=X_{11}+X_{12}+X_{13}+X_{14},X_{2[1]}=X_{21}+X_{22}+X_{23}+X_{24}$. 
		Relay $\text{Rx}_{1[1]}$ is able to decode $W_{11},W_{12},W_{13},W_{21}$ successively with the corresponding SINR values $\sim P^{d_{11}},\sim P^{d_{12}},\sim P^{d_{13}},\sim P^{d_{21}}$.  Relay $\text{Rx}_{1[1]}$ then reconstructs the codewords $X_{11},X_{12},X_{13},X_{21}$ and removes their contribution from its received signal. The remaining signal above the noise floor is  a linear combination of $X_{14}$ and $X_{22}$, which is denoted as $\mathcal{L}_1(X_{14},X_{22})$. Relay $\text{Rx}_{1[1]}$ amplifies the remaining signals by power $P^{-1}$, such that the power is $\E|\mathcal{L}_1|^2=P^{-1+d_{14}}$. Then, Relay $\text{Rx}_{1[1]}$ splits the messages $W_{11},W_{21}$ into $W_{11}=(W_{11}^1,W_{11}^2), W_{21}=(W_{21}^1,W_{21}^2)$, with the corresponding GDoF value $d_{11}^1=d_{11}^2=d_{21}^1=d_{21}^2=\frac{d_{11}}{2}=\frac{\alpha}{3}$. After that, Relay $\text{Tx}_{1[2]}\equiv \text{Rx}_{1[1]}$ re-encodes $W_{11}^1,W_{12}^1,W_{12},W_{13},W_{21}^1$ into codewords $X_{11}^1,X_{12}^1,X_{12},X_{13},X_{21}^1$ by assigning power $\E|X_{12}|^2=1-P^{-d_{12}}$, $\E|X_{11}^1|^2=P^{-d_{12}}-P^{-d_{12}-d_{11}^1}$, $\E|X_{21}^1|^2=P^{-d_{12}-d_{11}^1}-P^{-d_{12}-d_{11}^1-d_{21}^1},\E|X_{21}|^2=P^{-d_{12}-d_{11}^1-d_{21}^1}-P^{-d_{12}-d_{11}^1-d_{21}^1-d_{21}}$, $\E|X_{13}|^2=P^{-d_{12}-d_{11}^1-d_{21}^1-d_{21}}-P^{-1+d_{14}+d_{11}^2}$, $\E|X_{11}^2|^2=P^{-1+d_{14}+d_{11}^2}-P^{-1+d_{14}}$. Relay $\text{Tx}_{2[2]}\equiv \text{Rx}_{2[1]}$ proceeds similarly. The transmitted signals at relays are $X_{1[2]}=X_{12}+X_{11}^1+X_{21}^1+X_{13}+X_{11}^2+\mathcal{L}_1,X_{2[2]}=X_{22}+X_{11}^1+X_{21}^1+X_{23}+X_{21}^2+\mathcal{L}_2$.  Destination $\text{Rx}_{1[2]}$ is able to decode $W_{12},W_{11}^1,W_{21}^1,W_{13},W_{22},W_{11}^2,W_{14}$ successively, with the corresponding SINR values $\sim P^{d_{12}},\sim P^{d_{11}^1},\sim P^{d_{21}^1},\sim P^{d_{13}},\sim P^{d_{22}},\sim P^{d_{11}^2},\sim P^{d_{14}}$. Destination $\text{Rx}_{2[2]}$ proceeds similarly to decode $W_{22},W_{11}^1,W_{21}^1,W_{23},W_{12},W_{21}^2,W_{24}$ successively. See Figure \ref{fig4} for an illustration.
		\begin{figure}[!t]
			\centering
			\begin{tikzpicture}[scale=1.4]
			\draw[thick] (2,2) circle(1mm) node [above] {\small $\text{Tx}_{1[1]}$};
			\draw[thick] (3,2) circle(1mm) node [above] {\small $\text{Rx}_{1[1]}$};
			\draw[thick] (2,-2.5) circle(1mm) node [below] {\small $\text{Tx}_{2[1]}$};
			\draw[thick] (3,-2.5) circle(1mm) node [below] {\small $\text{Rx}_{2[1]}$};
			
			\begin{scope}[shift={(0.8,0)}]
			\draw[thick,dashed] (-0.5,2)--(1.0,2);
			\draw (0.25, 2) node[above]{${X}_{1[1]}$};
			\draw[ thick, fill=white] (0,0) rectangle (0.5,2);
			\draw[ thick, fill=cyan!30!white, pattern color=blue] (0,0) rectangle (0.5,2/9) node [pos=.5] {\tiny $X_{14}$};
			\draw[arrows=<->] (-0.15,0)--(-0.15,2/9)node[left,pos=0.5]{\tiny $\frac{\alpha}{3}$};
			\draw[ thick, fill=yellow!30!white, pattern color=blue] (0,2/3) rectangle (0.5,4/3) node [pos=.5] {\tiny $X_{13}$};
			\draw[arrows=<->] (-0.15,2/3)--(-0.15,4/3)node[left,pos=0.5]{\tiny $1-2\alpha$};
			\draw[ thick, fill=blue!30!white, pattern color=blue] (0,4/3) rectangle (0.5,14/9) node [pos=.5] {\tiny $X_{12}$};
			\draw[arrows=<->] (-0.15,4/3)--(-0.15,14/9)node[left,pos=0.5]{\tiny $\frac{\alpha}{3}$};
			\draw[ thick, fill=green!30!white, pattern color=green] (0,14/9) rectangle (0.5,2) node [pos=.5] {\tiny $X_{11}$};
			\draw[arrows=<->] (-0.15,14/9)--(-0.15,2)node[left,pos=0.5]{\tiny $\frac{2\alpha}{3}$};
			\end{scope}
			
			\begin{scope}[shift={(0.8,-2.5)}]
			\draw[thick,dashed] (-0.5,2)--(1.0,2);
			\draw (0.25, 2) node[above]{${X}_{2[1]}$};
			\draw[ thick, fill=white] (0,0) rectangle (0.5,2);
			\draw[ thick, fill=red!50!white, pattern color=blue]  (0,0) rectangle (0.5,2/9) node [pos=.5] {\tiny $X_{24}$};
			\draw[arrows=<->] (-0.15,0)--(-0.15,2/9)node[left,pos=0.5]{\tiny $\frac{\alpha}{3}$};
			\draw[ thick, fill=orange!30!white, pattern color=blue]  (0,2/3)rectangle (0.5,4/3) node [pos=.5] {\tiny $X_{23}$};
			\draw[arrows=<->] (-0.15,2/3)--(-0.15,4/3)node[left,pos=0.5]{\tiny $1-2\alpha$};
			\draw[ thick, fill=green!30!yellow, pattern color=blue] (0,4/3)rectangle (0.5,14/9) node [pos=.5] {\tiny $X_{22}$};
			\draw[arrows=<->] (-0.15,4/3)--(-0.15,14/9)node[left,pos=0.5]{\tiny $\frac{\alpha}{3}$};
			\draw[ thick, fill=purple!30!white, pattern color=blue] (0,14/9) rectangle (0.5,2) node [pos=.5] {\tiny $X_{21}$};
			\draw[arrows=<->] (-0.15,14/9)--(-0.15,2)node[left,pos=0.5]{\tiny $\frac{2\alpha}{3}$};
			\end{scope}
			
			\draw[thick,arrowmid] (2.1,2)--(2.9,2)node[above,pos=0.5]{$1$};
			\draw[thick,arrowmid] (2.1,-2.5)--(2.9,-2.5)node[below,pos=0.5]{$1$};
			\draw[thick,arrowmid] (2.1,2)--(2.9,-2.5)node[right,pos=0.8]{$\alpha$};
			\draw[thick,arrowmid] (2.1,-2.5)--(2.9,2)node[right,pos=0.8]{$\alpha$};

			\begin{scope}[shift={(4.0,0)}]
			\draw[thick,dashed] (-0.5,0)--(1.5,0);
			\draw[thick] (0.5,0) node[below]{$Y_{1[1]}$};
			\draw[ thick, fill=white] (0,0) rectangle (0.5,2);
			\draw[ thick, fill=cyan!30!white, pattern color=blue] (0,0) rectangle (0.5,2/9) node [pos=.5] {\tiny $X_{14}$};
			\draw[arrows=<->] (-0.15,0)--(-0.15,2/9)node[left,pos=0.5]{\tiny $\frac{\alpha}{3}$};
			\draw[ thick, fill=yellow!30!white, pattern color=blue] (0,2/3) rectangle (0.5,4/3) node [pos=.5] {\tiny $X_{13}$};
			\draw[arrows=<->] (-0.15,2/3)--(-0.15,4/3)node[left,pos=0.5]{\tiny $1-2\alpha$};
			\draw[ thick, fill=blue!30!white, pattern color=blue] (0,4/3) rectangle (0.5,14/9) node [pos=.5] {\tiny $X_{12}$};
			\draw[arrows=<->] (-0.15,4/3)--(-0.15,14/9)node[left,pos=0.5]{\tiny $\frac{\alpha}{3}$};
			\draw[ thick, fill=green!30!white, pattern color=green] (0,14/9) rectangle (0.5,2) node [pos=.5] {\tiny $X_{11}$};
			\draw[arrows=<->] (-0.15,14/9)--(-0.15,2)node[left,pos=0.5]{\tiny $\frac{2\alpha}{3}$};
			
			\draw[ thick, fill=green!30!yellow, pattern color=blue] (0.5,0)rectangle (1,2/9) node [pos=.5] {\tiny $X_{22}$};
			\draw[arrows=<->] (1.15,0)--(1.15,2/9)node[right,pos=0.5]{\tiny $\frac{\alpha}{3}$};
			\draw[ thick, fill=purple!30!white, pattern color=blue] (0.5,2/9) rectangle (1,2/3) node [pos=.5] {\tiny $X_{21}$};
			\draw[arrows=<->] (1.15,2/9)--(1.15,2/3)node[right,pos=0.5]{\tiny $\frac{2\alpha}{3}$};
			\end{scope}

			\begin{scope}[shift={(4.0,-2.5)}]
			\draw[thick,dashed] (-0.5,0)--(1.5,0);
			\draw[thick] (0.5,0) node[below]{$Y_{2[1]}$};
			\draw[ thick, fill=blue!30!white, pattern color=blue] (0,0) rectangle (0.5,2/9) node [pos=.5] {\tiny $X_{12}$};
			\draw[arrows=<->] (-0.15,0)--(-0.15,2/9)node[left,pos=0.5]{\tiny $\frac{\alpha}{3}$};
			\draw[ thick, fill=green!30!white, pattern color=green] (0,2/9) rectangle (0.5,2/3) node [pos=.5] {\tiny $X_{11}$};
			\draw[arrows=<->] (-0.15,2/9)--(-0.15,2/3)node[left,pos=0.5]{\tiny $\frac{2\alpha}{3}$};
			
			\draw[ thick, fill=white] (0.5,0) rectangle (1,2);
			\draw[ thick, fill=red!50!white, pattern color=blue]  (0.5,0) rectangle (1,2/9) node [pos=.5] {\tiny $X_{24}$};
			\draw[arrows=<->] (1.15,0)--(1.15,2/9)node[right,pos=0.5]{\tiny $\frac{\alpha}{3}$};
			\draw[ thick, fill=orange!30!white, pattern color=blue]  (0.5,2/3)rectangle (1,4/3) node [pos=.5] {\tiny $X_{23}$};
			\draw[arrows=<->] (1.15,2/3)--(1.15,4/3)node[right,pos=0.5]{\tiny $1-2\alpha$};
			\draw[ thick, fill=green!30!yellow, pattern color=blue] (0.5,4/3)rectangle (1,14/9) node [pos=.5] {\tiny $X_{22}$};
			\draw[arrows=<->] (1.15,4/3)--(1.15,14/9)node[right,pos=0.5]{\tiny $\frac{\alpha}{3}$};
			\draw[ thick, fill=purple!30!white, pattern color=blue] (0.5,14/9) rectangle (1,2) node [pos=.5] {\tiny $X_{21}$};
			\draw[arrows=<->] (1.15,14/9)--(1.15,2)node[right,pos=0.5]{\tiny $\frac{2\alpha}{3}$};
			\end{scope}
			
			\begin{scope}[shift={(6.5,0)}]
			\draw[thick,dashed] (-0.5,2)--(1.0,2);
			\draw (0.25, 2) node[above]{${X}_{1[2]}$};
			\draw[ thick, fill=white] (0,0) rectangle (0.5,2);
			\draw[ thick, fill=cyan!30!white, pattern color=blue] (0,0) rectangle (0.5,2/9) node [pos=.5] {\tiny $\mathcal{L}_1$};
			\draw[arrows=<->] (-0.15,0)--(-0.15,2/9)node[left,pos=0.5]{\tiny $\frac{\alpha}{3}$};
			\draw[ thick, fill=green!30!white, pattern color=green] (0,2/9) rectangle (0.5,4/9) node [pos=.5] {\tiny $X_{11}^2$};
			\draw[arrows=<->] (-0.15,2/9)--(-0.15,4/9)node[left,pos=0.5]{\tiny $\frac{\alpha}{3}$};
			\draw[ thick, fill=yellow!30!white, pattern color=blue] (0,2/3) rectangle (0.5,4/3) node [pos=.5] {\tiny $X_{13}$};
			\draw[arrows=<->] (-0.15,2/3)--(-0.15,4/3)node[left,pos=0.5]{\tiny $1-2\alpha$};
			
			\draw[ thick, fill=purple!50!white, pattern color=blue] (0,4/3) rectangle (0.5,14/9) node [pos=.5] {\tiny $X_{21}^1$};
			\draw[ thick, fill=green!50!white, pattern color=green] (0,14/9) rectangle (0.5,16/9) node [pos=.5] {\tiny $X_{11}^1$};
			\draw[arrows=<->] (-0.15,12/9)--(-0.15,16/9)node[left,pos=0.5]{\tiny $\frac{2\alpha}{3}$};
			
			\draw[ thick, fill=blue!30!white, pattern color=blue] (0,16/9) rectangle (0.5,2) node [pos=.5] {\tiny $X_{12}$};
			\draw[arrows=<->] (-0.15,16/9)--(-0.15,2)node[left,pos=0.5]{\tiny $\frac{\alpha}{3}$};
			\end{scope}
			
			\begin{scope}[shift={(6.5,-2.5)}]
			\draw[thick,dashed] (-0.5,2)--(1.0,2);
			\draw (0.25, 2) node[above]{${X}_{2[2]}$};
			\draw[ thick, fill=white] (0,0) rectangle (0.5,1);
			\draw[ thick, fill=red!50!white, pattern color=blue] (0,0) rectangle (0.5,2/9) node [pos=.5] {\tiny $\mathcal{L}_2$};
			\draw[arrows=<->] (-0.15,0)--(-0.15,2/9)node[left,pos=0.5]{\tiny $\frac{\alpha}{3}$};
			\draw[ thick, fill=purple!30!white, pattern color=blue] (0,2/9) rectangle (0.5,4/9) node [pos=.5] {\tiny $X_{21}^2$};
			\draw[arrows=<->] (-0.15,2/9)--(-0.15,4/9)node[left,pos=0.5]{\tiny $\frac{\alpha}{3}$};
			\draw[thick, fill=orange!30!white, pattern color=blue] (0,2/3) rectangle (0.5,4/3) node [pos=.5] {\tiny $X_{23}$};
			\draw[arrows=<->] (-0.15,2/3)--(-0.15,4/3)node[left,pos=0.3]{\tiny $1-2\alpha$};
			
			\draw[ thick, fill=purple!50!white, pattern color=blue] (0,4/3) rectangle (0.5,14/9) node [pos=.5] {\tiny $X_{21}^1$};
			\draw[ thick, fill=green!50!white, pattern color=green] (0,14/9) rectangle (0.5,16/9) node [pos=.5] {\tiny $X_{11}^1$};
			\draw[arrows=<->] (-0.15,12/9)--(-0.15,16/9)node[left,pos=0.5]{\tiny $\frac{2\alpha}{3}$};
			
			\draw[thick, fill=green!30!yellow, pattern color=blue] (0,16/9) rectangle (0.5,2) node [pos=.5] {\tiny $X_{22}$};
			\draw[arrows=<->] (-0.15,16/9)--(-0.15,2)node[left,pos=0.5]{\tiny $\frac{\alpha}{3}$};
			\end{scope}
			
			\begin{scope}[shift={(0.3,0)}]
			\draw[thick] (7.5,2) circle(1mm)node [above] {\small $\text{Tx}_{1[2]}$};
			\draw[thick] (8.5,2) circle(1mm)node [above] {\small $\text{Rx}_{1[2]}$};
			\draw[thick] (7.5,-2.5) circle(1mm)node [below] {\small $\text{Tx}_{2[2]}$};
			\draw[thick] (8.5,-2.5) circle(1mm)node [below] {\small $\text{Rx}_{2[2]}$};
			
			\draw[thick,arrowmid] (7.6,2)--(8.4,2)node[above,pos=0.5]{$1$};
			\draw[thick,arrowmid] (7.6,-2.5)--(8.4,-2.5)node[below,pos=0.5]{$1$};
			\draw[thick,arrowmid] (7.6,2)--(8.4,-2.5)node[right,pos=0.8]{$\alpha$};
			\draw[thick,arrowmid] (7.6,-2.5)--(8.4,2)node[right,pos=0.8]{$\alpha$};
			\end{scope}

			\begin{scope}[shift={(9.8,0)}]
			\draw[thick,dashed] (-0.5,0)--(1.5,0);
			\draw[thick] (0.5,0) node[below]{$Y_{1[2]}$};
			\draw[ thick, fill=white] (0,0) rectangle (0.5,2);
			\draw[ thick, fill=cyan!30!white, pattern color=blue] (0,0) rectangle (0.5,2/9) node [pos=.5] {\tiny $\mathcal{L}_1$};
			\draw[arrows=<->] (-0.15,0)--(-0.15,2/9)node[left,pos=0.5]{\tiny $\frac{\alpha}{3}$};
			\draw[ thick, fill=green!30!white, pattern color=green] (0,2/9) rectangle (0.5,4/9) node [pos=.5] {\tiny $X_{11}^2$};
			\draw[arrows=<->] (-0.15,2/9)--(-0.15,4/9)node[left,pos=0.5]{\tiny $\frac{\alpha}{3}$};
			\draw[ thick, fill=yellow!30!white, pattern color=blue] (0,2/3) rectangle (0.5,4/3) node [pos=.5] {\tiny $X_{13}$};
			\draw[arrows=<->] (-0.15,2/3)--(-0.15,4/3)node[left,pos=0.5]{\tiny $1-2\alpha$};
			
			\draw[ thick, fill=purple!50!white, pattern color=blue] (0,4/3) rectangle (0.5,14/9) node [pos=.5] {\tiny $X_{21}^1$};
			\draw[ thick, fill=green!50!white, pattern color=green] (0,14/9) rectangle (0.5,16/9) node [pos=.5] {\tiny $X_{11}^1$};
			\draw[arrows=<->] (-0.15,12/9)--(-0.15,16/9)node[left,pos=0.5]{\tiny $\frac{2\alpha}{3}$};
			
			\draw[ thick, fill=blue!30!white, pattern color=blue] (0,16/9) rectangle (0.5,2) node [pos=.5] {\tiny $X_{12}$};
			\draw[arrows=<->] (-0.15,16/9)--(-0.15,2)node[left,pos=0.5]{\tiny $\frac{\alpha}{3}$};
			
			\draw[ thick, fill=purple!50!white, pattern color=blue] (0.5,0) rectangle (1,2/9) node [pos=.5] {\tiny $X_{21}^1$};
			\draw[ thick, fill=green!50!white, pattern color=green] (0.5,2/9) rectangle (1,4/9) node [pos=.5] {\tiny $X_{11}^1$};
			\draw[arrows=<->] (1.15,0)--(1.15,4/9)node[right,pos=0.5]{\tiny $\frac{2\alpha}{3}$};
			
			\draw[thick, fill=green!30!yellow, pattern color=blue] (0.5,4/9) rectangle (1,2/3) node [pos=.5] {\tiny $X_{22}$};
			\draw[arrows=<->] (1.15,4/9)--(1.15,2/3)node[right,pos=0.5]{\tiny $\frac{\alpha}{3}$};
			\end{scope}

			\begin{scope}[shift={(9.8,-2.5)}]
			\draw[thick,dashed] (-0.5,0)--(1.5,0);
			\draw[thick] (0.5,0) node[below]{$Y_{2[2]}$};
			\draw[ thick, fill=purple!50!white, pattern color=blue] (0,0) rectangle (0.5,2/9) node [pos=.5] {\tiny $X_{21}^1$};
			\draw[ thick, fill=green!50!white, pattern color=green] (0,2/9) rectangle (0.5,4/9) node [pos=.5] {\tiny $X_{11}^1$};
			\draw[arrows=<->] (-0.15,0)--(-0.15,4/9)node[left,pos=0.5]{\tiny $\frac{2\alpha}{3}$};
			
			\draw[ thick, fill=blue!30!white, pattern color=blue] (0,4/9) rectangle (0.5,2/3) node [pos=.5] {\tiny $X_{12}$};
			\draw[arrows=<->] (-0.15,4/9)--(-0.15,2/3)node[left,pos=0.5]{\tiny $\frac{\alpha}{3}$};

			\draw[ thick, fill=white] (0.5,0) rectangle (1,2);
			\draw[ thick, fill=red!50!white, pattern color=blue] (0.5,0) rectangle (1,2/9) node [pos=.5] {\tiny $\mathcal{L}_2$};
			\draw[arrows=<->] (1.15,0)--(1.15,2/9)node[right,pos=0.5]{\tiny $\frac{\alpha}{3}$};
			\draw[ thick, fill=purple!30!white, pattern color=blue] (0.5,2/9) rectangle (1,4/9) node [pos=.5] {\tiny $X_{21}^2$};
			\draw[arrows=<->] (1.15,2/9)--(1.15,4/9)node[right,pos=0.5]{\tiny $\frac{\alpha}{3}$};
			\draw[thick, fill=orange!30!white, pattern color=blue] (0.5,2/3) rectangle (1,4/3) node [pos=.5] {\tiny $X_{23}$};
			\draw[arrows=<->] (1.15,2/3)--(1.15,4/3)node[right,pos=0.5]{\tiny $1-2\alpha$};
			
			\draw[ thick, fill=purple!50!white, pattern color=blue] (0.5,4/3) rectangle (1,14/9) node [pos=.5] {\tiny $X_{21}^1$};
			\draw[ thick, fill=green!50!white, pattern color=green] (0.5,14/9) rectangle (1,16/9) node [pos=.5] {\tiny $X_{11}^1$};
			\draw[arrows=<->] (1.15,12/9)--(1.15,16/9)node[right,pos=0.5]{\tiny $\frac{2\alpha}{3}$};
			
			\draw[thick, fill=green!30!yellow, pattern color=blue] (0.5,16/9) rectangle (1,2) node [pos=.5] {\tiny $X_{22}$};
			\draw[arrows=<->] (1.15,16/9)--(1.15,2)node[right,pos=0.5]{\tiny $\frac{\alpha}{3}$};
			\end{scope}
			\end{tikzpicture}
			\caption{\it\small Achievable scheme for $\alpha\leq\frac{1}{2}$. The dashed line at the receivers represents the noise floor, at the transmitters it represents unit power. $\mathcal{L}_1, \mathcal{L}_2$ are short for $\mathcal{L}_1(X_{14},X_{22}),\mathcal{L}_2(X_{24},X_{12})$, respectively. The left figure is the first hop while the right figure represents the second hop.}\label{fig4}
		\end{figure}
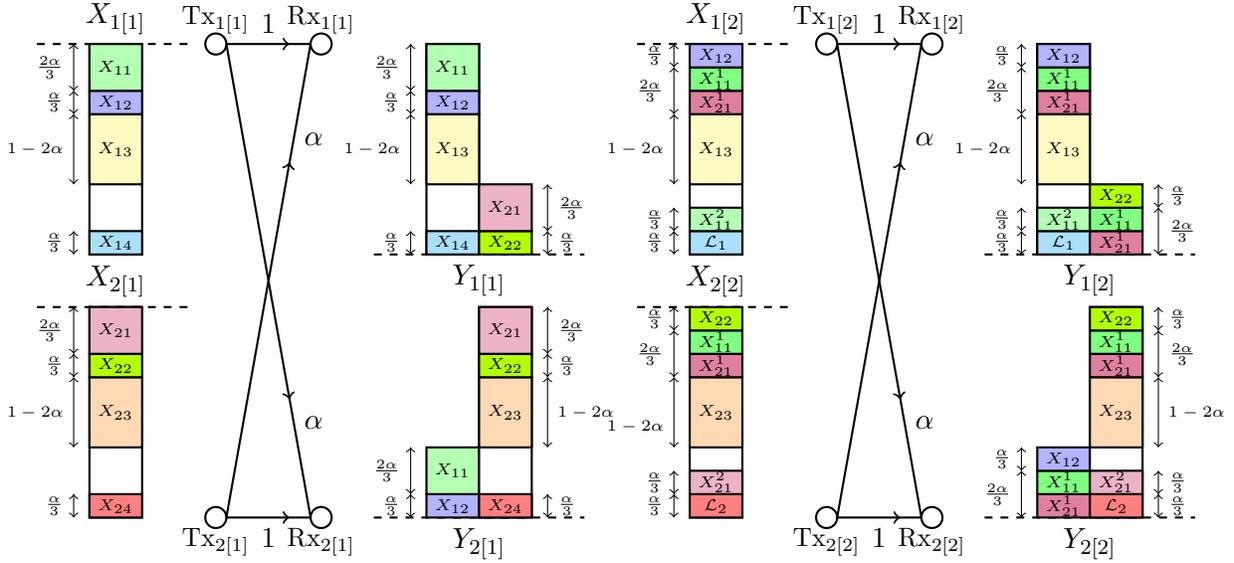	
		
		\item $\frac{1}{2}\leq\alpha\leq\frac{4}{7}$\\
		The achievable scheme is as follows: $W_i$ is split into $3$ sub-messages, i.e., $W_i=(W_{i1},W_{i2},W_{i3})$, whose corresponding GDoF are $d_{i1}=\frac{2-2\alpha}{3},d_{i2}=\frac{1-\alpha}{3},d_{i3}=\frac{5\alpha-2}{3}$. They are encoded into independent Gaussian codebooks producing codewords $X_{i1},X_{i2},X_{i3}$ with assigned power $\E|X_{i1}|^2=1-P^{-d_{i1}},\E|X_{i2}|^2=P^{-d_{i1}}-P^{-1+d_{i3}},\E|X_{i3}|^2=P^{-1+d_{i3}}$. The transmitted signals at the sources are $X_{1[1]}=X_{11}+X_{12}+X_{13},X_{2[1]}=X_{21}+X_{22}+X_{23}$. Relay $\text{Rx}_{1[1]}$ is able to decode $W_{11},W_{12},W_{21}$ successively by treating everything else as noise. Then Relay $\text{Rx}_{1[1]}$ reconstructs and subtracts the codewords $X_{11},X_{12},X_{21}$ from its received signal. The remaining signal above the noise floor at Relay $\text{Rx}_{1[1]}$ is a linear combination of $X_{13},X_{22}$, which we denote as $\mathcal{L}_1$. Relay $\text{Rx}_{1[1]}$ amplifies the remaining signals such that they carry power $\E|\mathcal{L}_1|^2=P^{-1+d_{13}}$. Then, the relay $\text{Rx}_{1[1]}$ splits $W_{11},W_{21}$ into $W_{11}=(W_{11}^1,W_{11}^2), W_{21}=(W_{21}^1,W_{21}^2)$, with the corresponding GDoF $d_{11}^1=d_{11}^2=d_{22}^1=d_{22}^2=\frac{d_{11}}{2}=\frac{1-\alpha}{3}$. Relay $\text{Tx}_{1[2]}\equiv\text{Rx}_{1[1]}$ then re-encodes $W_{12},W_{11}^1,W_{11}^2,W_{22}^1$ into independent Gaussian codebooks producing codewords $X_{12},X_{11}^1,X_{11}^2,X_{22}^1$ with power $\E|X_{12}|^2=1-P^{-d_{12}},\E|X_{11}^1|^2=P^{-d_{12}}-P^{-d_{12}-d_{11}^1},\E|X_{21}^1|^2=P^{-d_{12}-d_{11}^1}-P^{-1+d_{11}^2+d_{13}},\E|X_{11}^2|^2=P^{-1+d_{11}^2+d_{13}}-P^{-1+d_{13}}$. Relay $\text{Tx}_{2[2]}\equiv\text{Rx}_{2[1]}$  proceeds similarly. The transmitted signals at the relays are $X_{1[2]}=X_{12}+X_{11}^1+X_{21}^1+X_{11}^2+\mathcal{L}_1,X_{2[2]}=X_{22}+X_{11}^1+X_{21}^1+X_{21}^2+\mathcal{L}_2$. Then destination node $\text{Rx}_{1[2]}$ is able to decode $W_{12},W_{11}^1,W_{21}^1,W_{22},W_{11}^2,W_{13}$ successively, while treating the other signals as noise, with the corresponding SINR values $\sim P^{d_{12}},\sim P^{d_{11}^1},\sim P^{d_{21}^1},\sim P^{d_{22}},\sim P^{d_{11}^2},\sim P^{d_{13}}$. Destination $\text{Rx}_{2[2]}$ proceeds similarly by decoding  $W_{22},W_{11}^1,W_{21}^1,W_{12},W_{21}^2,W_{23}$ successively. See Figure \ref{fig:sym} for an illustration.
		\begin{figure}[!t]
			\centering
			\begin{tikzpicture}[scale=1.4]
			\draw[thick] (2,2) circle(1mm)node [above] {\small $\text{Tx}_{1[1]}$};
			\draw[thick] (3,2) circle(1mm)node [above] {\small $\text{Rx}_{1[1]}$};
			\draw[thick] (2,-2.5) circle(1mm)node [below] {\small $\text{Tx}_{2[1]}$};
			\draw[thick] (3,-2.5) circle(1mm)node [below] {\small $\text{Rx}_{2[1]}$};
			
			\begin{scope}[shift={(0.8,0)}]
			\draw[thick,dashed] (-0.5,2)--(1.0,2);
			\draw (0.25, 2) node[above]{${X}_{1[1]}$};
			\draw[ thick, fill=white] (0,0) rectangle (0.5,2);
			\draw[ thick, fill=yellow!30!white, pattern color=blue] (0,0) rectangle (0.5,19/42) node [pos=.5] {\tiny $X_{13}$};
			\draw[arrows=<->] (-0.15,0)--(-0.15,19/42)node[left,pos=0.5]{\tiny $\frac{5\alpha-2}{3}$};
			\draw[ thick, fill=blue!30!white, pattern color=blue] (0,15/14) rectangle (0.5,29/21) node [pos=.5] {\tiny $X_{12}$};
			\draw[arrows=<->] (-0.15,15/14)--(-0.15,29/21)node[left,pos=0.5]{\tiny $\frac{1-\alpha}{3}$};
			\draw[ thick, fill=green!30!white, pattern color=green] (0,29/21) rectangle (0.5,2) node [pos=.5] {\tiny $X_{11}$};
			\draw[arrows=<->] (-0.15,29/21)--(-0.15,2)node[left,pos=0.5]{\tiny $\frac{2-2\alpha}{3}$};
			\end{scope}
			
			\begin{scope}[shift={(0.8,-2.5)}]
			\draw[thick,dashed] (-0.5,2)--(1.0,2);
			\draw (0.25, 2) node[above]{${X}_{2[1]}$};
			\draw[ thick, fill=white] (0,0) rectangle (0.5,2);
			\draw[ thick, fill=orange!30!white, pattern color=blue] (0,0) rectangle (0.5,19/42) node [pos=.5] {\tiny $X_{23}$};
			\draw[arrows=<->] (-0.15,0)--(-0.15,19/42)node[left,pos=0.5]{\tiny $\frac{5\alpha-2}{3}$};
			\draw[ thick, fill=green!30!yellow, pattern color=blue] (0,15/14) rectangle (0.5,29/21) node [pos=.5] {\tiny $X_{22}$};
			\draw[arrows=<->] (-0.15,15/14)--(-0.15,29/21)node[left,pos=0.5]{\tiny $\frac{1-\alpha}{3}$};
			\draw[ thick, fill=purple!30!white, pattern color=blue] (0,29/21) rectangle (0.5,2) node [pos=.5] {\tiny $X_{21}$};
			\draw[arrows=<->] (-0.15,29/21)--(-0.15,2)node[left,pos=0.5]{\tiny $\frac{2-2\alpha}{3}$};
			\end{scope}
			
			\draw[thick,arrowmid] (2.1,2)--(2.9,2)node[above,pos=0.5]{$1$};
			\draw[thick,arrowmid] (2.1,-2.5)--(2.9,-2.5)node[below,pos=0.5]{$1$};
			\draw[thick,arrowmid] (2.1,2)--(2.9,-2.5)node[right,pos=0.8]{$\alpha$};
			\draw[thick,arrowmid] (2.1,-2.5)--(2.9,2)node[right,pos=0.8]{$\alpha$};

			\begin{scope}[shift={(4.0,0)}]
			\draw[thick,dashed] (-0.5,0)--(1.5,0);
			\draw[thick] (0.5,0) node[below]{$Y_{1[1]}$};
			\draw[ thick, fill=white] (0,0) rectangle (0.5,2);
			\draw[ thick, fill=yellow!30!white, pattern color=blue] (0,0) rectangle (0.5,19/42) node [pos=.5] {\tiny $X_{13}$};
			\draw[arrows=<->] (-0.15,0)--(-0.15,19/42)node[left,pos=0.5]{\tiny $\frac{5\alpha-2}{3}$};
			\draw[ thick, fill=blue!30!white, pattern color=blue] (0,15/14) rectangle (0.5,29/21) node [pos=.5] {\tiny $X_{12}$};
			\draw[arrows=<->] (-0.15,15/14)--(-0.15,29/21)node[left,pos=0.5]{\tiny $\frac{1-\alpha}{3}$};
			\draw[ thick, fill=green!30!white, pattern color=green] (0,29/21) rectangle (0.5,2) node [pos=.5] {\tiny $X_{11}$};
			\draw[arrows=<->] (-0.15,29/21)--(-0.15,2)node[left,pos=0.5]{\tiny $\frac{2-2\alpha}{3}$};
			\draw[ thick, fill=white] (0.5,0) rectangle (1,15/14);
			\draw[ thick, fill=green!30!yellow, pattern color=blue] (0.5,1/7) rectangle (1,19/42) node [pos=.5] {\tiny $X_{22}$};
			\draw[arrows=<->] (1.15,1/7)--(1.15,19/42)node[right,pos=0.5]{\tiny $\frac{1-\alpha}{3}$};
			\draw[ thick, fill=purple!30!white, pattern color=blue] (0.5,19/42) rectangle (1,15/14) node [pos=.5] {\tiny $X_{21}$};
			\draw[arrows=<->] (1.15,19/42)--(1.15,15/14)node[right,pos=0.5]{\tiny $\frac{2-2\alpha}{3}$};
			\end{scope}

			\begin{scope}[shift={(4.0,-2.5)}]
			\draw[thick,dashed] (-0.5,0)--(1.5,0);
			\draw[thick] (0.5,0) node[below]{$Y_{2[1]}$};
			\draw[ thick, fill=white] (0,0) rectangle (0.5,15/28);
			\draw[ thick, fill=blue!30!white, pattern color=blue] (0,1/7) rectangle (0.5,19/42) node [pos=.5] {\tiny $X_{12}$};
			\draw[arrows=<->] (-0.15,1/7)--(-0.15,19/42)node[left,pos=0.5]{\tiny $\frac{1-\alpha}{3}$};
			\draw[ thick, fill=green!30!white, pattern color=green] (0,19/42) rectangle (0.5,15/14) node [pos=.5] {\tiny $X_{11}$};
			\draw[arrows=<->] (-0.15,19/42)--(-0.15,15/14)node[left,pos=0.5]{\tiny $\frac{2-\alpha}{3}$};
			
			\draw[ thick, fill=white] (0.5,0) rectangle (1,2);
			\draw[ thick, fill=orange!30!white, pattern color=blue] (0.5,0) rectangle (1,19/42) node [pos=.5] {\tiny $X_{23}$};
			\draw[arrows=<->] (1.15,0)--(1.15,19/42)node[right,pos=0.5]{\tiny $\frac{5\alpha-2}{3}$};
			\draw[ thick, fill=green!30!yellow, pattern color=blue] (0.5,15/14) rectangle (1,29/21) node [pos=.5] {\tiny $X_{22}$};
			\draw[arrows=<->] (1.15,15/14)--(1.15,29/21)node[right,pos=0.5]{\tiny $\frac{1-\alpha}{3}$};
			\draw[ thick, fill=purple!30!white, pattern color=blue] (0.5,29/21) rectangle (1,2) node [pos=.5] {\tiny $X_{21}$};
			\draw[arrows=<->] (1.15,29/21)--(1.15,2)node[right,pos=0.5]{\tiny $\frac{2-2\alpha}{3}$};
			\end{scope}
			
			\begin{scope}[shift={(6.5,0)}]
			\draw[thick,dashed] (-0.5,2)--(1.0,2);
			\draw (0.25, 2) node[above]{${X}_{1[2]}$};
			\draw[ thick, fill=white] (0,0) rectangle (0.5,2);
			\draw[ thick, fill=cyan!30!white, pattern color=blue] (0,0) rectangle (0.5,19/42) node [pos=.5] {\tiny $\mathcal{L}_1$};
			\draw[arrows=<->] (-0.15,0)--(-0.15,19/42)node[left,pos=0.5]{\tiny $\frac{5\alpha-2}{3}$};
			\draw[ thick, fill=green!30!white, pattern color=green] (0,19/42) rectangle (0.5,16/21) node [pos=.5] {\tiny $X_{11}^2$};
			\draw[arrows=<->] (-0.15,19/42)--(-0.15,16/21)node[left,pos=0.5]{\tiny $\frac{1-\alpha}{3}$};
			\draw[ thick, fill=yellow!30!white, pattern color=blue] (0,15/14) rectangle (0.5,29/21) node [pos=.5] {\tiny $X_{21}^1$};
			\draw[ thick, fill=yellow!50!white, pattern color=blue] (0,29/21) rectangle (0.5,71/42) node [pos=.5] {\tiny $X_{11}^1$};
			\draw[arrows=<->] (-0.15,15/14)--(-0.15,71/42)node[left,pos=0.5]{\tiny $\frac{2-2\alpha}{3}$};
			\draw[ thick, fill=blue!30!white, pattern color=blue] (0,71/42) rectangle (0.5,2) node [pos=.5] {\tiny $X_{12}$};
			\draw[arrows=<->] (-0.15,71/42)--(-0.15,2)node[left,pos=0.5]{\tiny $\frac{1-\alpha}{3}$};
			\end{scope}
			
			\begin{scope}[shift={(6.5,-2.5)}]
			\draw[thick,dashed] (-0.5,2)--(1.0,2);
			\draw (0.25, 2) node[above]{${X}_{2[2]}$};
			\draw[ thick, fill=white] (0,0) rectangle (0.5,2);
			\draw[ thick, fill=cyan!30!yellow, pattern color=blue](0,0) rectangle (0.5,19/42) node [pos=.5] {\tiny $\mathcal{L}_2$};
			\draw[arrows=<->] (-0.15,0)--(-0.15,19/42)node[left,pos=0.5]{\tiny $\frac{5\alpha-2}{3}$};
			\draw[ thick, fill=purple!30!white, pattern color=blue](0,19/42) rectangle (0.5,16/21)node [pos=.5] {\tiny $X_{21}^2$};
			\draw[arrows=<->] (-0.15,19/42)--(-0.15,16/21)node[left,pos=0.5]{\tiny $\frac{1-\alpha}{3}$};
			\draw[ thick, fill=yellow!30!white, pattern color=blue] (0,15/14) rectangle (0.5,29/21) node [pos=.5] {\tiny $X_{21}^1$};
			\draw[ thick, fill=yellow!50!white, pattern color=blue] (0,29/21) rectangle (0.5,71/42) node [pos=.5] {\tiny $X_{11}^1$};
			\draw[arrows=<->] (-0.15,15/14)--(-0.15,71/42)node[left,pos=0.5]{\tiny $\frac{2-2\alpha}{3}$};
			\draw[ thick, fill=green!30!yellow, pattern color=blue] (0,71/42) rectangle (0.5,2) node [pos=.5] {\tiny $X_{22}$};
			\draw[arrows=<->] (-0.15,71/42)--(-0.15,2)node[left,pos=0.5]{\tiny $\frac{1-\alpha}{3}$};
			\end{scope}
			\begin{scope}[shift={(0.3,0)}]
			\draw[thick] (7.5,2) circle(1mm)node [above] {\small $\text{Tx}_{1[2]}$};
			\draw[thick] (8.5,2) circle(1mm)node [above] {\small $\text{Rx}_{1[2]}$};
			\draw[thick] (7.5,-2.5) circle(1mm)node [below] {\small $\text{Tx}_{2[2]}$};
			\draw[thick] (8.5,-2.5) circle(1mm)node [below] {\small $\text{Rx}_{2[2]}$};
			
			\draw[thick,arrowmid] (7.6,2)--(8.4,2)node[above,pos=0.5]{$1$};
			\draw[thick,arrowmid] (7.6,-2.5)--(8.4,-2.5)node[below,pos=0.5]{$1$};
			\draw[thick,arrowmid] (7.6,2)--(8.4,-2.5)node[right,pos=0.8]{$\alpha$};
			\draw[thick,arrowmid] (7.6,-2.5)--(8.4,2)node[right,pos=0.8]{$\alpha$};
			\end{scope}
			
			\begin{scope}[shift={(9.8,0)}]
			\draw[thick] (0.5,0) node[below]{$Y_{1[2]}$};
			\draw[thick,dashed] (-0.5,0)--(1.5,0);
			\draw[ thick, fill=white] (0,0) rectangle (0.5,2);
			\draw[ thick, fill=cyan!30!white, pattern color=blue] (0,0) rectangle (0.5,19/42) node [pos=.5] {\tiny $\mathcal{L}_1$};
			\draw[arrows=<->] (-0.15,0)--(-0.15,19/42)node[left,pos=0.5]{\tiny $\frac{5\alpha-2}{3}$};
			\draw[ thick, fill=green!30!white, pattern color=green] (0,19/42) rectangle (0.5,16/21) node [pos=.5] {\tiny $X_{11}^2$};
			\draw[arrows=<->] (-0.15,19/42)--(-0.15,16/21)node[left,pos=0.5]{\tiny $\frac{1-\alpha}{3}$};
			\draw[ thick, fill=yellow!30!white, pattern color=blue] (0,15/14) rectangle (0.5,29/21) node [pos=.5] {\tiny $X_{21}^1$};
			\draw[ thick, fill=yellow!50!white, pattern color=blue] (0,29/21) rectangle (0.5,71/42) node [pos=.5] {\tiny $X_{11}^1$};
			\draw[arrows=<->] (-0.15,15/14)--(-0.15,71/42)node[left,pos=0.5]{\tiny $\frac{2-2\alpha}{3}$};
			\draw[ thick, fill=blue!30!white, pattern color=blue] (0,71/42) rectangle (0.5,2) node [pos=.5] {\tiny $X_{12}$};
			\draw[arrows=<->] (-0.15,71/42)--(-0.15,2)node[left,pos=0.5]{\tiny $\frac{1-\alpha}{3}$};
			\draw[ thick, fill=white] (0.5,0) rectangle (1,1/7);
			\draw[ thick, fill=yellow!30!white, pattern color=blue] (0.5,1/7) rectangle (1,19/42) node [pos=.5] {\tiny $X_{21}^1$};
			\draw[ thick, fill=yellow!50!white, pattern color=blue] (0.5,19/42) rectangle (1,16/21) node [pos=.5] {\tiny $X_{11}^1$};
			\draw[arrows=<->] (1.15,1/7)--(1.15,16/21)node[right,pos=0.5]{\tiny $\frac{2-2\alpha}{3}$};
			\draw[ thick, fill=green!30!yellow, pattern color=blue] (0.5,16/21) rectangle (1,15/14) node [pos=.5] {\tiny $X_{22}$};
			\draw[arrows=<->] (1.15,16/21)--(1.15,15/14)node[right,pos=0.5]{\tiny $\frac{1-\alpha}{3}$};
			
			\end{scope}

			\begin{scope}[shift={(9.8,-2.5)}]
			\draw[thick,dashed] (-0.5,0)--(1.5,0);
			\draw[thick] (0.5,0) node[below]{$Y_{2[2]}$};
			\draw[ thick, fill=white] (0,0) rectangle (0.5,1/7);
			\draw[ thick, fill=yellow!50!white, pattern color=blue] (0,1/7) rectangle (0.5,19/42) node [pos=.5] {\tiny $X_{21}^1$};
			\draw[ thick, fill=yellow!50!white, pattern color=blue] (0,19/42) rectangle (0.5,16/21) node [pos=.5] {\tiny $X_{11}^1$};
			\draw[arrows=<->] (-0.15,1/7)--(-0.15,16/21)node[left,pos=0.5]{\tiny $\frac{2-2\alpha}{3}$};
			\draw[ thick, fill=blue!30!white, pattern color=blue] (0,16/21) rectangle (0.5,15/14) node [pos=.5] {\tiny $X_{12}$};
			\draw[arrows=<->] (-0.15,16/21)--(-0.15,15/14)node[left,pos=0.5]{\tiny $\frac{1-\alpha}{3}$};
			
			\draw[ thick, fill=white] (0.5,0) rectangle (1,2);
			\draw[ thick, fill=cyan!30!yellow, pattern color=blue](0.5,0) rectangle (1,19/42) node [pos=.5] {\tiny $\mathcal{L}_2$};
			\draw[arrows=<->] (1.15,0)--(1.15,19/42)node[right,pos=0.5]{\tiny $\frac{5\alpha-2}{3}$};
			\draw[ thick, fill=purple!30!white, pattern color=blue](0.5,19/42) rectangle (1,16/21) node [pos=.5] {\tiny $X_{21}^2$};
			\draw[arrows=<->] (1.15,19/42)--(1.15,16/21)node[right,pos=0.5]{\tiny $\frac{1-\alpha}{3}$};
			\draw[ thick, fill=yellow!50!white, pattern color=blue] (0.5,15/14) rectangle (1,29/21) node [pos=.5] {\tiny $X_{21}^1$};
			\draw[ thick, fill=yellow!50!white, pattern color=blue] (0.5,29/21) rectangle (1,71/42) node [pos=.5] {\tiny $X_{11}^1$};
			\draw[arrows=<->] (1.15,15/14)--(1.15,71/42)node[right,pos=0.5]{\tiny $\frac{2-2\alpha}{3}$};
			\draw[ thick, fill=green!30!yellow, pattern color=blue] (0.5,71/42) rectangle (1,2) node [pos=.5] {\tiny $X_{22}$};
			\draw[arrows=<->] (1.15,71/42)--(1.15,2)node[right,pos=0.5]{\tiny $\frac{1-\alpha}{3}$};
			\end{scope}
			\end{tikzpicture}
			\caption{\it\small Achievable scheme for $\frac{1}{2}\leq\alpha\leq\frac{4}{7}$. $\mathcal{L}_1=\mathcal{L}(X_{13},X_{22}),\mathcal{L}_2=\mathcal{L}(X_{23},X_{12})$. The left figure is the first hop while the right figure represents the second hop.}\label{fig:sym}
		\end{figure}
		
		\item $\alpha\in[4/7,2/3]$\\
		In the regime $\alpha\in[4/7,2/3]$, we have the bound $d_1+d_2=2-\alpha$. The achievable scheme is similar to the case $\alpha\in[1/2,4/7]$ except that $d_{i1}=\frac{2-2\alpha}{3}, d_{i2}=\frac{1-\alpha}{3}, d_{i3}=\frac{\alpha}{2}$.
		\item $\frac{2}{3}\leq\alpha\leq 1$\\
As noted previously, the achievable scheme in this regime is quite simple, the $2$-hop network simply operates as a concatenation of two interference channels using decode-and-forward.
	\end{itemize}
	\subsection{Proof of Achievability for Theorem \ref{thm:Lhop}}\label{sec:Lhopach}
	Building on the insights from the $2$-hop solution, the achievable scheme for $L$-hop setting makes use of rate-splitting at the sources and partial decode-and-forward combined with amplify and forward at the relays. It can be visualized as incremental peeling off of interfered layers, such that the relays of the next layer can decode one more interfered layer compared to the relays in the previous layer. The main ideas are illustrated through the following example of a $3$-hop network.
	
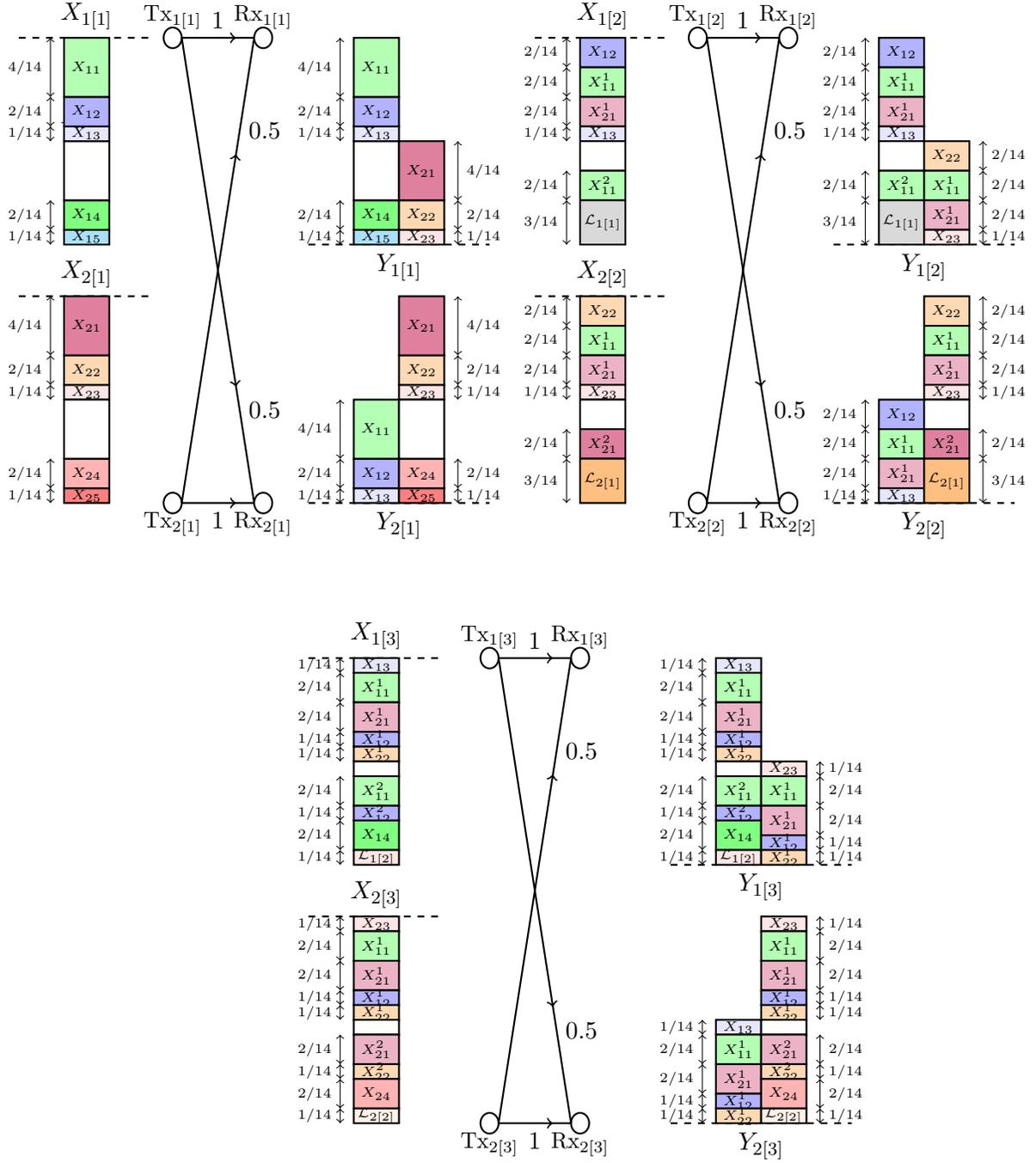
\begin{figure}[!t]
	\centering
	\begin{tikzpicture}[xscale=1.4,yscale=1.6]
	\draw[thick] (2,2) circle(1mm)node [above] {\small $\text{Tx}_{1[1]}$};
	\draw[thick] (3,2) circle(1mm)node [above] {\small $\text{Rx}_{1[1]}$};
	\draw[thick] (2,-2.5) circle(1mm)node [below] {\small $\text{Tx}_{2[1]}$};
	\draw[thick] (3,-2.5) circle(1mm)node [below] {\small $\text{Rx}_{2[1]}$};
	
	\begin{scope}[shift={(0.8,0)}]
	\draw (0.25, 2) node[above]{${X}_{1[1]}$};
	\draw[thick,dashed] (-0.5,2)--(1.0,2);
	\draw (0.25, 1) node[above]{${X}_{1[1]}$};
	\draw[ thick, fill=cyan!30!white, pattern color=blue] (0,0) rectangle (0.5,2/14) node [pos=.5] {\tiny $X_{15}$};
	\draw[arrows=<->] (-0.15,0)--(-0.15,2/14)node[left,pos=0.5]{\tiny $1/14$};
	\draw[ thick, fill=green!50!white, pattern color=yellow] (0,2/14) rectangle (0.5,6/14) node [pos=.5] {\tiny $X_{14}$};
	\draw[arrows=<->] (-0.15,2/14)--(-0.15,6/14)node[left,pos=0.5]{\tiny $2/14$};
	\draw[ thick, fill=white] (0,6/14) rectangle (0.5,1);
	\draw[ thick, fill=blue!10!white, pattern color=blue] (0,1) rectangle (0.5,16/14) node [pos=.5] {\tiny $X_{13}$};
	\draw[arrows=<->] (-0.15,14/14)--(-0.15,16/14)node[left,pos=0.5]{\tiny $1/14$};
	\draw[ thick, fill=blue!30!white, pattern color=blue] (0,16/14) rectangle (0.5,20/14) node [pos=.5] {\tiny $X_{12}$};
	\draw[arrows=<->] (-0.15,16/14)--(-0.15,20/14)node[left,pos=0.5]{\tiny $2/14$};
	\draw[ thick, fill=green!30!white, pattern color=green] (0,20/14) rectangle (0.5,2) node [pos=.5] {\tiny $X_{11}$};
	\draw[arrows=<->] (-0.15,20/14)--(-0.15,2)node[left,pos=0.5]{\tiny $4/14$};
	\end{scope}

	\begin{scope}[shift={(0.8,-2.5)}]
	\draw[thick,dashed] (-0.5,2)--(1.0,2);
	\draw (0.25, 2) node[above]{${X}_{2[1]}$};
	\draw[ thick, fill=red!50!white, pattern color=blue] (0,0) rectangle (0.5,2/14) node [pos=.5] {\tiny $X_{25}$};
	\draw[arrows=<->] (-0.15,0)--(-0.15,2/14)node[left,pos=0.5]{\tiny $1/14$};
	\draw[ thick, fill=red!30!white, pattern color=yellow] (0,2/14) rectangle (0.5,6/14) node [pos=.5] {\tiny $X_{24}$};
	\draw[arrows=<->] (-0.15,2/14)--(-0.15,6/14)node[left,pos=0.5]{\tiny $2/14$};
	\draw[ thick, fill=red!10!white, pattern color=blue] (0,14/14) rectangle (0.5,16/14) node [pos=.5] {\tiny $X_{23}$};
	\draw[arrows=<->] (-0.15,14/14)--(-0.15,16/14)node[left,pos=0.5]{\tiny $1/14$};
	\draw[ thick, fill=white] (0,6/14) rectangle (0.5,14/14);
	\draw[ thick, fill=orange!30!white, pattern color=blue] (0,16/14) rectangle (0.5,20/14) node [pos=.5] {\tiny $X_{22}$};
	\draw[arrows=<->] (-0.15,16/14)--(-0.15,20/14)node[left,pos=0.5]{\tiny $2/14$};

	\draw[ thick, fill=purple!50!white, pattern color=blue] (0,20/14) rectangle (0.5,2) node [pos=.5] {\tiny $X_{21}$};
	\draw[arrows=<->] (-0.15,20/14)--(-0.15,2)node[left,pos=0.5]{\tiny $4/14$};
	\end{scope}

	\draw[thick,arrowmid] (2.1,2)--(2.9,2)node[above,pos=0.5]{$1$};
	\draw[thick,arrowmid] (2.1,-2.5)--(2.9,-2.5)node[below,pos=0.5]{$1$};
	\draw[thick,arrowmid] (2.1,2)--(2.9,-2.5)node[right,pos=0.8]{$0.5$};
	\draw[thick,arrowmid] (2.1,-2.5)--(2.9,2)node[right,pos=0.8]{$0.5$};

	\begin{scope}[shift={(4.0,0)}]
	\draw[thick,dashed] (-0.5,0)--(1.5,0);
	\draw[thick] (0.5,0) node[below]{$Y_{1[1]}$};
	\draw[ thick, fill=cyan!30!white, pattern color=blue] (0,0) rectangle (0.5,2/14) node [pos=.5] {\tiny $X_{15}$};
	\draw[arrows=<->] (-0.15,0)--(-0.15,2/14)node[left,pos=0.5]{\tiny $1/14$};
	\draw[ thick, fill=green!50!white, pattern color=yellow] (0,2/14) rectangle (0.5,6/14) node [pos=.5] {\tiny $X_{14}$};
	\draw[arrows=<->] (-0.15,2/14)--(-0.15,6/14)node[left,pos=0.5]{\tiny $2/14$};
	\draw[ thick, fill=white] (0,6/14) rectangle (0.5,14/14);
	\draw[ thick, fill=blue!10!white, pattern color=blue] (0,14/14) rectangle (0.5,16/14) node [pos=.5] {\tiny $X_{13}$};
	\draw[arrows=<->] (-0.15,14/14)--(-0.15,16/14)node[left,pos=0.5]{\tiny $1/14$};
	\draw[ thick, fill=blue!30!white, pattern color=blue] (0,16/14) rectangle (0.5,20/14) node [pos=.5] {\tiny $X_{12}$};
	\draw[arrows=<->] (-0.15,16/14)--(-0.15,20/14)node[left,pos=0.5]{\tiny $2/14$};
	\draw[ thick, fill=green!30!white, pattern color=green] (0,20/14) rectangle (0.5,2) node [pos=.5] {\tiny $X_{11}$};
	\draw[arrows=<->] (-0.15,20/14)--(-0.15,2)node[left,pos=0.5]{\tiny $4/14$};		
	\draw[ thick, fill=red!10!white, pattern color=blue] (0.5,0) rectangle (1,2/14) node [pos=.5] {\tiny $X_{23}$};
	\draw[arrows=<->] (1.15,0)--(1.15,2/14)node[right,pos=0.5]{\tiny $1/14$};
	\draw[ thick, fill=orange!30!white, pattern color=blue] (0.5,2/14) rectangle (1,6/14) node [pos=.5] {\tiny $X_{22}$};
	\draw[arrows=<->] (1.15,2/14)--(1.15,6/14)node[right,pos=0.5]{\tiny $2/14$};
	\draw[ thick, fill=purple!50!white, pattern color=blue] (0.5,6/14) rectangle (1,14/14) node [pos=.5] {\tiny $X_{21}$};
	\draw[arrows=<->] (1.15,6/14)--(1.15,14/14)node[right,pos=0.5]{\tiny $4/14$};
	\end{scope}

	\begin{scope}[shift={(4.0,-2.5)}]
	\draw[thick,dashed] (-0.5,0)--(1.5,0);
	\draw[thick] (0.5,0) node[below]{$Y_{2[1]}$};
	\draw[ thick, fill=blue!10!white, pattern color=blue] (0,0) rectangle (0.5,2/14) node [pos=.5] {\tiny $X_{13}$};
	\draw[arrows=<->] (-0.15,0)--(-0.15,2/14)node[left,pos=0.5]{\tiny $1/14$};
	\draw[ thick, fill=blue!30!white, pattern color=blue] (0,2/14) rectangle (0.5,6/14) node [pos=.5] {\tiny $X_{12}$};
	\draw[arrows=<->] (-0.15,2/14)--(-0.15,6/14)node[left,pos=0.5]{\tiny $2/14$};
	\draw[ thick, fill=green!30!white, pattern color=green] (0,6/14) rectangle (0.5,14/14) node [pos=.5] {\tiny $X_{11}$};
	\draw[arrows=<->] (-0.15,6/14)--(-0.15,14/14)node[left,pos=0.5]{\tiny $4/14$};
	\draw[ thick, fill=red!50!white, pattern color=blue] (0.5,0) rectangle (1,2/14) node [pos=.5] {\tiny $X_{25}$};
	\draw[arrows=<->] (1.15,0)--(1.15,2/14)node[right,pos=0.5]{\tiny $1/14$};
	\draw[ thick, fill=red!30!white, pattern color=yellow] (0.5,2/14) rectangle (1,6/14) node [pos=.5] {\tiny $X_{24}$};
	\draw[arrows=<->] (1.15,2/14)--(1.15,6/14)node[right,pos=0.5]{\tiny $2/14$};
	\draw[ thick, fill=red!10!white, pattern color=blue] (0.5,14/14) rectangle (1,16/14) node [pos=.5] {\tiny $X_{23}$};
	\draw[arrows=<->] (1.15,14/14)--(1.15,16/14)node[right,pos=0.5]{\tiny $1/14$};
	\draw[ thick, fill=white] (0.5,6/14) rectangle (1,14/14);
	\draw[ thick, fill=orange!30!white, pattern color=blue] (0.5,16/14) rectangle (1,20/14) node [pos=.5] {\tiny $X_{22}$};
	\draw[arrows=<->] (1.15,16/14)--(1.15,20/14)node[right,pos=0.5]{\tiny $2/14$};

	\draw[ thick, fill=purple!50!white, pattern color=blue] (0.5,20/14) rectangle (1,2) node [pos=.5] {\tiny $X_{21}$};
	\draw[arrows=<->] (1.15,20/14)--(1.15,2)node[right,pos=0.5]{\tiny $4/14$};
	\end{scope}

	\begin{scope}[shift={(6.5,0)}]
	\draw[thick,dashed] (-0.5,2)--(1.0,2);
	\draw (0.25, 2) node[above]{${X}_{1[2]}$};
	\draw[ thick, fill=blue!30!white, pattern color=blue] (0,24/14) rectangle (0.5,2) node [pos=.5] {\tiny $X_{12}$};
	\draw[arrows=<->] (-0.15,24/14)--(-0.15,2)node[left,pos=0.5]{\tiny $2/14$};
	\draw[ thick, fill=green!30!white, pattern color=red] (0,20/14) rectangle (0.5,24/14) node [pos=.5] {\tiny $X_{11}^1$};
	\draw[arrows=<->] (-0.15,20/14)--(-0.15,24/14)node[left,pos=0.5]{\tiny $2/14$};
	\draw[ thick, fill=purple!30!white, pattern color=red] (0,16/14) rectangle (0.5,20/14) node [pos=.5] {\tiny $X_{21}^1$};
	\draw[arrows=<->] (-0.15,16/14)--(-0.15,20/14)node[left,pos=0.5]{\tiny $2/14$};
	\draw[ thick, fill=blue!10!white, pattern color=blue] (0,14/14) rectangle (0.5,16/14) node [pos=.5] {\tiny $X_{13}$};
	\draw[arrows=<->] (-0.15,14/14)--(-0.15,16/14)node[left,pos=0.5]{\tiny $1/14$};
	\draw[ thick, fill=white] (0,10/14) rectangle (0.5,14/14);
	\draw[ thick, fill=green!30!white, pattern color=blue] (0,6/14) rectangle (0.5,10/14) node [pos=.5] {\tiny $X_{11}^2$};
	\draw[arrows=<->] (-0.15,6/14)--(-0.15,10/14)node[left,pos=0.5]{\tiny $2/14$};
	\draw[ thick, fill=gray!30!white, pattern color=blue] (0,0) rectangle (0.5,6/14) node [pos=.5] {\tiny $\mathcal{L}_{1[1]}$};
	\draw[arrows=<->] (-0.15,0)--(-0.15,6/14)node[left,pos=0.5]{\tiny $3/14$};
	\end{scope}

	\begin{scope}[shift={(6.5,-2.5)}]
	\draw[thick,dashed] (-0.5,2)--(1.0,2);
	\draw (0.25, 2) node[above]{${X}_{2[2]}$};
	\draw[ thick, fill=orange!30!white, pattern color=blue](0,24/14) rectangle (0.5,2) node [pos=.5] {\tiny $X_{22}$};
	\draw[arrows=<->] (-0.15,24/14)--(-0.15,2)node[left,pos=0.5]{\tiny $2/14$};
	\draw[ thick, fill=green!30!white, pattern color=red] (0,20/14) rectangle (0.5,24/14) node [pos=.5] {\tiny $X_{11}^1$};
	\draw[arrows=<->] (-0.15,20/14)--(-0.15,24/14)node[left,pos=0.5]{\tiny $2/14$};
	\draw[ thick, fill=purple!30!white, pattern color=red] (0,16/14) rectangle (0.5,20/14) node [pos=.5] {\tiny $X_{21}^1$};
	\draw[arrows=<->] (-0.15,16/14)--(-0.15,20/14)node[left,pos=0.5]{\tiny $2/14$};
	\draw[ thick, fill=red!10!white, pattern color=blue] (0,14/14) rectangle (0.5,16/14) node [pos=.5] {\tiny $X_{23}$};
	\draw[arrows=<->] (-0.15,14/14)--(-0.15,16/14)node[left,pos=0.5]{\tiny $1/14$};
	\draw[ thick, fill=white] (0,10/14) rectangle (0.5,14/14);
	\draw[ thick, fill=purple!50!white, pattern color=blue] (0,6/14) rectangle (0.5,10/14) node [pos=.5] {\tiny $X_{21}^2$};
	\draw[arrows=<->] (-0.15,6/14)--(-0.15,10/14)node[left,pos=0.5]{\tiny $2/14$};
	\draw[ thick, fill=orange!50!white, pattern color=blue] (0,0) rectangle (0.5,6/14) node [pos=.5] {\tiny $\mathcal{L}_{2[1]}$};
    \draw[arrows=<->] (-0.15,0)--(-0.15,6/14)node[left,pos=0.5]{\tiny $3/14$};
	\end{scope}

			\begin{scope}[shift={(0.3,0)}]
		\draw[thick] (7.5,2) circle(1mm)node [above] {\small $\text{Tx}_{1[2]}$};
		\draw[thick] (8.5,2) circle(1mm)node [above] {\small $\text{Rx}_{1[2]}$};
		\draw[thick] (7.5,-2.5) circle(1mm)node [below] {\small $\text{Tx}_{2[2]}$};
		\draw[thick] (8.5,-2.5) circle(1mm)node [below] {\small $\text{Rx}_{2[2]}$};
		
		\draw[thick,arrowmid] (7.6,2)--(8.4,2)node[above,pos=0.5]{$1$};
		\draw[thick,arrowmid] (7.6,-2.5)--(8.4,-2.5)node[below,pos=0.5]{$1$};
		\draw[thick,arrowmid] (7.6,2)--(8.4,-2.5)node[right,pos=0.8]{$0.5$};
		\draw[thick,arrowmid] (7.6,-2.5)--(8.4,2)node[right,pos=0.8]{$0.5$};
		\end{scope}

	\begin{scope}[shift={(9.8,0)}]
	\draw[thick,dashed] (-0.5,0)--(1.5,0);
	\draw[thick] (0.5,0) node[below]{$Y_{1[2]}$};
	\draw[ thick, fill=blue!30!white, pattern color=blue] (0,24/14) rectangle (0.5,2) node [pos=.5] {\tiny $X_{12}$};
	\draw[arrows=<->] (-0.15,24/14)--(-0.15,2)node[left,pos=0.5]{\tiny $2/14$};
	\draw[ thick, fill=green!30!white, pattern color=red] (0,20/14) rectangle (0.5,24/14) node [pos=.5] {\tiny $X_{11}^1$};
	\draw[arrows=<->] (-0.15,20/14)--(-0.15,24/14)node[left,pos=0.5]{\tiny $2/14$};
	\draw[ thick, fill=purple!30!white, pattern color=red] (0,16/14) rectangle (0.5,20/14) node [pos=.5] {\tiny $X_{21}^1$};
	\draw[arrows=<->] (-0.15,16/14)--(-0.15,20/14)node[left,pos=0.5]{\tiny $2/14$};
	\draw[ thick, fill=blue!10!white, pattern color=blue] (0,14/14) rectangle (0.5,16/14) node [pos=.5] {\tiny $X_{13}$};
	\draw[arrows=<->] (-0.15,14/14)--(-0.15,16/14)node[left,pos=0.5]{\tiny $1/14$};
	\draw[ thick, fill=white] (0,10/14) rectangle (0.5,14/14);
	\draw[ thick, fill=green!30!white, pattern color=blue] (0,6/14) rectangle (0.5,10/14) node [pos=.5] {\tiny $X_{11}^2$};
	\draw[arrows=<->] (-0.15,6/14)--(-0.15,10/14)node[left,pos=0.5]{\tiny $2/14$};
	\draw[ thick, fill=gray!30!white, pattern color=blue] (0,0) rectangle (0.5,6/14) node [pos=.5] {\tiny $\mathcal{L}_{1[1]}$};
	\draw[arrows=<->] (-0.15,0)--(-0.15,6/14)node[left,pos=0.5]{\tiny $3/14$};
	
	\draw[ thick, fill=orange!30!white, pattern color=blue](0.5,10/14) rectangle (1,1) node [pos=.5] {\tiny $X_{22}$};
	\draw[arrows=<->] (1.15,10/14)--(1.15,14/14)node[right,pos=0.5]{\tiny $2/14$};
	\draw[ thick, fill=green!30!white, pattern color=red] (0.5,6/14) rectangle (1,10/14) node [pos=.5] {\tiny $X_{11}^1$};
	\draw[arrows=<->] (1.15,6/14)--(1.15,10/14)node[right,pos=0.5]{\tiny $2/14$};
	\draw[ thick, fill=purple!30!white, pattern color=red] (0.5,2/14) rectangle (1,6/14) node [pos=.5] {\tiny $X_{21}^1$};
	\draw[arrows=<->] (1.15,2/14)--(1.15,6/14)node[right,pos=0.5]{\tiny $2/14$};
	\draw[ thick, fill=red!10!white, pattern color=blue] (0.5,0) rectangle (1,2/14) node [pos=.5] {\tiny $X_{23}$};
	\draw[arrows=<->] (1.15,0)--(1.15,2/14)node[right,pos=0.5]{\tiny $1/14$};
	\end{scope}

	\begin{scope}[shift={(9.8,-2.5)}]
	\draw[thick,dashed] (-0.5,0)--(1.5,0);
	\draw[thick] (0.5,0) node[below]{$Y_{2[2]}$};
	\draw[ thick, fill=blue!30!white, pattern color=blue] (0,10/14) rectangle (0.5,14/14) node [pos=.5] {\tiny $X_{12}$};
	\draw[arrows=<->] (-0.15,10/14)--(-0.15,14/14)node[left,pos=0.5]{\tiny $2/14$};
	\draw[ thick, fill=green!30!white, pattern color=red] (0,6/14) rectangle (0.5,10/14) node [pos=.5] {\tiny $X_{11}^1$};
	\draw[arrows=<->] (-0.15,6/14)--(-0.15,10/14)node[left,pos=0.5]{\tiny $2/14$};
	\draw[ thick, fill=purple!30!white, pattern color=red] (0,2/14) rectangle (0.5,6/14) node [pos=.5] {\tiny $X_{21}^1$};
	\draw[arrows=<->] (-0.15,2/14)--(-0.15,6/14)node[left,pos=0.5]{\tiny $2/14$};
	\draw[ thick, fill=blue!10!white, pattern color=blue] (0,0) rectangle (0.5,2/14) node [pos=.5] {\tiny $X_{13}$};
	\draw[arrows=<->] (-0.15,0)--(-0.15,2/14)node[left,pos=0.5]{\tiny $1/14$};
	
	\draw[ thick, fill=orange!30!white, pattern color=blue](0.5,24/14) rectangle (1,2) node [pos=.5] {\tiny $X_{22}$};
	\draw[arrows=<->] (1.15,24/14)--(1.15,2)node[right,pos=0.5]{\tiny $2/14$};
	\draw[ thick, fill=green!30!white, pattern color=red] (0.5,20/14) rectangle (1,24/14) node [pos=.5] {\tiny $X_{11}^1$};
	\draw[arrows=<->] (1.15,20/14)--(1.15,24/14)node[right,pos=0.5]{\tiny $2/14$};
	\draw[ thick, fill=purple!30!white, pattern color=red] (0.5,16/14) rectangle (1,20/14) node [pos=.5] {\tiny $X_{21}^1$};
	\draw[arrows=<->] (1.15,16/14)--(1.15,20/14)node[right,pos=0.5]{\tiny $2/14$};
	\draw[ thick, fill=red!10!white, pattern color=blue] (0.5,14/14) rectangle (1,16/14) node [pos=.5] {\tiny $X_{23}$};
	\draw[arrows=<->] (1.15,14/14)--(1.15,16/14)node[right,pos=0.5]{\tiny $1/14$};
	\draw[ thick, fill=white] (0.5,10/14) rectangle (1,14/14);
	\draw[ thick, fill=purple!50!white, pattern color=blue] (0.5,6/14) rectangle (1,10/14) node [pos=.5] {\tiny $X_{21}^2$};
	\draw[arrows=<->] (1.15,6/14)--(1.15,10/14)node[right,pos=0.5]{\tiny $2/14$};
	\draw[ thick, fill=orange!50!white, pattern color=blue] (0.5,0) rectangle (1,6/14) node [pos=.5] {\tiny $\mathcal{L}_{2[1]}$};
	\draw[arrows=<->] (1.15,0)--(1.15,6/14)node[right,pos=0.5]{\tiny $3/14$};
	\end{scope}
	
	\begin{scope}[shift={(4,-6)}]
	\draw[thick,dashed] (-0.5,2)--(1.0,2);
	\draw (0.25, 2) node[above]{${X}_{1[3]}$};
	\draw[ thick, fill=blue!10!white, pattern color=blue] (0,26/14) rectangle (0.5,2) node [pos=.5] {\tiny $X_{13}$};
	\draw[arrows=<->] (-0.15,26/14)--(-0.15,2)node[left,pos=0.5]{\tiny $1/14$};
	\draw[ thick, fill=green!30!white, pattern color=red] (0,22/14) rectangle (0.5,26/14) node [pos=.5] {\tiny $X_{11}^1$};
	\draw[arrows=<->] (-0.15,22/14)--(-0.15,26/14)node[left,pos=0.5]{\tiny $2/14$};
	\draw[ thick, fill=purple!30!white, pattern color=red] (0,18/14) rectangle (0.5,22/14) node [pos=.5] {\tiny $X_{21}^1$};
	\draw[arrows=<->] (-0.15,18/14)--(-0.15,22/14)node[left,pos=0.5]{\tiny $2/14$};
	\draw[ thick, fill=blue!30!white, pattern color=red] (0,16/14) rectangle (0.5,18/14) node [pos=.5] {\tiny $X_{12}^1$};
	\draw[arrows=<->] (-0.15,16/14)--(-0.15,18/14)node[left,pos=0.5]{\tiny $1/14$};
	\draw[ thick, fill=orange!30!white, pattern color=blue] (0,14/14) rectangle (0.5,16/14) node [pos=.5] {\tiny $X_{22}^1$};
	\draw[arrows=<->] (-0.15,14/14)--(-0.15,16/14)node[left,pos=0.5]{\tiny $1/14$};
	\draw[ thick, fill=white] (0,12/14) rectangle (0.5,1);
	\draw[ thick, fill=green!30!white, pattern color=blue] (0,8/14) rectangle (0.5,12/14) node [pos=.5] {\tiny $X_{11}^2$};
	\draw[arrows=<->] (-0.15,8/14)--(-0.15,12/14)node[left,pos=0.5]{\tiny $2/14$};
	\draw[ thick, fill=blue!30!white, pattern color=blue](0,6/14) rectangle (0.5,8/14) node [pos=.5] {\tiny $X_{12}^2$};
	\draw[arrows=<->] (-0.15,6/14)--(-0.15,8/14)node[left,pos=0.5]{\tiny $1/14$};
	\draw[ thick, fill=green!50!white, pattern color=yellow] (0,2/14) rectangle (0.5,6/14) node [pos=.5] {\tiny $X_{14}$};
	\draw[arrows=<->] (-0.15,2/14)--(-0.15,6/14)node[left,pos=0.5]{\tiny $2/14$};
	\draw[ thick, fill=red!10!white, pattern color=blue] (0,0) rectangle (0.5,2/14) node [pos=.5] {\tiny $\mathcal{L}_{1[2]}$};
	\draw[arrows=<->] (-0.15,0)--(-0.15,2/14)node[left,pos=0.5]{\tiny $1/14$};
	\end{scope}
	
	\begin{scope}[shift={(4,-8.5)}]
	\draw[thick,dashed] (-0.5,2)--(1.0,2);
	\draw (0.25, 2) node[above]{${X}_{2[3]}$};
	\draw[ thick, fill=red!10!white, pattern color=blue] (0,26/14) rectangle (0.5,2) node [pos=.5] {\tiny $X_{23}$};
	\draw[arrows=<->] (-0.15,26/14)--(-0.15,2)node[left,pos=0.5]{\tiny $1/14$};
	\draw[ thick, fill=green!30!white, pattern color=red] (0,22/14) rectangle (0.5,26/14) node [pos=.5] {\tiny $X_{11}^1$};
	\draw[arrows=<->] (-0.15,22/14)--(-0.15,26/14)node[left,pos=0.5]{\tiny $2/14$};
	\draw[ thick, fill=purple!30!white, pattern color=red] (0,18/14) rectangle (0.5,22/14) node [pos=.5] {\tiny $X_{21}^1$};
	\draw[arrows=<->] (-0.15,18/14)--(-0.15,22/14)node[left,pos=0.5]{\tiny $2/14$};
	\draw[ thick, fill=blue!30!white, pattern color=red] (0,16/14) rectangle (0.5,18/14) node [pos=.5] {\tiny $X_{12}^1$};
	\draw[arrows=<->] (-0.15,16/14)--(-0.15,18/14)node[left,pos=0.5]{\tiny $1/14$};
	\draw[ thick, fill=orange!30!white, pattern color=blue] (0,14/14) rectangle (0.5,16/14) node [pos=.5] {\tiny $X_{22}^1$};
	\draw[arrows=<->] (-0.15,14/14)--(-0.15,16/14)node[left,pos=0.5]{\tiny $1/14$};
	\draw[ thick, fill=white] (0,12/14) rectangle (0.5,1);
	\draw[ thick, fill=purple!30!white, pattern color=blue] (0,8/14) rectangle (0.5,12/14) node [pos=.5] {\tiny $X_{21}^2$};
	\draw[arrows=<->] (-0.15,8/14)--(-0.15,12/14)node[left,pos=0.5]{\tiny $2/14$};
	\draw[ thick, fill=orange!30!white, pattern color=blue](0,6/14) rectangle (0.5,8/14) node [pos=.5] {\tiny $X_{22}^2$};
	\draw[arrows=<->] (-0.15,6/14)--(-0.15,8/14)node[left,pos=0.5]{\tiny $1/14$};
	\draw[ thick, fill=red!30!white, pattern color=yellow] (0,2/14) rectangle (0.5,6/14) node [pos=.5] {\tiny $X_{24}$};
	\draw[arrows=<->] (-0.15,2/14)--(-0.15,6/14)node[left,pos=0.5]{\tiny $2/14$};
	\draw[ thick, fill=orange!10!white, pattern color=blue] (0,0) rectangle (0.5,2/14) node [pos=.5] {\tiny $\mathcal{L}_{2[2]}$};
	\draw[arrows=<->] (-0.15,0)--(-0.15,2/14)node[left,pos=0.5]{\tiny $1/14$};
	\end{scope}
	
		\begin{scope}[shift={(-2,-6)}]
		\draw[thick] (7.5,2) circle(1mm)node [above] {\small $\text{Tx}_{1[3]}$};
		\draw[thick] (8.5,2) circle(1mm)node [above] {\small $\text{Rx}_{1[3]}$};
		\draw[thick] (7.5,-2.5) circle(1mm)node [below] {\small $\text{Tx}_{2[3]}$};
		\draw[thick] (8.5,-2.5) circle(1mm)node [below] {\small $\text{Rx}_{2[3]}$};
		
		\draw[thick,arrowmid] (7.6,2)--(8.4,2)node[above,pos=0.5]{$1$};
		\draw[thick,arrowmid] (7.6,-2.5)--(8.4,-2.5)node[below,pos=0.5]{$1$};
		\draw[thick,arrowmid] (7.6,2)--(8.4,-2.5)node[right,pos=0.8]{$0.5$};
		\draw[thick,arrowmid] (7.6,-2.5)--(8.4,2)node[right,pos=0.8]{$0.5$};
		\end{scope}
	\begin{scope}[shift={(8,-6)}]
	\draw[thick,dashed] (-0.5,0)--(1.5,0);
	\draw[thick] (0.5,0) node[below]{$Y_{1[3]}$};
	\draw[ thick, fill=blue!10!white, pattern color=blue] (0,26/14) rectangle (0.5,2) node [pos=.5] {\tiny $X_{13}$};
	\draw[arrows=<->] (-0.15,26/14)--(-0.15,2)node[left,pos=0.5]{\tiny $1/14$};
	\draw[ thick, fill=green!30!white, pattern color=red] (0,22/14) rectangle (0.5,26/14) node [pos=.5] {\tiny $X_{11}^1$};
	\draw[arrows=<->] (-0.15,22/14)--(-0.15,26/14)node[left,pos=0.5]{\tiny $2/14$};
	\draw[ thick, fill=purple!30!white, pattern color=red] (0,18/14) rectangle (0.5,22/14) node [pos=.5] {\tiny $X_{21}^1$};
	\draw[arrows=<->] (-0.15,18/14)--(-0.15,22/14)node[left,pos=0.5]{\tiny $2/14$};
	\draw[ thick, fill=blue!30!white, pattern color=red] (0,16/14) rectangle (0.5,18/14) node [pos=.5] {\tiny $X_{12}^1$};
	\draw[arrows=<->] (-0.15,16/14)--(-0.15,18/14)node[left,pos=0.5]{\tiny $1/14$};
	\draw[ thick, fill=orange!30!white, pattern color=blue] (0,14/14) rectangle (0.5,16/14) node [pos=.5] {\tiny $X_{22}^1$};
	\draw[arrows=<->] (-0.15,14/14)--(-0.15,16/14)node[left,pos=0.5]{\tiny $1/14$};
	\draw[ thick, fill=white] (0,12/14) rectangle (0.5,1);
	\draw[ thick, fill=green!30!white, pattern color=blue] (0,8/14) rectangle (0.5,12/14) node [pos=.5] {\tiny $X_{11}^2$};
	\draw[arrows=<->] (-0.15,8/14)--(-0.15,12/14)node[left,pos=0.5]{\tiny $2/14$};
	\draw[ thick, fill=blue!30!white, pattern color=blue](0,6/14) rectangle (0.5,8/14) node [pos=.5] {\tiny $X_{12}^2$};
	\draw[arrows=<->] (-0.15,6/14)--(-0.15,8/14)node[left,pos=0.5]{\tiny $1/14$};
	\draw[ thick, fill=green!50!white, pattern color=yellow] (0,2/14) rectangle (0.5,6/14) node [pos=.5] {\tiny $X_{14}$};
	\draw[arrows=<->] (-0.15,2/14)--(-0.15,6/14)node[left,pos=0.5]{\tiny $2/14$};
	\draw[ thick, fill=red!10!white, pattern color=blue] (0,0) rectangle (0.5,2/14) node [pos=.5] {\tiny $\mathcal{L}_{1[2]}$};
	\draw[arrows=<->] (-0.15,0)--(-0.15,2/14)node[left,pos=0.5]{\tiny $1/14$};
	
	\draw[ thick, fill=red!10!white, pattern color=blue] (0.5,12/14) rectangle (1,1) node [pos=.5] {\tiny $X_{23}$};
	\draw[arrows=<->] (1.15,12/14)--(1.15,1)node[right,pos=0.5]{\tiny $1/14$};
	\draw[ thick, fill=green!30!white, pattern color=red] (0.5,8/14) rectangle (1,12/14) node [pos=.5] {\tiny $X_{11}^1$};
	\draw[arrows=<->] (1.15,8/14)--(1.15,12/14)node[right,pos=0.5]{\tiny $2/14$};
	\draw[ thick, fill=purple!30!white, pattern color=red] (0.5,4/14) rectangle (1,8/14) node [pos=.5] {\tiny $X_{21}^1$};
	\draw[arrows=<->] (1.15,4/14)--(1.15,8/14)node[right,pos=0.5]{\tiny $2/14$};
	\draw[ thick, fill=blue!30!white, pattern color=red] (0.5,2/14) rectangle (1,4/14) node [pos=.5] {\tiny $X_{12}^1$};
	\draw[arrows=<->] (1.15,2/14)--(1.15,4/14)node[right,pos=0.5]{\tiny $1/14$};
	\draw[ thick, fill=orange!30!white, pattern color=blue] (0.5,0) rectangle (1,2/14) node [pos=.5] {\tiny $X_{22}^1$};
	\draw[arrows=<->] (1.15,0)--(1.15,2/14)node[right,pos=0.5]{\tiny $1/14$};
	\end{scope}

	\begin{scope}[shift={(8.0,-8.5)}]
	\draw[thick,dashed] (-0.5,0)--(1.5,0);
	\draw[thick] (0.5,0) node[below]{$Y_{2[3]}$};
	\draw[ thick, fill=blue!10!white, pattern color=blue] (0,12/14) rectangle (0.5,1) node [pos=.5] {\tiny $X_{13}$};
	\draw[arrows=<->] (-0.15,12/14)--(-0.15,1)node[left,pos=0.5]{\tiny $1/14$};
	\draw[ thick, fill=green!30!white, pattern color=red] (0,8/14) rectangle (0.5,12/14) node [pos=.5] {\tiny $X_{11}^1$};
	\draw[arrows=<->] (-0.15,8/14)--(-0.15,12/14)node[left,pos=0.5]{\tiny $2/14$};
	\draw[ thick, fill=purple!30!white, pattern color=red] (0,4/14) rectangle (0.5,8/14) node [pos=.5] {\tiny $X_{21}^1$};
	\draw[arrows=<->] (-0.15,4/14)--(-0.15,8/14)node[left,pos=0.5]{\tiny $2/14$};
	\draw[ thick, fill=blue!30!white, pattern color=red] (0,2/14) rectangle (0.5,4/14) node [pos=.5] {\tiny $X_{12}^1$};
	\draw[arrows=<->] (-0.15,2/14)--(-0.15,4/14)node[left,pos=0.5]{\tiny $1/14$};
	\draw[ thick, fill=orange!30!white, pattern color=blue] (0,0) rectangle (0.5,2/14) node [pos=.5] {\tiny $X_{22}^1$};
	\draw[arrows=<->] (-0.15,0)--(-0.15,2/14)node[left,pos=0.5]{\tiny $1/14$};
	
	\draw[ thick, fill=red!10!white, pattern color=blue] (0.5,26/14) rectangle (1,2) node [pos=.5] {\tiny $X_{23}$};
	\draw[arrows=<->] (1.15,26/14)--(1.15,2)node[right,pos=0.5]{\tiny $1/14$};
	\draw[ thick, fill=green!30!white, pattern color=red] (0.5,22/14) rectangle (1,26/14) node [pos=.5] {\tiny $X_{11}^1$};
	\draw[arrows=<->] (1.15,22/14)--(1.15,26/14)node[right,pos=0.5]{\tiny $2/14$};
	\draw[ thick, fill=purple!30!white, pattern color=red] (0.5,18/14) rectangle (1,22/14) node [pos=.5] {\tiny $X_{21}^1$};
	\draw[arrows=<->] (1.15,18/14)--(1.15,22/14)node[right,pos=0.5]{\tiny $2/14$};
	\draw[ thick, fill=blue!30!white, pattern color=red] (0.5,16/14) rectangle (1,18/14) node [pos=.5] {\tiny $X_{12}^1$};
	\draw[arrows=<->] (1.15,16/14)--(1.15,18/14)node[right,pos=0.5]{\tiny $1/14$};
	\draw[ thick, fill=orange!30!white, pattern color=blue] (0.5,14/14) rectangle (1,16/14) node [pos=.5] {\tiny $X_{22}^1$};
	\draw[arrows=<->] (1.15,14/14)--(1.15,16/14)node[right,pos=0.5]{\tiny $1/14$};
	\draw[ thick, fill=white] (0.5,12/14) rectangle (1,1);
	\draw[ thick, fill=purple!30!white, pattern color=blue] (0.5,8/14) rectangle (1,12/14) node [pos=.5] {\tiny $X_{21}^2$};
	\draw[arrows=<->] (1.15,8/14)--(1.15,12/14)node[right,pos=0.5]{\tiny $2/14$};
	\draw[ thick, fill=orange!30!white, pattern color=blue](0.5,6/14) rectangle (1,8/14) node [pos=.5] {\tiny $X_{22}^2$};
	\draw[arrows=<->] (1.15,6/14)--(1.15,8/14)node[right,pos=0.5]{\tiny $1/14$};
	\draw[ thick, fill=red!30!white, pattern color=yellow] (0.5,2/14) rectangle (1,6/14) node [pos=.5] {\tiny $X_{24}$};
	\draw[arrows=<->] (1.15,2/14)--(1.15,6/14)node[right,pos=0.5]{\tiny $2/14$};
	\draw[ thick, fill=orange!10!white, pattern color=blue] (0.5,0) rectangle (1,2/14) node [pos=.5] {\tiny $\mathcal{L}_{2[2]}$};
	\draw[arrows=<->] (1.15,0)--(1.15,2/14)node[right,pos=0.5]{\tiny $1/14$};
	\end{scope}
\end{tikzpicture}
\caption{\it\small Achievable scheme for $L=3,\alpha=1/2$. $\mathcal{L}_{1[1]}=\mathcal{L}(X_{14},X_{22},X_{15},X_{23}),\mathcal{L}_{1[2]}=\mathcal{L}(X_{24},X_{12},X_{25},X_{13}), \mathcal{L}_{1[2]}=\mathcal{L}(X_{15},X_{23}),\mathcal{L}_{2[2]}=\mathcal{L}(X_{25},X_{13})$. The top left, top right, bottom figures are the $1^{st},2^{nd},3^{rd}$ hop respectively. The interfered layer at $\text{Rx}_{i1}$ are $X_{i4},X_{\overline{i}2},X_{i5},X_{\overline{i}3}$. It can do nothing but amplify and forward this layer. Then, the relay at the next hop, $\text{Rx}_{i[2]}$ is able to decode $W_{i4},W_{\overline{i}2}$, such that the interfered layer becomes $X_{i5},X_{\overline{i}3}$. After that, $\text{Rx}_{i[3]}$ is able to decode $W_{i5},W_{\overline{i}3}$.}\label{fig:3hop}
\end{figure}

		{\example Suppose $L=3,\alpha=\frac{1}{2}$. Then we have the sum-GDoF value $\mathcal{D}_{\Sigma,4}^{\tiny f.p.} = 20/7$. The achievable scheme is illustrated in Figure \ref{fig:3hop}. }

	In general, the achievable schemes have the following four sub-cases when $\alpha\leq1$. 
	\begin{itemize}[leftmargin=*]
	\item $\alpha\leq\frac{1}{2}$	\\
	
		In this regime, message $W_i$ is split into $2L$ sub-messages, i.e., $W_i=(W_{i1},W_{i2},\cdots,W_{i(2L)})$, which carry $
	d_{i1}=\frac{\alpha\times2^L}{2(2^L-1)}$, $d_{i2}=\frac{\alpha\times2^{L-1}}{2(2^L-1)}$, $\cdots$, $d_{iL}=\frac{\alpha\times2^{1}}{2(2^L-1)}$, $d_{iL+1}=1-2\alpha$, $d_{iL+2}=\frac{\alpha\times2^{L-1}}{2(2^L-1)}$, $\cdots$, $d_{i(2L)}=\frac{\alpha\times2^{1}}{2(2^L-1)}$ GDoF respectively. These sub-messages are encoded into independent Gaussian codebooks with codewords $X_{i1},X_{i2},\cdots,X_{i(2L)}$ with powers $\E|X_{i1}|^2=1$, $\E|X_{i2}|^2=P^{-d_{i1}}$, $\E|X_{i3}|^2=P^{-d_{i1}-d_{i2}}$, $\cdots$, $\E|X_{iL}|^2=P^{-d_{i1}-d_{i2}-\cdots-d_{i(L-1)}}$, $\E|X_{i(L+1)}|^2=P^{-d_{i1}-d_{i2}-\cdots-d_{i(L)}}$, $\E|X_{i(L+2)}|^2=P^{d_{i(L+1)}+d_{i(L+2)}+\cdots+d_{i(2L)}-1}$,$\cdots$, $\E|X_{i(2L-1)}|^2=P^{d_{i(2L-1)}+d_{i(2L)}-1}$, $E|X_{i(2L)}|^2=P^{d_{i(2L)}-1}$  respectively, upto scaling by an $O(1)$ constant to ensure a sum-power of unity. 
	The  relay  node $\text{Rx}_{1[1]}$ is able to decode $X_{11}$, $X_{12}$, $\cdots$, $X_{1L}$, $X_{1(L+1)}$, $X_{21}$ successively. The remaining signal above the noise floor of relay $\text{Rx}_{1[1]}$ is the combination of codewords: $(X_{1(L+2)}$, $\cdots$, $X_{1(2L)}$, $X_{22}$, $\cdots$, $X_{2(L)})$, and we denote the combination as $\mathcal{L}_{1[1]}$. Relay $\text{Rx}_{1[1]}$ scales this combination (amplify and forward)  such that the power of this combination is $P^{d_{1(L+2)}+\cdots+d_{1(2L)}-1}=P^{\frac{\alpha(2^L-2)}{2(2^L-1)}-1}$. Since both relays know messages $W_{11},W_{21}$, they split these two messages into two sub-messages: $W_{11}=(W_{11}^1,W_{11}^2),W_{21}=(W_{21}^1,W_{21}^2)$, where $d_{i1}^1=d_{i1}^2=\frac{d_{i1}}{2}$. Relay $\text{Tx}_{1[2]}\equiv\text{Rx}_{1[1]}$ re-encodes message $W_{12}$, $W_{11}^1$, $W_{21}^1$, $W_{13}$, $W_{14}$, $\cdots$, $W_{1L}$, $W_{1(L+1)}$, $W_{11}^2$ into independent Gaussian codebooks producing codewords $X_{12}$, $X_{11}^1$, $X_{21}^1$, $X_{13}$, $X_{14}$,$\cdots$, $X_{1L}$, $X_{1(L+1)}$, $X_{11}^2$ with powers $\E|X_{12}|^2=1$, $\E|X_{11}^1|^2=P^{-d_{12}}$, $\E|X_{21}^1|^2=P^{-d_{12}-d_{11}^1}$, $\E|X_{13}|^2=P^{-d_{12}-d_{11}^1-d_{21}^1}$, 	$\cdots$, $\E|X_{1(L+1)}|^2=P^{-d_{12}-d_{11}^1-d_{21}^1-d_{13}-\cdots-d_{1L}}$, $\E|X_{11}^2|^2=P^{d_{11}^2+d_{1(L+2)}+\cdots+d_{1(2L)}-1}-P^{d_{1(L+2)}+\cdots+d_{1(2L)}-1}$. Relay $\text{Tx}_{2[2]}\equiv\text{Rx}_{2[1]}$ proceeds similarly. The transmitted signals are $X_{1[2]}=X_{12}+X_{11}^1+X_{21}^1+X_{13}+X_{14}+\cdots+X_{1L}+X_{1(L+1)}+X_{11}^2+\mathcal{L}_{1[1]},X_{1[2]}=X_{22}+X_{11}^1+X_{21}^1+X_{23}+X_{24}+\cdots+X_{2L}+X_{2(L+1)}+X_{21}^2+\mathcal{L}_{2[1]}$. Then, the relay at the next layer: Relay $\text{Rx}_{1[2]}$ is able to decode $W_{12},W_{11}^1,W_{21}^1,W_{13},W_{14},\cdots,W_{1L},W_{1(L+1)},W_{22},W_{11}^2,W_{1(L+2)}$ successively. Compared to Relay $\text{Rx}_{1[1]}$, an additional sub-message $W_{1(L+2)}$ can also be decoded at Relay $\text{Rx}_{1[2]}$. The remaining signal above the noise floor of relay $\text{Rx}_{1[2]}$ is the combination of codewords: $(X_{1(L+3)},\cdots,X_{1(2L)},X_{23},\cdots,X_{2(L)})$, which is denoted as $\mathcal{L}_{1[2]}$ and amplified with power $P^{-1}$. Then relays $\text{Rx}_{1[2]},\text{Rx}_{2[2]}$ split $W_{12},W_{22}$, i.e., $W_{12}=(W_{12}^1,W_{12}^2),W_{22}=(W_{12}^1,W_{12}^2)$, with the corresponding GDoF $d_{i2}^1=d_{i2}^2=\frac{d_{i2}}{2}$. Relay $\text{Rx}_{1[2]}\equiv\text{Tx}_{1[3]}$ re-encodes message $W_{13}$, $W_{11}^1$, $W_{21}^1$, $W_{12}^1$, $W_{22}^1$, $W_{13}$, $W_{14},$ $\cdots,W_{1L},$ $W_{1(L+1)}$, $W_{11}^2$, $W_{12}^2$ into independent Gaussian codebooks producing codewords $X_{12}$, $X_{11}^1$, $X_{21}^1$, $X_{12}^1$, $X_{22}^1$, $X_{13}$, $X_{14},$ $\cdots,X_{1L},X_{1(L+1)},X_{11}^2,X_{12}^2$. The transmitted signals at Relay $\text{Rx}_{1[2]}\equiv\text{Tx}_{1[3]}$ is $X_{1[3]}=X_{13}+X_{11}^1+X_{21}^1+X_{12}^1+X_{22}^1+X_{14}+X_{15}+\cdots+X_{1L}+X_{1(L+1)}+X_{11}^2+X_{12}^2+X_{1(L+2)}+\mathcal{L}_{1[2]}$. This idea applies to all the subsequent layers, such that Relay $\text{Rx}_{1[l]}$ is able to decode one more sub-message $W_{i(L+l)}$ compared to the Relay  $\text{Rx}_{1[l-1]}$. In other words, the relays at the next layer can decode one more sub-message compared to the relays in the preceding layer. In the $\ell^{th}$ hop, Relay $\text{Tx}_{1[l+1]}\equiv\text{Rx}_{1[l]}$ splits the message $W_{i(i-1)}$, i.e., $W_{i(i-1)}=(W_{i(i-1)}^1,W_{i(i-1)}^2)$. The transmitted signal at Relay $\text{Tx}_{1[l+1]}\equiv\text{Rx}_{1[l]}$ is $X_{1[l+1]}=X_{1(l+1)}+X_{11}^1+X_{21}^1+X_{12}^1+X_{22}^1+\cdots +X_{1(l)}^1+X_{2(l)}^1+X_{1(l+2)}+\cdots+X_{1L}+X_{2(L+1)}+X_{11}^2+X_{12}^2+\cdots X_{1(l)}^2+\mathcal{L}_{1[l]}$. Therefore, in the last hop, Destination $\text{Rx}_{1[L]}$ can decode $W_{1(L)},W_{11}^1,W_{21}^1,W_{12}^1,W_{22}^1,\cdots,W_{1(L-1)}^1,W_{2(L-1)}^1,W_{1(L+1)},W_{11}^2,\cdots,W_{1(L-1)}^2,W_{1(2L)}$ successively.
	\item $\frac{1}{2}\leq\alpha\leq\frac{2^L}{2^{L+1}-1}$\\
	In the achievable scheme, $W_i$ is split into $(2L-1)$ sub-messages, i.e., $W_i=(W_{i1},W_{i2},\cdots,W_{i(2L-1)})$, which carry $
	\frac{(1-\alpha)\times2^L}{2(2^L-1)},\frac{(1-\alpha)\times2^{L-1}}{2(2^L-1)},\cdots,\frac{(1-\alpha)\times2^{1}}{2(2^L-1)},\frac{\alpha\times2^{L-1}}{2(2^L-1)},\cdots,\frac{\alpha\times2^{1}}{2(2^L-1)}, 2\alpha-1+\frac{1-\alpha}{2^L-1}$ GDoF respectively. The idea of message splitting,  successive decoding, partial decode-and-forward and amplify-and-forward at the relays is similar to the case $\alpha\leq\frac{1}{2}$.
	\item $\frac{2^L}{2^{L+1}-1}\leq \alpha\leq\frac{2}{3}$	\\
	In the achievable scheme, $W_i$ is split into $(2L-1)$ sub-messages, i.e., $W_i=(W_{i1},W_{i2},\cdots,W_{i(2L-1)})$, which carry $
	\frac{(1-\alpha)\times2^L}{2(2^L-1)},\frac{(1-\alpha)\times2^{L-1}}{2(2^L-1)},\cdots,\frac{(1-\alpha)\times2^{1}}{2(2^L-1)},\frac{\alpha\times2^{L-1}}{2(2^L-1)},\cdots,\frac{\alpha\times2^{1}}{2(2^L-1)}, \frac{\alpha}{2}-\frac{(1-\alpha)(2^{L-1}-2)}{2^L-1}$ GDoF respectively. The idea of message splitting,  successive decoding, partial decode-and-forward and amplify-and-forward at the relays is similar to the case $\alpha\leq\frac{1}{2}$.
	\item $\frac{2}{3}\leq\alpha\leq 1$
	The achievable scheme for this case is simple as the bound equals to $2-\alpha$, so each hop acts as the interference channel, and a simple decode and forward strategy suffices.
	\end{itemize}
	 
\subsection{Proof of Achievability for Theorem \ref{thm:Lhopvstrong}}\label{sec:Lhopvstrongach}
\begin{figure}[!t]
	\centering
\input{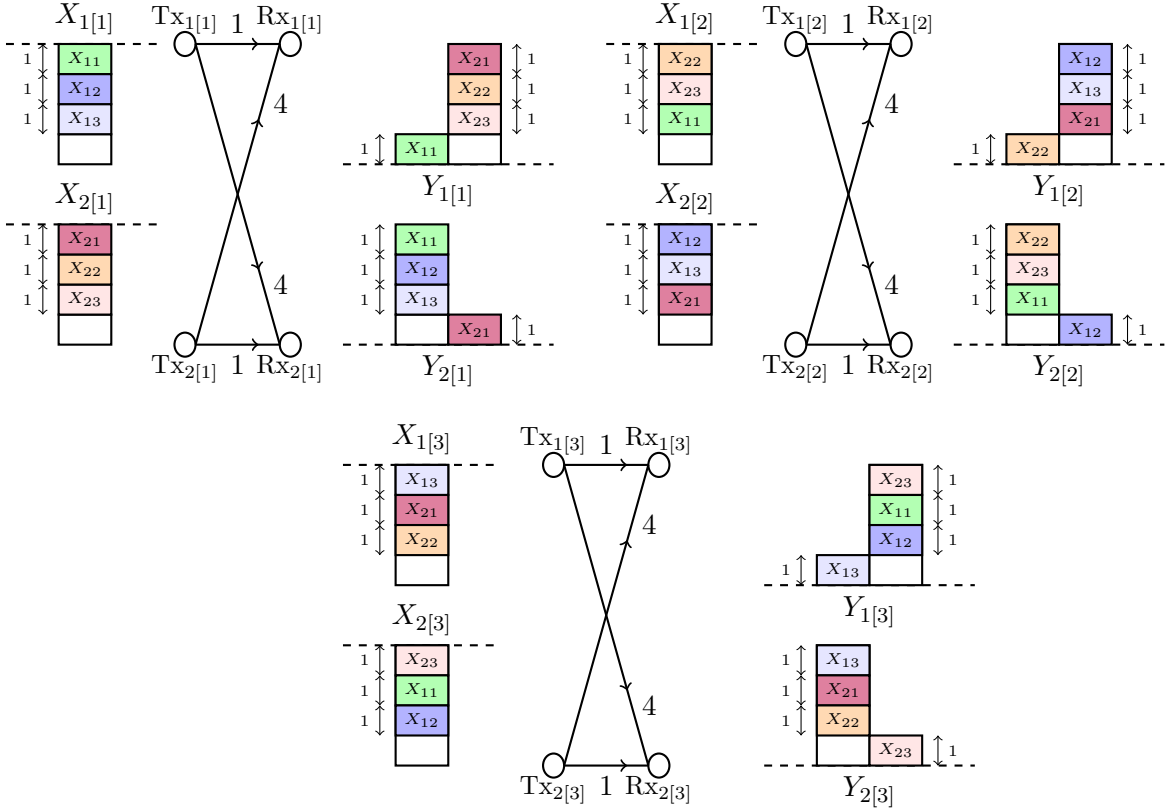}
	\caption{\it\small Achievable scheme for $L=3,\alpha=4$. }\label{fig:Lodd}
\end{figure}
The lower bound is achieved as follows: $W_1,W_2$ are split into $L$ sub-messages: $W_1$ $=$ $(W_{11}$, $W_{12}$, $\cdots$, $W_{1L})$, $W_2=(W_{21}$, $W_{22}$, $\cdots$, $W_{2L})$, and each sub-message carries $1$ GDoF. At Source $\text{Tx}_{i[1]}$,  the sub-messages $W_{i1}$, $W_{i2}$, $\cdots$, $W_{iL}$ are encoded into independent Gaussian codebooks producing codewords $X_{i1}$, $X_{i2}$, $\cdots$, $X_{iL}$ with powers $\E|X_{i1}|^2=1$, $\E|X_{i2}|^2=P^{-1}$, $\cdots$, $\E|X_{iL}|=P^{-L+1}$, upto an $O(1)$ scaling factor to ensure the sum power of unity. Then the relay node $\text{Rx}_{i[1]}$ is able to decode $W_{\bar{i}1}$, $W_{\bar{i}2}$, $\cdots$, $W_{\bar{i}L}$, $W_{i1}$ by successive decoding. Next, the relay acts as transmitter $\text{Tx}_{i[2]}$ and re-encodes $W_{\bar{i}2},\cdots,W_{\bar{i}L},W_{i1}$ into independent Gaussian codebooks producing codewords $X_{\bar{i}2}$, $\cdots$, $X_{\bar{i}L}$, $X_{i1}$
with power $\E|X_{\bar{i}2}|^2=1$, $\E|X_{\bar{i}3}|^2=P^{-1}$, $\cdots$, $\E|X_{\bar{i}L}|^2=P^{-L+2}$, $\E|X_{i1}|^2=P^{-L+1}$, respectively, upto an $O(1)$ normalizing factor. There is a simple interpretation for this scheme: the relays remove (decode and subtract) the sub-message that has the maximum power (topmost layer) and re-transmit the remaining sub-messages by keeping their original layers and scaling power to meet the maximum allowed power level. Hence, if $\ell$ is even, $\text{Rx}_{i[\ell]}$ is able to decode $W_{i\ell},\cdots,W_{iL},W_{\bar{i}1},\cdots,W_{\bar{i}L}$ successively. Then as a transmitter, $\text{Tx}_{i[\ell+1]}$ transmits   $W_{i(\ell+1)},\cdots,W_{iL},W_{\bar{i}1},\cdots,W_{\bar{i}L}$. If $\ell$ is odd, $\text{Rx}_{i[\ell]}$ is able to decode $W_{\bar{i}\ell},\cdots,W_{\bar{i}L},W_{i1},\cdots,W_{i\ell}$ successively, and then as a transmitter, $\text{Tx}_{i[\ell+1]}$ transmits   $W_{\bar{i}(\ell+1)},\cdots,W_{\bar{i}L},W_{i1},\cdots,W_{i\ell}$. Hence, at the last hop, where $L$ is odd, destination $\text{Rx}_{i[L]}$ is able to decode $W_{\bar{i}L},W_{i1},\cdots,W_{iL}$ successively. An example of $L=3,\alpha=4$ is illustrated in Figure \ref{fig:Lodd}.

	\section{Conclusion}\label{sec:con}
	Motivated by the need to understand the robust information-theoretic limits of multihop communication networks, in this work we initiated the study of the sum-GDoF of layered symmetric $L$-hop $2\times 2\times \cdots\times 2$ networks comprised of $2$ nodes in each layer, under finite precision CSIT. As our main contribution, recently introduced sum-set inequalities \cite{Arash_Jafar_sumset} that build upon Aligned Images bounds of \cite{Arash_Jafar} were shown to be sufficient to settle the sum-GDoF of this symmetric setting. Notable technical issues that surfaced in our study include the challenge of applying deterministic transformations that were developed for one-hop communication to multihop settings, as well as the dependence of coding functions on CSIT that may be available for previous hops and need not satisfy the finite precision assumption. These challenges were overcome through recursive reasoning that applies the deterministic transformation to only one hop at a time. In terms of optimal solutions, under finite precision CSIT we found that ideas such as Interference Neutralization \cite{Cao_Chen,Simeone_mesh, Mohajer_Diggavi_Fragouli_Tse, Issa_Fong_Avestimehr_linear_twohop, Shomorony_Avestimehr}, Aligned Interference Neutralization \cite{Gou_Wang_Jafar_Jeon_Chung, Wang_Gou_Jafar_MU, Gou_Wang_Jafar_NL} and Network Diagonalization \cite{Shomorony_Avestimehr_K} are too fragile to retain their GDoF benefits, and instead rate-splitting solutions that combine amplify-and-forward and decode-and-forward principles, along with careful layering (superposition) of messages that allows each successive stage of relays to acquire more common information, are sum-GDoF optimal. The compact expressions obtained from converse bounds prove insightful in designing the optimal achievable schemes. From the big-picture perspective, a  takeaway message from the sum-GDoF characterizations is that, on  one hand, optimal robust solutions tend to not improve much upon basic alternatives (in this case, the trivial decode-and-forward solution) when all channels are of comparable strength, but on the other hand, when the channel strengths are sufficiently different then significant gains over basic alternatives are possible by optimizing robust solutions. The latter is particularly important  for the high-frequency communication networks that motivated this work, where due to high path loss, blockages,  and often due to directional transmission, there tends to be a much higher spatial variance in channel strengths than in conventional cellular networks. Thus, the results of this work, while clearly limited by the simplifying assumptions of layered structure and symmetric gains, nevertheless indicate that significant robust gains  are possible by optimizing multihop communication for the types of richly diverse topologies that would be typical in high-frequency directed communication networks. From the information theoretic perspective, a promising path forward would be through more elaborate GDoF studies that explore asymmetric and/or unlayered topologies, perhaps by utilizing extremal network theory principles as in \cite{Yoga_Junge_Jafar} to counter the problem of parameter explosion.

	\section{Appendix A}
	In this section, let us summarize the proof of the bound $I(W_1;Y_{1[\ell]}^{[N]}\mid \mathcal{G}_{[1:L]})
	\leq I(W_1,\overline{Y}_{1[\ell]}^{[N]}\mid \mathcal{G}_{[1:L]})$, which converts from the original canonical channel to the deterministic channel with conditioning on $\mathcal{G}_{[1:L]}$ instead of $\mathcal{G}_{[\ell]}$.\footnote{Since the channels $\mathcal{G}_{[\ell+1:L]}$ do not appear until after the $\ell^{th}$ hop, the conditioning can be reduced to $\mathcal{G}_{[1:\ell]}$ trivially.} The other term $I(W_2;Y_{2[\ell]}^{[N]}\mid \mathcal{G}_{[1:L]})
	\leq I(W_2,\overline{Y}_{2[\ell]}^{[N]}\mid \mathcal{G}_{[1:L]})$ follows symmetrically. Following  the proof that appears in the Appendix section of \cite{Arash_Jafar}, the first step is to limit the input and output to integers. Let an intermediate deterministic channel model of the $\ell^{th}$ hop have integer inputs $\lfloor X_{1[\ell]}^{[N]}\rfloor,\lfloor X_{2[\ell]}^{[N]}\rfloor$, and integer outputs\footnote{By integer-valued inputs and outputs, we mean that the real and imaginary parts of these inputs and outputs are integer valued.}
	\begin{align}
		\bar{\bar{Y}}_{1[\ell]}(n)=\lfloor \overline{P}^{1-\max(1,\alpha)}G_{11[\ell]}\lfloor X_{1[\ell]}(n)\rfloor\rfloor+\lfloor \overline{P}^{\alpha-\max(1,\alpha)}G_{12[\ell]}\lfloor X_{2[\ell]}(n)\rfloor\rfloor
	\end{align}
	while the  original canonical channel is
	\begin{align}
		Y_{1[\ell]}(n)=\overline{P}^{1-\max(1,\alpha)}G_{11[\ell]}X_{1[\ell]}(n)+\overline{P}^{\alpha-\max(1,\alpha)}G_{12[\ell]}X_{2[\ell]}(n)+Z_{1[\ell]}(n)
	\end{align}
	Define \begin{align}
	E_{1[\ell]}^{[N]}&=Y_{1[\ell]}^{[N]}-\bar{\bar{Y}}_{1[\ell]}^{[N]}\notag\\
	&=\overline{P}^{1-\max(1,\alpha)}G_{11[\ell]}^{[N]}X_{1[\ell]}^{[N]}-\lfloor \overline{P}^{1-\max(1,\alpha)}G_{11[\ell]}\lfloor X_{1[\ell]}^{[N]}\rfloor\rfloor\notag\\
	&\hspace{1cm}+\overline{P}^{\alpha-\max(1,\alpha)}G_{12[\ell]}^{[N]}X_{2[\ell]}^{[N]}-\lfloor \overline{P}^{\alpha-\max(1,\alpha)}G_{12[\ell]}\lfloor X_{2[\ell]}^{[N]}\rfloor\rfloor+Z_{1[\ell]}^{[N]}
	\end{align}
	Then we have,
	\begin{align}
		I(W_1;Y_{1[\ell]}^{[N]}\mid \mathcal{G}_{[1:L]})&=I(W_1;\bar{\bar{Y}}_{1[\ell]}^{[N]}+E_{1[\ell]}^{[N]}\mid \mathcal{G}_{[1:L]})\\
		&\leq I(W_1;\bar{\bar{Y}}_{1[\ell]}^{[N]}\mid \mathcal{G}_{[1:L]})+I(W_1;E_{1[\ell]}^{[N]}\mid \bar{\bar{Y}}_{1[\ell]}^{[N]}, \mathcal{G}_{[1:L]})\\
		&\leq I(W_1;\bar{\bar{Y}}_{1[\ell]}^{[N]}\mid \mathcal{G}_{[1:L]})+h(E_{1[\ell]}^{[N]}\mid \mathcal{G}_{[1:L]})-h(Z_{1[\ell]}^{[N]})\label{ineq:11}\\
		&\leq I(W_1;\bar{\bar{Y}}_{1[\ell]}^{[N]}\mid \mathcal{G}_{[1:L]})+\sum_{n=1}^{N}\big[h(E_{1[\ell]}(n)\mid\mathcal{G}_{[1:L]})-h(Z_{1[\ell]}(n))\big]\\
		&\leq I(W_1;\bar{\bar{Y}}_{1[\ell]}^{[N]}\mid \mathcal{G}_{[1:L]})+\sum_{n=1}^{N}\big[h(E_{1[\ell]}\mid\mathcal{G}_{[\ell]})-h(Z_{1[\ell]})\big]\\
		&\leq I(W_1;\bar{\bar{Y}}_{1[\ell]}^{[N]}\mid \mathcal{G}_{[1:L]})+\sum_{n=1}^{N}E_{G_{11[\ell]},G_{12[\ell]}}\big[\log\big((G_{11}(n)+G_{12}(n))^2+1\big)\big]
	\end{align}
	where \eqref{ineq:11} requires $I(Z_{1[\ell]};\mathcal{G}_{[1:L]})=0$, which naturally holds as noise is independent of all messages and channel coefficients. So the difference between $I(W_1;Y_{1[N]}^{[N]}\mid \mathcal{G}_{[1:L]})$ and $I(W_1;\bar{\bar{Y}}_{1[N]}^{[N]}+E_{1[\ell]}^{[N]}\mid \mathcal{G}_{[1:L]})$ approaches $0$ when normalized by $N\log(P)$ as $G_{11[\ell]}(n),G_{12[\ell]}(n)$ are bounded above by some constant $\Delta$. Thus, the integer input and output channel with the per-codeword power constraints,
	\begin{align}
		\sum_{t=1}^{N}\big((\lfloor X_{1[\ell]}\rfloor)^2\big)&\leq NP^{\max(1,\alpha)}\\
		\sum_{t=1}^{N}\big((\lfloor X_{2[\ell]}\rfloor)^2\big)&\leq NP^{\max(1,\alpha)}
	\end{align}
	achieves at least the same GDoF as the original canonical channel model. 
	
	The next step is to convert the per-codeword power constraints into per-symbol power constraints. Let us define 
	\begin{align}
		\overline{X}_{1[\ell]}(n)&=\lfloor X_{1[\ell]}(n)\rfloor \mod \lceil \overline{P}^{\max(1,\alpha)}\rceil\label{symbolcons1}\\
		\overline{X}_{2[\ell]}(n)&=\lfloor X_{2[\ell]}(n)\rfloor \mod \lceil \overline{P}^{\max(1,\alpha)}\rceil\label{symbolcons2}\\
		\overline{Y}_{1[\ell]}(n)&=\lfloor \overline{P}^{1-\max(1,\alpha)}G_{11[\ell]}(n)\overline{X}_{1[\ell]}(n)\rfloor+\lfloor \overline{P}^{\alpha-\max(1,\alpha)}G_{12[\ell]}(n)\overline{X}_{2[\ell]}\rfloor\label{symbolcons3}\\
		\hat{Y}_{1[\ell]}(n)&=\bar{\bar{Y}}_{1[\ell]}(n)-\overline{Y}_{1[\ell]}(n)\label{symbolcons4}
	\end{align}
	Now we have,
	\begin{align}
		I(W_1;\bar{\bar{Y}}_{1[\ell]}^{[N]}\mid \mathcal{G}_{[1:L]}^{[N]})&=I(W_1;\overline{Y}_{1[\ell]}^{[N]}+\hat{Y}_{1[\ell]}^{[N]}\mid\mathcal{G}_{[1:L]})\\
		&\leq I(W_1;\overline{Y}_{1[\ell]}^{[N]},\hat{Y}_{1[\ell]}^{[N]}\mid\mathcal{G}_{[1:L]})\\
		&\leq I(W_1;\overline{Y}_{1[\ell]}^{[N]}\mid\mathcal{G}_{[1:L]})+H(\hat{Y}_{1[\ell]}^{[N]}\mid\mathcal{G}_{[1:L]})\\
		&\leq I(W_1;\overline{Y}_{1[\ell]}^{[N]}\mid\mathcal{G}_{[1:L]})+H(\hat{Y}_{1[\ell]}^{[N]}\mid\mathcal{G}_{[\ell]})
	\end{align}
	It can now be shown that the term $H(\hat{Y}_{1[\ell]}^{[N]}\mid\mathcal{G}_{[\ell]})$ is negligible in the GDoF sense following the same proof as in  \cite[eq. (124)-(149)]{Arash_Jafar}. Therefore, at the $\ell^{th}$ hop, replacing the long-term (per-codeword) power constraint with short-term (per-symbol) power constraint will not reduce the GDoF value.
	\section{Appendix B}\label{vectordeter}
	In this section we are going to prove $I(W_1,W_2;Y_{1[\ell]}^{[N]},Y_{2[\ell]}^{[N]}\mid \mathcal{G}_{[1:\ell]})
	\leq I(W_1,W_2,\overline{Y}_{1[\ell]}^{[N]},\overline{Y}_{2[\ell]}^{[N]}\mid \mathcal{G}_{[1:\ell]})$. Let the deterministic channel model of the $\ell^{th}$ hop have integer inputs $\lfloor X_{1[\ell]}^{[N]}\rfloor,\lfloor X_{2[\ell]}^{[N]}\rfloor$, and output
	\begin{align}
	\bar{\bar{Y}}_{1[\ell]}(n)&=\lfloor \overline{P}^{1-\max(1,\alpha)}G_{11[\ell]}\lfloor X_{1[\ell]}(n)\rfloor\rfloor+\lfloor \overline{P}^{\alpha-\max(1,\alpha)}G_{12[\ell]}\lfloor X_{2[\ell]}(n)\rfloor\rfloor\\
	\bar{\bar{Y}}_{2[\ell]}(n)&=\lfloor \overline{P}^{\alpha-\max(1,\alpha)}G_{21[\ell]}\lfloor X_{1[\ell]}(n)\rfloor\rfloor+\lfloor \overline{P}^{1-\max(1,\alpha)}G_{22[\ell]}\lfloor X_{2[\ell]}(n)\rfloor\rfloor
	\end{align}
	while the  original canonical channel is
	\begin{align}
	Y_{1[\ell]}(n)&=\overline{P}^{1-\max(1,\alpha)}G_{11[\ell]}X_{1[\ell]}(n)+\overline{P}^{\alpha-\max(1,\alpha)}G_{12[\ell]}X_{2[\ell]}(n)+Z_{1[\ell]}(n)\\
	Y_{2[\ell]}(n)&=\overline{P}^{\alpha-\max(1,\alpha)}G_{21[\ell]}X_{1[\ell]}(n)+\overline{P}^{1-\max(1,\alpha)}G_{22[\ell]}X_{2[\ell]}(n)+Z_{2[\ell]}(n)
	\end{align}
	Then, defining $E_{i[\ell]}^{[N]}=Y_{i[\ell]}^{[N]}-\bar{\bar{Y}}_{i[\ell]}^{[N]}$, we have
	\begin{align}
	&I(W_1,W_2;Y_{1[\ell]}^{[N]},Y_{2[\ell]}^{[N]}\mid \mathcal{G}_{[1:\ell]})\notag\\&=I(W_1,W_2;\bar{\bar{Y}}_{1[\ell]}^{[N]}+E_{1[\ell]}^{[N]},\bar{\bar{Y}}_{2[\ell]}^{[N]}+E_{2[\ell]}^{[N]}\mid \mathcal{G}_{[1:\ell]})\\
	&\leq I(W_1,W_2;\bar{\bar{Y}}_{1[\ell]}^{[N]},E_{1[\ell]}^{[N]},\bar{\bar{Y}}_{2[\ell]}^{[N]},E_{2[\ell]}^{[N]}\mid \mathcal{G}_{[1:\ell]})\\
	&= I(W_1,W_2;\bar{\bar{Y}}_{1[\ell]}^{[N]},\bar{\bar{Y}}_{2[\ell]}^{[N]}\mid \mathcal{G}_{[1:\ell]})+I(W_1,W_2;E_{1[\ell]}^{[N]},E_{2[\ell]}^{[N]}\mid \bar{\bar{Y}}_{1[\ell]}^{[N]},\bar{\bar{Y}}_{2[\ell]}^{[N]}, \mathcal{G}_{[1:\ell]})\\
	&=I(W_1,W_2;\bar{\bar{Y}}_{1[\ell]}^{[N]},\bar{\bar{Y}}_{2[\ell]}^{[N]}\mid \mathcal{G}_{[1:\ell]})+h(E_{1[\ell]}^{[N]},E_{2[\ell]}^{[N]}\mid \bar{\bar{Y}}_{1[\ell]}^{[N]},\bar{\bar{Y}}_{2[\ell]}^{[N]}, \mathcal{G}_{[1:\ell]})\notag\\&\hspace{2cm}-h(E_{1[\ell]}^{[N]},E_{2[\ell]}^{[N]}\mid W_1,W_2,\bar{\bar{Y}}_{1[\ell]}^{[N]},\bar{\bar{Y}}_{2[\ell]}^{[N]}, \mathcal{G}_{[1:\ell]})\\
	&\leq I(W_1,W_2;\bar{\bar{Y}}_{1[\ell]}^{[N]},\bar{\bar{Y}}_{2[\ell]}^{[N]}\mid \mathcal{G}_{[1:\ell]})+h(E_{1[\ell]}^{[N]},E_{2[\ell]}^{[N]}\mid\mathcal{G}_{[1:\ell]})-h(Z_{1[\ell]}^{[N]},Z_{2[\ell]}^{[N]})\label{ineq:21}\\
	&\leq I(W_1,W_2;\bar{\bar{Y}}_{1[\ell]}^{[N]},\bar{\bar{Y}}_{2[\ell]}^{[N]}\mid \mathcal{G}_{[1:\ell]})+h(E_{1[\ell]}^{[N]}\mid\mathcal{G}_{[1:\ell]})-h(Z_{1[\ell]}^{[N]})+h(E_{2[\ell]}^{[N]}\mid\mathcal{G}_{[1:\ell]})-h(Z_{2[\ell]}^{[N]})\label{ineq:22}\\
	&= I(W_1,W_2;\bar{\bar{Y}}_{1[\ell]}^{[N]},\bar{\bar{Y}}_{2[\ell]}^{[N]}\mid \mathcal{G}_{[1:\ell]})+\sum_{i=1}^{2}\sum_{n=1}^{N}\big[h(E_{i[\ell]}\mid\mathcal{G}_{[1:\ell]})-h(Z_{i[\ell]})\big]\\
	&\leq I(W_1,W_2;\bar{\bar{Y}}_{1[\ell]}^{[N]},\bar{\bar{Y}}_{2[\ell]}^{[N]}\mid \mathcal{G}_{[1:\ell]})+\sum_{i=1}^{2}\sum_{n=1}^{N}E_{G_{i1[\ell]},G_{i2[\ell]}}\big[\log\big((G_{i1[\ell]}(n)+G_{i2[\ell]}(n))^2+1\big)\big]
	\end{align}
	where \eqref{ineq:21} requires $I(Z_{1[\ell]},Z_{2[\ell]};\mathcal{G}_{[1:\ell]})=0$, which naturally holds as noise is independent of the channel coefficients. \eqref{ineq:22} holds because the noise terms are independent. So the difference between $I(W_1,W_2;Y_{1[\ell]}^{[N]},Y_{2[N]}^{[N]}\mid \mathcal{G}_{[1:\ell]})$ and $I(W_1,W_2,\bar{\bar{Y}}_{1[\ell]}^{[N]},\bar{\bar{Y}}_{2[\ell]}^{[N]}\mid \mathcal{G}_{[1:\ell]})$ approaches $0$ when normalized by $N\log(P)$ as $G_{i1[\ell]}(n),G_{i2[\ell]}(n)$ are bounded above by some constant $\Delta$. 
	The next step is to convert the per-codeword power constraints into per-symbol power constraints, using the definitions 
	\begin{align}
	\overline{X}_{1[\ell]}(n)&=\lfloor X_{1[\ell]}(n)\rfloor \mod \lceil \overline{P}^{\max(1,\alpha)}\rceil\\
	\overline{X}_{2[\ell]}(n)&=\lfloor X_{2[\ell]}(n)\rfloor \mod \lceil \overline{P}^{\max(1,\alpha)}\rceil\\
	\overline{Y}_{1[\ell]}(n)&=\lfloor \overline{P}^{1-\max(1,\alpha)}G_{11[\ell]}(n)\overline{X}_{1[\ell]}(n)\rfloor+\lfloor \overline{P}^{\alpha-\max(1,\alpha)}G_{12[\ell]}(n)\overline{X}_{2[\ell]}\rfloor\\
	\overline{Y}_{2[\ell]}(n)&=\lfloor \overline{P}^{\alpha-\max(1,\alpha)}G_{21[\ell]}(n)\overline{X}_{1[\ell]}(n)\rfloor+\lfloor \overline{P}^{1-\max(1,\alpha)}G_{22[\ell]}(n)\overline{X}_{2[\ell]}\rfloor\\
	\hat{Y}_{1[\ell]}(n)&=\bar{\bar{Y}}_{1[\ell]}(n)-\overline{Y}_{1[\ell]}(n)\\
	\hat{Y}_{2[\ell]}(n)&=\bar{\bar{Y}}_{2[\ell]}(n)-\overline{Y}_{2[\ell]}(n)
	\end{align}
	Now we have, 
	\begin{align}
	I(W_1,W_2;\bar{\bar{Y}}_{1[\ell]}^{[N]},\bar{\bar{Y}}_{2[\ell]}^{[N]}\mid \mathcal{G}_{[1:\ell]}^{[N]})&=I(W_1,W_2;\overline{Y}_{1[\ell]}^{[N]}+\hat{Y}_{1[\ell]}^{[N]},\overline{Y}_{2[\ell]}^{[N]}+\hat{Y}_{2[\ell]}^{[N]}\mid\mathcal{G}_{[1:\ell]})\\
	&\leq I(W_1,W_2;\overline{Y}_{1[\ell]}^{[N]},\hat{Y}_{1[\ell]}^{[N]},\overline{Y}_{2[\ell]}^{[N]},\hat{Y}_{2[\ell]}^{[N]}\mid\mathcal{G}_{[1:\ell]})\\
	&\leq I(W_1,W_2;\overline{Y}_{1[\ell]}^{[N]},\overline{Y}_{2[\ell]}^{[N]}\mid\mathcal{G}_{[1:\ell]})+H(\hat{Y}_{1[\ell]}^{[N]},\hat{Y}_{1[\ell]}^{[N]}\mid\mathcal{G}_{[1:\ell]})\\
	&\leq I(W_1,W_2;\overline{Y}_{1[\ell]}^{[N]},\overline{Y}_{2[\ell]}^{[N]}\mid\mathcal{G}_{[1:\ell]})+\sum_{i=1}^{2}H(\hat{Y}_{1[\ell]}^{[N]}\mid\mathcal{G}_{[\ell]})
	\end{align}
	It can now be shown that the term $H(\hat{Y}_{i[\ell]}^{[N]}\mid\mathcal{G}_{[\ell]})$ is negligible in the GDoF sense following the same idea as in  \cite[eq. (124)-(149)]{Arash_Jafar}, which completes the proof.
	
	\bibliography{Thesis}

\end{document}